%% file: arxiv_apla.tex
\documentclass[letter,11pt]{article}
\input{macros.tex}

\newif\iffigures
\figurestrue 

\usepackage{rotating}
\usepackage{booktabs}

\usepackage{amssymb,upgreek,nicefrac,amsmath}
\usepackage{sgame}
\usepackage[mathscr]{eucal}
\usepackage[dvips]{epsfig}
\usepackage[dvips]{graphicx}
\usepackage[section]{placeins}
\usepackage{lipsum}
\usepackage{amsfonts}
\usepackage{graphicx}
\usepackage{epstopdf}
\usepackage{algorithm} 
\usepackage{algorithmic}
\usepackage{tikz,verbatim}
\usepackage{pgfplots}
\usepackage{subcaption}
\usepgfplotslibrary{fillbetween}
\usetikzlibrary{arrows}
\usetikzlibrary{shapes}
\usetikzlibrary{positioning}
\usetikzlibrary{calc}
\usetikzlibrary{intersections}
\usetikzlibrary{decorations.markings}
\usepackage{multirow} 
\usepackage{setspace}
\usepackage{soul}
\usepackage{authblk}

\ifpdf
  \DeclareGraphicsExtensions{.eps,.pdf,.png,.jpg}
\else
  \DeclareGraphicsExtensions{.eps}
\fi

\newif\ifproofs
\proofstrue

\begin{document}

\title{Aspiration-based Perturbed Learning Automata in Weakly-Acyclic Games with Noisy Utility Measurements\thanks{This work has been partially supported by the European Union grant EU H2020-ICT-2014-1 project RePhrase (No. 644235). The research reported in this paper has been partly funded by the Federal Ministry for Innovation, Mobility and Infrastructure (BMIMI), the Federal Ministry for Economy, Energy and Tourism (BMWET), and the State of Upper Austria in the frame of the SCCH competence center INTEGRATE [(FFG grant no. 892418)] in the COMET - Competence Centers for Excellent Technologies Programme managed by Austrian Research Promotion Agency FFG.}}
    
    \author[1]{Georgios C. Chasparis\thanks{Corresponding Author: georgios.chasparis@scch.at}

   \affil[1]{{\small Software Competence Center Hagenberg GmbH, Softwarepark 21, A$-$4232 Hagenberg, Austria.}}
}

\date{November 23, 2025 \\ August 29th, 2026 (revised)}

\maketitle

\begin{abstract}
Reinforcement-based learning dynamics may exhibit several limitations when applied in a distributed setup. In (repeatedly-played) multi-player/action strategic-form games, and when each player applies an independent copy of the learning dynamics, convergence to (usually desirable) pure Nash equilibria cannot be guaranteed. Prior work has only focused on a small class of games, namely potential and coordination games. Furthermore, strong convergence guarantees (i.e., almost sure convergence or weak convergence) are mostly restricted to two-player games. To address this main limitation of reinforcement-based learning in repeatedly-played strategic-form games, this paper introduces a novel payoff-based learning scheme for distributed optimization in multi-player/action strategic-form games. We present an extension of perturbed learning automata (PLA), namely aspiration-based perturbed learning automata (APLA), in which each player's probability distribution for selecting actions is reinforced both by repeated selection and an aspiration factor that captures the player's satisfaction level. We provide a stochastic stability analysis of APLA in multi-player positive-utility weakly-acyclic games under the presence of noisy observations. We provide conditions under which convergence is attained (in weak sense) to the set of pure Nash equilibria. A methodology is also derived for calculating the stochastically stable equilibria through the aspiration action functional minimization, which simplifies the derivation of stochastically stable states. Conditions can then be derived for convergence to the Pareto efficient Nash equilibria. To the best of our knowledge, this is the first reinforcement-based learning scheme that provides global convergence guarantees in weakly-acyclic games and in a fully-distributed setup. A Monte-Carlo simulation study validates the derived conclusions.
\end{abstract}

\textbf{Keywords:} distributed optimization, learning automata, multi-agent systems, noisy observations, reinforcement learning, weakly-acyclic games

\section{Introduction} \label{sec:Introduction}

In several cases, large scale optimization problems need to be tackled through distributed or decentralized formulations. This might be due to communication limitations (i.e., sharing the global state of the problem might be prohibitive) or due to the high computational complexity of centralized optimization problems. Multi-agent formulations can be used to tackle optimization problems in a distributed fashion, where the original group objective is decomposed into multiple local (agent-based) objectives. However, due to the interdependencies among agents' utility functions, \emph{local} (or \emph{distributed}) optimization does not necessarily imply \emph{global} (or \emph{centralized}) optimization (i.e., maximization of the sum of all agents' utilities). The problem becomes even more challenging when the utility function of each agent is unknown, and only measurements of this function (possibly corrupted by noise) are available. 

Naturally, several such distributed optimization problems can be formulated as strategic-form games. A rather common objective is then to derive conditions under which convergence to \emph{efficient Nash equilibria} can be achieved, i.e., locally stable outcomes that also maximize a centralized objective. One large class of payoff-based learning dynamics that has been utilized for convergence to Nash equilibria is \emph{reinforcement-based learning}. It may appear under alternative forms, including \emph{discrete-time replicator dynamics} \cite{Arthur93}, \emph{learning automata} \cite{Tsetlin73,Narendra89} or \emph{approximate policy iteration} or \emph{$Q$-learning} \cite{hu_nash_2003}. It is highly attractive to several engineering applications, since agents do not need to know neither the actions of other agents, nor their own utility function. For example, it has been utilized for system identification and pattern recognition \cite{ThathacharSastry04}, distributed network formation and resource-allocation problems \cite{ChasparisShamma11_DGA}.

In reinforcement-based learning, deriving conditions for convergence to Nash equilibria may not be a trivial task, especially in the case of large number of agents. In particular and in the context of \emph{coordination games} (e.g.,~\cite{ChasparisAriShamma13_SIAM}), two main difficulties arise: a) excluding convergence to pure strategies that are \emph{not} Nash equilibria, and b) excluding convergence to mixed strategy profiles. Recent work by the author on \emph{perturbed learning automata} (PLA) \cite{chasparis_stochastic_2019}, overcame these limitations by directly characterizing the stochastically stable states of the induced Markov chain (independently of the number of players or actions). This type of analysis allows for acquiring convergence guarantees in multi-player \emph{coordination games} (thus, extending previous results in reinforcement-based learning restricted only to potential games). 

The convergence guarantees presented in \cite{chasparis_stochastic_2019} apply to games with strict structural properties, namely the \emph{coordination property}. To extend such convergence guarantees to a larger class of games, a novel class of dynamics is presented in this paper, namely \emph{aspiration-based perturbed learning automata} (briefly, APLA). While in standard reinforcement-based learning, actions are reinforced by repeated selection and proportionally to the received reward, in APLA, the reinforcement level is also scaled by agent's own satisfaction level. For this class of reinforcement-based learning, we provide convergence guarantees for any weakly-acyclic game satisfying the positive utility property (independently of the number of players or actions). Furthermore, we present a methodology for efficiently computing the stochastically stable states, through which conditions for convergence to the Pareto efficient Nash equilibria can also be derived. The presented analysis also addresses the possibility of utilities perturbed by small bounded noise. Lastly, we provide a specialization of the results to the classical Stag-Hunt game, together with a simulation study. In the supplementary material, an additional simulation study is also provided in the class of load-balancing games.

In the remainder of the paper, Section~\ref{sec:APLA} presents the APLA learning dynamics. Section~\ref{sec:RelatedWorkContributions} discusses related work and the main contributions of this paper.  Section~\ref{sec:WeaklyAcyclicGames} defines weakly-acyclic games. Section~\ref{sec:StochasticStability} presents the stochastic stability analysis background, following \cite{chasparis_stochastic_2019}. Section~\ref{sec:ApproximationFiniteMarkovProcess} then presents an approximation of the induced finite-state Markov process and an approximation of the computation of stochastically stable states through the notion of resistance of one-step transitions.  Section~\ref{sec:StochasticStabilityWeaklyAcyclicGames} specializes the stochastic stability analysis in the context of weakly-acyclic games, also introducing the notion of aspiration action-functional. Section~\ref{sec:StagHuntSimulationStudy} presents a simulation study together with a specialization of the analysis in the Stag-Hunt coordination game. To further provide the effectiveness of the method in a real-world application scenario, an extensive simulation analysis is presented for load-balancing games in Sections~\ref{sec:SM:LoadBalancingGames}, \ref{sec:SM:SimulationsLoadBalancingGames} of the Supplementary Material. Finally, Section~\ref{sec:Conclusions} presents concluding remarks.

{\bf Notation:}
\begin{itemize}
\item For a Euclidean topological space $\cZ\subset\mathbb{R}^{n}$, let $\Neigh{\delta}{x}$ be the $\delta$-neighborhood of $x\in\cZ$, i.e., $\Neigh{\delta}{x} \df \{y\in\cZ:|x-y|<\delta\},$
where $|\cdot|$ denotes the Euclidean distance.
\item $e_j$ denotes the \emph{unit vector} in $\mathbb{R}^{n}$ where its $j$th entry is equal to 1 and all other entries is equal to 0.
\item $\Delta(n)$ denotes the \emph{probability simplex} of dimension $n$, i.e.,
$\Delta(n) \df \left\{ x\in\mathbb{R}^{n} : x\geq{0}, \mathbf{1}\tr x=1 \right\}.$

\item $\Dirac{x}$ denotes the Dirac measure at $x$.

\item For a finite set $A$, $\magn{A}$ denotes its cardinality.

\item Let $\sigma\in\Delta(\magn{A})$ be a finite probability distribution for some finite set $A$. The random selection of an element of $A$ will be denoted ${\rm rand}_{\sigma}[A]$. If $\sigma=(\nicefrac{1}{\magn{A}},...,\nicefrac{1}{\magn{A}})$, i.e., it corresponds to the uniform distribution, the random selection will be denoted by ${\rm rand}_{\rm unif}[A]$.

\item For any set $A$ in a probability space, $A^c$ denotes its complement.

\end{itemize}

\section{Aspiration-based Perturbed Learning Automata (APLA)}	\label{sec:APLA}

\subsection{Algorithm}

The proposed dynamics is presented in Algorithm~\ref{Al:APLA}. This class of dynamics is an advancement of the \emph{perturbed learning automata} (PLA) introduced in \cite{ChasparisShamma11_DGA,ChasparisShammaRantzer15}.

\begin{algorithm}		
\caption{Aspiration-based Perturbed Learning Automata (APLA)}
\label{Al:APLA}
\begin{algorithmic}
\STATE{Define $\alpha_i(t)=\alpha_i$, $x_i(t)=x_i$ and $\rho_i(t)=\rho_i$ as the current action, strategy and aspiration-level of agent/player $i$, respectively, at time $t\in\mathbb{N}$. The following steps are executed recursively by each agent.}
\FOR{$t=1,2,...$}

\STATE{(\emph{\textbf{action update}}) Agent $i$ selects a new action as follows: 
\begin{eqnarray}	\label{eq:ActionUpdate}
\alpha_i(t+1) = \begin{cases}
{\rm rand}_{x_i}[\mathcal{A}_i], & \mbox{ with probability } 1-\lambda, \cr
{\rm rand}_{\rm unif}[\mathcal{A}_i], & \mbox{ with probability } \lambda,
\end{cases}
\end{eqnarray} 
for some small perturbation factor $\lambda>0$.
}

\STATE{(\emph{\textbf{evaluation}}) Agent $i$ applies $\alpha_i^+\df\alpha_i(t+1)$ and receives a measurement of its utility function $\rewardpert{i}=\rewardpert{i}(\alpha^+)>0$, which depends on the action profile $\alpha^+=(\alpha_1^+,...,\alpha_{n}^+)$ of all agents. The utility is assumed to be a noisy measurement of a nominal utility function of the form
\begin{equation}	\label{eq:UtilityPerturbed}
\rewardpert{i}(\alpha^+) = \reward{i}(\alpha^+) + \noise_i
\end{equation}
for a bounded noise term $\noise_i\in[-\snoise,\snoise]$ for all $i\in\cI$.}

\STATE{(\emph{\textbf{strategy update}}) Agent $i$ revises its strategy vector $x_i\in\Delta(\magn{\mathcal{A}_i})$ to a new strategy $x_i^+\df x_i(t+1)$ as follows: 
\begin{eqnarray}	\label{eq:ReinforcementLearningModel}
x_i^+ & = & x_i + \epsilon \cdot (e_{\alpha_i^+} - x_i ) \cdot \phi_i\left(\rewardpert{i},\rewardpert{i}-\rho_i\right)
\end{eqnarray}
for some constant step size $\epsilon>0$. The term $\phi_i\in\mathbb{R}_+$ is the \emph{aspiration factor} of player $i$, defined as follows 
\begin{eqnarray}	\label{eq:AspirationTerm}
\phi_{i}(\rewardpert{i},y) \df \begin{cases}
\rewardpert{i} , & y \geq{0} \\
\max\{h,\rewardpert{i} + \zeta y\}, & y < 0
\end{cases} 
\end{eqnarray}
for some positive constants $h>0,$ $\zeta>0$.}

\STATE{(\emph{\textbf{aspiration-level update}}) Agent $i$ revises its aspiration level $\rho_i\in[\underline{\rho},\overline{\rho}]$ to a new aspiration-level $\rho_i^+\df\rho_i(t+1)$ as follows:
\begin{eqnarray}	\label{eq:AspirationUpdate}
\rho_i^+ & = & \rho_i + \epsilon \nu(\epsilon) \cdot \left( \rewardpert{i}-\rho_i \right).
\end{eqnarray}
for some constant step size $\nu=\nu(\epsilon)>0$.}

\ENDFOR
\end{algorithmic}
\end{algorithm}

At periodic instances denoted by $t=1,2,...$, each agent $i$ selects an action according to a finite probability distribution or \emph{strategy} $x_i(t)\in\cX_i\df\Delta(\magn{\cA_i})$, which captures its current beliefs about the most rewarding action. Its selection is slightly perturbed by a \emph{perturbation} (or \emph{mutations}) \emph{factor} $\lambda>0$, such that, with a small probability $\lambda$ agent $i$ follows a uniform strategy (or, it \emph{trembles}).  At the second step, agent $i$ evaluates its new selection by receiving a utility measurement from the environment, also influenced by the actions of the other agents. This new ``experience'' is evaluated with respect to both the utility received $\rewardpert{i}$ but also its difference with agent $i$'s ``benchmark'', namely its aspiration/satisfaction level $\rho_i$. The exact way the aspiration level influences the update of the strategy $x_i$ is captured by the \emph{aspiration factor} $\phi_i$. Given this new experience, agent $i$ updates its strategy vector, according to (\ref{eq:ReinforcementLearningModel}), and its aspiration level $\rho_i\in[\underline{\rho},\overline{\rho}]$ according to (\ref{eq:AspirationUpdate}).

\begin{figure}[t!]	
\centering
\includegraphics[scale=1]{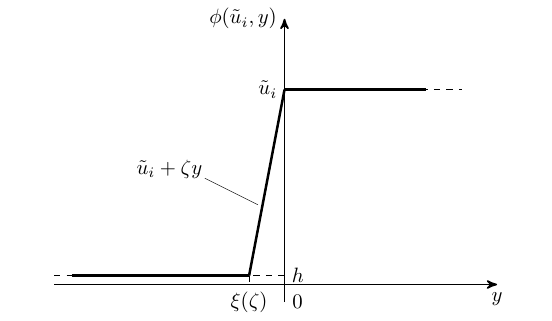}
\caption{Aspiration factor $\phi(\rewardpert{i},y)$ for fixed $\rewardpert{i}>0$.}
\label{fig:AspirationFactor}
\end{figure}

We identify actions $\mathcal{A}_i$ with vertices of the simplex, $\{e_1,...,e_{\magn{\mathcal{A}_i}}\}$. For example, if agent $i$ selects its $j$th action at time $t$, then $e_{\alpha_i(t)}\equiv e_j$. To better clarify how the strategies evolve, consider the following toy example. Let the current strategy of player $i$ be $x_i(t) = (\begin{array}{cc} \nicefrac{1}{2} & \nicefrac{1}{2} \end{array})\tr$, i.e., player $i$ has two actions, each assigned probability $\nicefrac{1}{2}$. Let also $\alpha_i(t+1)=\alpha_i=1$, i.e., player $i$ selects the first action according to 
(\ref{eq:ActionUpdate}). Then, the new strategy vector for agent $i$, updated according to (\ref{eq:ReinforcementLearningModel}), is:
\begin{eqnarray*}
x_i(t+1) = \nicefrac{1}{2} \left(\begin{array}{c} 1  + \epsilon\cdot \phi_{i}(\rewardpert{i},\rewardpert{i} -\rho_i)  \\ 1  - \epsilon \cdot \phi_{i}(\rewardpert{i},\rewardpert{i}-\rho_i)  \end{array}\right)
\end{eqnarray*}
where $\rewardpert{i}=\rewardpert{i}(\alpha(t+1))$ denotes the reward received and $\rho_i(t)=\rho_i$ denotes the current aspiration level. The strategy of the selected first action would be reinforced by $\epsilon \phi_{i}/2>{0}$, while the strategy of the second action would be suppressed by $-\epsilon \phi_{i}/2<{0}$. In other words, repeated selection is reinforced, but the reinforcement is scaled by $\phi_i$. 

The aspiration factor $\phi_i$ defined by Equation~(\ref{eq:AspirationTerm}) is schematically shown in Figure~\ref{fig:AspirationFactor}. We have considered here a rather simple form of aspiration factor motivated by \cite{ChasparisAriShamma13_SIAM}, but alternative forms could also be defined, such as the sigmoid function of \cite{Karandikar98}. The purpose of introducing this factor is to improve the stability properties of PLA in the context of repeatedly played strategic-form games towards attaining group efficiency, which is often a design criterion for multi-agent coordination problems \cite{shamma_dimensions_2007}. 

The introduction of the aspiration factor into the strategy update of each agent, constitutes the main novelty of the proposed learning scheme. In the original class of PLA dynamics \cite{chasparis_stochastic_2019} (which can be derived from APLA by setting $h={\zeta}={0}$ and considering positive utilities), an agent reinforces \emph{repeated selection} by an amount that is proportional to the received reward, i.e., $\phi_i\equiv \reward{i}$ (under the assumption of noiseless rewards). The higher the reward of the selected action, the higher the amount of the reinforcement. Instead, in the case of APLA the amount of reinforcement may differ depending also on the current running average reward (or \emph{benchmark} performance). In case of a \emph{satisfactory action} (i.e., $\reward{i}(\alpha^+)>\rho_i$), the aspiration term is equal to $\phi_i\equiv\rewardpert{i}(\alpha^+)$, while in the case of an \emph{unsatisfactory action} (i.e., $\rewardpert{i}(\alpha^+)<\rho_i$) the aspiration factor becomes $\phi_i=\max\{h,\rewardpert{i}(\alpha^+)+ \zeta (\rewardpert{i}(\alpha^+)-\rho_i)\}<\rewardpert{i}(\alpha^+)$ which reduces the reinforcement term by an amount that is proportional to the level of satisfaction experienced by the player. The role of the aspiration factor is also explained schematically in Section~\ref{sec:SM:SampleStrategyEvolution} of the Supplementary Material.  

For reasons that will become apparent in the forthcoming sections, we set the aspiration level to evolve at a slower rate than the strategy vector. \emph{\textbf{For the remainder of the paper}}, we will assume the following design property.
\begin{property} \label{P:AspirationLevelRate}
Given $\epsilon>0$, we set the step size of the aspiration factor $\epsilon\nu(\epsilon)$ such that, 
$$\lim_{\epsilon\downarrow{0}}\frac{\epsilon\nu(\epsilon)}{\epsilon} = 0\,.$$
\end{property}
This property dictates that the aspiration level updates at a slower time-scale in comparison with the strategy update. For example, by setting $\nu(\epsilon)=\epsilon$, the step size satisfies Property~\ref{P:AspirationLevelRate}. Informally, the aspiration level intends on capturing a running average performance with a smaller frequency compared to the frequency of the strategy update. Thus, it acts as a low-pass filter for the players' performance and maintains the memory of the last actions for longer time in order to provide opportunities for exploring better actions in the upcoming steps.

\subsection{Motivation}

In comparison with standard (variable-structure) learning automata, cf.,~\cite{Narendra89}, there are two forms of reinforcement terms employed here, namely \emph{repeated selection} and \emph{aspiration}. In standard \emph{repeated selection} with positive rewards, e.g., the PLA dynamics \cite{chasparis_stochastic_2019}, the strategies of the selected actions are always reinforced proportionally to the reward received. The reason for considering this type of dynamics as a basis of the proposed scheme is the use of strategies in the selection of actions. Instead of agents making decisions directly based on the received rewards, in learning automata agents make decisions based on a probability distribution captured by their strategy $x_i$. The selection of actions through strategies provides an indirect filtering mechanism of possibly perturbed or noisy observations, while it does not allow for rapid changes in actions. In this paper, this filtering advantage will become evident when considering noisy observations.

As expected, standard learning automata schemes are utilizing only the currently experienced rewards, thus exploiting only a rather myopic information. By introducing an \emph{aspiration or satisficing} term, through the aspiration factor $\phi_i$, we also incorporate the most recent reward history into the decision making process. This provides additional possibilities in terms of influencing the asymptotic stability properties of the dynamics.

%
\section{Related work and contributions} 	\label{sec:RelatedWorkContributions}

\begin{figure*}[th!]
\centering
\includegraphics[scale=0.9,keepaspectratio]{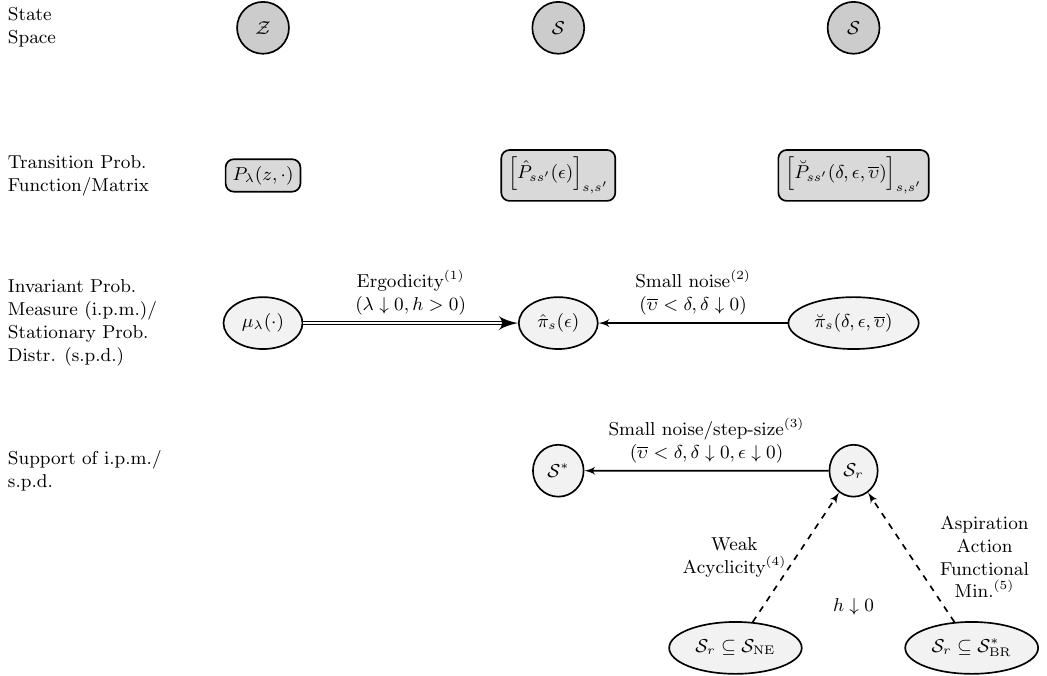}
\caption{Graphical sketch of contributions. It demonstrates the approximations performed on the transition probability function and the invariant probability measure of the induced Markov chain of the dynamics. (1) Theorem~\ref{Th:StochasticStability:PureStrategyStates}, (2) Lemma~\ref{Lm:StationaryDistributionApproximation}, (3) Theorem~\ref{Th:StochasticallyStableStatesMinimumResistance}, (4) Theorem~\ref{Th:StationaryDistributionInWeaklyAcyclicGames}, (5) Theorem~\ref{Th:AspirationActionFunctionalAndStochasticStability}.} 
\label{fig:TheoreticalContributions}
\end{figure*}

\subsection{Related work}	

In this section, we provide a short overview of alternative payoff-based learning schemes specifically designed for repeatedly-played strategic-form games with a \emph{finite} set of actions and a \emph{fixed} utility function for each player. A wide range of reinforcement-based learning dynamics has been investigated, including \emph{replicator-type processes}, \emph{learning automata}, \emph{$Q$-learning}, and \emph{aspiration-based learning}. In this line of research, the main objective is to investigate conditions under which convergence to Nash equilibria or efficient outcomes can be ensured. Note that payoff-based learning has also been applied to static games with continuous action sets, e.g., extremum-seeking control \cite{frihauf_nash_2012,ye_distributed_2015} or actor-critic reinforcement learning \cite{perkins_mixed-strategy_2017}. The focus here instead is only on \emph{finite} action sets. 

\emph{Replicator dynamics} and their discrete-time variants \cite{Hofbauer98,Arthur93,Erev98,HopkinsPosch05,BorgersSarin97,Leslie04} provide a natural interpretation of social behavior but may converge with positive probability to non-Nash outcomes. Perturbed learning automata (PLA) \cite{ChasparisShamma11_DGA}, which constitute discrete-time versions of replicator dynamics, addressed this limitation by introducing a perturbation factor that eliminates non-Nash absorbing states. While convergence to pure equilibria is possible only under restrictive conditions \cite{ChasparisShammaRantzer15}, stochastic stability analysis has established robustness in coordination games of arbitrary size \cite{chasparis_stochastic_2019}. Recently, equilibrium selection in continuous-time replicator dynamics has also been investigated in \cite{leeStochasticLatentActorCritic2020}, however the focus is only on centralized control laws.

\emph{Learning automata} \cite{Tsetlin73,Narendra89} offer an alternative, computationally lightweight approach. Their reinforcement rules resemble replicator dynamics, but convergence guarantees in games remain limited to special structures (zero-sum or identical interest games \cite{Sastry94}). Strong conditions such as absolute monotonicity are generally required, while modifications introducing coordinated exploration phases \cite{verbeeck_exploring_2007} improve performance at the expense though of decentralization.  

\emph{$Q$-learning methods} extend reinforcement learning to stochastic games by tracking value functions \cite{leslie_individual_2005}. While Nash-$Q$ learning \cite{hu_nash_2003} and related schemes \cite{chapman_convergent_2013} can guarantee convergence in potential or weakly-acyclic games, they often rely on stronger informational assumptions (e.g., observing joint actions or other players' rewards). Alternative weaker forms of side information have been considered in \cite{sylvestre_q-learning_2019}, namely a local/global utility split and information about the payoff of the unplayed actions. Independent $Q$-learning converges only in very restricted cases, and convergence in general multi-player games remains open. To address this, synchronized exploration phases have been introduced \cite{arslan_decentralized_2016}, thus partially dropping the decentralization assumption.  

Finally, \emph{aspiration-based learning} builds upon the notion of satisficing behavior in social systems \cite{simon_rational_1956,macy_learning_2002}. Recent game-theoretic models--such as benchmark-based \cite{marden_payoff_2009}, trial-and-error \cite{young_learning_2009}, mood-based \cite{marden_pareto_2014}, and aspiration learning \cite{ChasparisAriShamma13_SIAM}--have demonstrated convergence to Nash equilibria or efficient outcomes without strong monotonicity assumptions. Nevertheless, the treatment of noisy payoff observations is limited (e.g., through the introduction of synchronized exploration phases in \cite{marden_payoff_2009}).  

In summary, while significant progress has been made across these different lines of research, some limitations exist. Discrete-time replicator dynamics and learning automata provide convergence guarantees only under restrictive structural assumptions, and Q-learning demands for strong information requirements. On the other hand, strong convergence guarantees can be achieved under aspiration-based learning in generic multi-player games, however addressing noisy observations requires a form of coordination between agents.

\subsection{Contributions}

This paper presents a novel reinforcement-based learning scheme that is built upon the perturbed learning-automata schema (PLA) \cite{chasparis_stochastic_2019}, and augments it with an aspiration- or satisficing-based reasoning in updating players' strategies. Thus, it inherits the filtering capabilities of replicator processes in noisy environments, and exploits the stronger convergence properties of aspiration-based decision making. Such combination further allows retaining the distributed nature of the dynamics, since no synchronized explorations or other types of coordination is enforced. As a result, stronger convergence guarantees can be derived in large games even under noisy observations. 

More specifically, the contributions of the paper are as follows:
\begin{enumerate}
\item[(C1)] We derive a finite-dimensional approximation of the induced Markov-chain of the process, and we provide an approximation of the (one-step) transition probabilities. This simplification allows for a) \emph{an approximation of the invariant probability measure} (Theorem~\ref{Th:StochasticStability:PureStrategyStates}), and b) \emph{the characterization of the stochastically stable states through the notion of resistance of transition graphs} (Theorem~\ref{Th:StochasticallyStableStatesMinimumResistance}). 
\item[(C2)] We capitalize on the derived approximations in (C1) to provide \emph{an explicit characterization of the stochastically stable states in positive-utility weakly-acyclic games}. We further show that \emph{the stochastically stable states coincide with the set of (pure) Nash equilibria for sufficiently small step size and noise} (Theorem~\ref{Th:StationaryDistributionInWeaklyAcyclicGames}).
\item[(C3)] We further simplify the computation of stochastically stable states in weakly-acyclic games by \emph{establishing equivalence  with the solutions of the aspiration action-functional optimization criterion}, which constitutes a graph (or transience) property of the game (Theorem~\ref{Th:AspirationActionFunctionalAndStochasticStability}). Such optimization criterion can be used to derive conditions for convergence to Pareto efficient outcomes.
\item[(C4)] We demonstrate the analytical derivation and computation of stochastically-stable states in the case of the classical Stag-Hunt game, which may be of independent interest, accompanied with an extensive simulation study using the simulator \cite{chasparis2025apla}. An additional simulation investigation is also provided in the context of load-balancing games in the Supplementary Material (Section~\ref{sec:SM:SimulationsLoadBalancingGames}).
\end{enumerate}

A graphical visualization of the analytical contributions of this paper is also provided in Figure~\ref{fig:TheoreticalContributions}. 

An earlier version of this paper was presented in \cite{chasparis_aspiration_2018}, however the analysis there was restricted to a limited class of coordination games (which is a subset of weakly-acyclic games addressed here) and to noiseless  observations.

\section{Weakly Acyclic Games}	\label{sec:WeaklyAcyclicGames}

We consider the standard setup of finite strategic-form games. There is a finite set of \emph{agents} or \emph{players}, $\cI=\{1,2,\dotsc,n\}$, and each agent has a finite set of actions, $\cA_{i}$. The set of action profiles is the Cartesian product $\cA\df \cA_{1}\times\dotsb\times\cA_{n}$; $\alpha_{i}\in\cA_{i}$ denotes an \textit{action} of agent $i$; and
$\alpha=(\alpha_{1},\dotsc,\alpha_{n})\in\cA$ denotes the \textit{action profile} or \textit{joint action} of all agents. The \emph{payoff/utility function} of player $i$ is a mapping $u_{i}:\cA\rightarrow\mathbb{R}$. The tuple $\langle\mathcal{I},\mathcal{A},\{u_i\}_i\rangle$ defines a \emph{strategic-form game}.

\begin{definition}[Pure Nash Equilibrium] \label{def:PureNashEquilibrium}
An action profile $\alpha^*\in\cA$ is a \emph{pure Nash equilibrium} if, for each $i\in\cI$,
\begin{equation}\label{eq:NashCondition}
u_{i}(\alpha_{i}^*,\alpha_{-i}^*) \geq u_{i}(\alpha_{i}',\alpha_{-i}^*)
\end{equation}
for all $\alpha_{i}'\in\cA_{i}$, where $-i$ denotes the complementary set $\cI\setminus\{i\}$. We denote the set of pure Nash equilibria by $\cA_{\rm NE}$. 
\end{definition}

Before defining weakly acyclic games, we first need to define the notion of \emph{better reply}.
\begin{definition}[Better~Reply] \label{def:BestReply}
An action profile $\alpha'=(\alpha_i',\alpha_{-i})$ is a better reply of player $i$ to an action profile $\alpha=(\alpha_i,\alpha_{-i})$, if $u_i(\alpha') > u_i(\alpha).$ Let ${\rm BR}_i(\alpha)$ denote the set of better replies of player $i$ to action profile $\alpha$, i.e., $${\rm BR}_i(\alpha) \df \{\alpha'\in\cA: u_i(\alpha') > u_i(\alpha)\}.$$
\end{definition}

An improvement-path is then defined as:
\begin{definition}[Improvement Path] \label{def:ImprovementPath}
Given an action profile $\alpha\in\cA$ which is not a Nash equilibrium, an improvement path is a sequence of action profiles $\{\alpha^{(k)}\}_{k=0}^{K}\df\{\alpha=\alpha^{(0)},\alpha^{(1)},...,\alpha^{(K)}\}$ such that, for any $k=0,1,...,K-1$, there exists $i_k\in\cI$ for which $$\alpha^{(k+1)}\in{\rm BR}_{i_k}(\alpha^{(k)}).$$
\end{definition}
Naturally, an improvement path also defines a directed graph where the action profiles in $\cA$ are playing the role of the vertices of the graph and the better replies are playing the role of the directed edges. We will often use the terms \emph{improvement path} and \emph{improvement graph} interchangeably. Let also $G_{\alpha}$ denote an improvement path/graph starting from action profile $\alpha$ with vertices $\{\alpha=\alpha^{(0)},...,\alpha^{(K)}\}$ and directed edges $\{(\alpha^{(j)}\to\alpha^{(j+1)})\}_{j=0,...,K-1}$. Of course there might be more than one improvement paths starting from the same action profile $\alpha$. 

A \emph{weakly acyclic game} is defined as follows:
\begin{definition}[Weakly acyclic game]\label{def:WeaklyAcyclicGame}
A strategic-form game of two or more agents is a weakly acyclic game if, for every action profile $\alpha\in\cA$, there exists an improvement path $$\{\alpha=\alpha^{(0)},\alpha^{(1)},\alpha^{(2)},...,\alpha^{(K)}\}$$ for some finite $K\in\mathbb{N}$ such that $\alpha^{K}$ is a pure Nash equilibrium of the game. 
\end{definition}

We will also use the term of \emph{strict local stability} defined as follows: 
\begin{definition}[Strict local stability of Nash equilibria] \label{def:StrictLocalStability}
The set of pure Nash equilibria satisfy strict local stability if for any two action profiles $\alpha\in\cA_{\rm NE}$ and $\alpha'=(\alpha_i',\alpha_{-i})\notin\cA_{\rm NE}$, we have $u_i(\alpha) > u_i(\alpha')$.
\end{definition}
Furthermore, \emph{Pareto efficient action profiles} are defined as follows.
\begin{definition}[Pareto efficiency]
An action profile $\alpha^*\in\cA$ is Pareto efficient, if there is no $\alpha\neq\alpha^*$ such that $u_i(\alpha)>u_i(\alpha^*)$ for every $i\in\cI$. 
\end{definition}
Based on this definition, we will often refer to Nash equilibria that are Pareto efficient as \emph{Pareto Nash equilibria}, denoted by $\overline{\cA}_{\rm NE}$.

\textit{\textbf{For the remainder of the paper}}, we will be concerned with weakly-acyclic games that satisfy the \emph{\textbf{Positive-Utility Property}}. 

\begin{property}[Positive-Utility Property]		\label{P:PositiveUtilityProperty}
For any agent $i\in\mathcal{I}$ and any action profile $\alpha\in\mathcal{A}$, $\reward{i}(\alpha)>0$.
\end{property}

Several classes of games belong to the family of weakly-acyclic games. For example, it is straightforward to show that any potential game (cf.,~\cite{MondererShapley96}) will also be weakly-acyclic. Different types of coordination games, such as the class of coordination games introduced in \cite{ChasparisAriShamma13_SIAM}, will also be weakly-acyclic. Furthermore, several multi-agent engineering applications that can be formulated as strategic form games can be defined by strictly positive utility functions, such as the network-formation games of \cite{chasparis_network_2013}, the dynamic pinning of parallelized threads in \cite{chasparisEfficientDynamicPinning2017a} and the load-balancing game variation analyzed in Section~\ref{sec:SM:LoadBalancingGames} of the Supplementary Material. 

\section{Stochastic Stability Background} \label{sec:StochasticStability}

In this section, we present the necessary terminology and notation for characterizing stochastic stability. We also present an approximation of the invariant probability measure of the induced Markov chain of the dynamics. The analysis follows similar steps with the corresponding one for the PLA dynamics in \cite{chasparis_stochastic_2019} with some small adaptations due to the use of the aspiration factor and the presence of measurement noise. 

\subsection{Terminology and notation}	\label{sec:Terminology}

Let $\cZ\df \mathcal{A}\times \cX \times \mathcal{R} \times \dnoise$, where $\cX\df\cX_1\times\ldots\times\cX_n$, $\mathcal{R}\df[\underline{\rho},\overline{\rho}]^{n}$, and $\dnoise\df[-\snoise,\snoise]^{n}$, i.e., tuples of joint actions $\alpha$, strategy profiles $x=(x_1,...,x_n)$, aspiration-level profiles $\rho=(\rho_1,...,\rho_n)$ and the realization of the measurement noise profile $\noise=(\noise_1,...,\noise_n)$. We will denote the elements of the state space $\cZ$ by $z$. 

The set $\mathcal{A}$ is endowed with the discrete topology, $\cX$, $\mathcal{R}$ and $\dnoise$ with the usual Euclidean topology, and $\cZ$ with the corresponding product topology. We also let $\Bor(\cZ)$ denote the Borel $\sigma$-field of $\cZ$, and $\mathfrak{P}(\cZ)$ the set of \emph{probability measures} (p.m.) on $\Bor(\cZ)$ endowed with the Prohorov topology, i.e., the topology of weak convergence. The dynamics of Algorithm~\ref{Al:APLA} defines an $\cZ$-valued Markov chain. Let $P_{\lambda}:\cZ\times\Bor(\cZ)\to[0,1]$ denote its \emph{transition probability function} (t.p.f.), parameterized by $\lambda>0$. We will refer to this process as the \emph{perturbed process}, where the action selection process of each agent follows Equation~(\ref{eq:ActionUpdate}). In other words, under the perturbed t.p.f. $P_{\lambda}$, one or more agents may \textit{tremble} (i.e., select randomly an action according to the uniform distribution).  

Define also the t.p.f. $P:\cZ\times\Bor(\cZ)\to[0,1]$ where $\lambda=0$, i.e., \emph{no agent trembles}. We will refer to the corresponding process as the \emph{unperturbed process}.
%
%
The unperturbed process will be denoted by $Z=\{Z_{t} : t\ge0\}$. Let $\Omega\df\cZ^{\infty}$ denote the canonical path space, i.e., an element $\omega\in\Omega$ is a sequence $\{\omega(0),\omega(1),\dotsc\}$, with $\omega(t)= (\alpha(t),x(t),\rho(t),\noise(t))\in\cZ$. For simplicity, we use the same notation for the elements $(\alpha,x,\rho,\noise)$ of the space $\cZ$ and for the coordinates of the process $Z_{t}=(\alpha(t),x(t),\rho(t),\noise(t))$. Note that the characterization of unperturbed/perturbed process is only with respect to the mutation parameter $\lambda$.

Let also $\Prob_{z}[\cdot]$ denote the unique p.m. induced by the unperturbed t.p.f. $P$ on the product $\sigma$-field of $\cZ^{\infty}$ (i.e., the infinite-step unperturbed process), initialized at $z=(\alpha,x,\rho,\noise)$, and $\Exp_{z}[\cdot]$ be the corresponding expectation operator. Let also $\sF_{t} \df \sigma(Z_{\tau}\,,~ \tau\le t)\,,$ $t\geq{0}$, denote the $\sigma$-field of $\cZ^{\infty}$ generated by $\{Z_{\tau},~\tau\le{t}\}$.



The measure $\mu_{\lambda}\in\mathfrak{P}(\cZ)$ is called an \emph{invariant probability measure} (i.p.m.) for $P_{\lambda}$ if
\begin{equation*}
(\mu_{\lambda}P_{\lambda})(A) \df \int_{\cZ}\mu_{\lambda}(dz)P_{\lambda}(z,A) = \mu_{\lambda}(A), \qquad A\in\Bor(\cZ).
\end{equation*}

\subsection{Feller Property and Pure Strategy States}

First, note that both $P$ and $P_{\lambda}$ ($\lambda,h>0$) satisfy the \emph{strong-Feller property} (cf.,~\cite[Definition~4.4.2]{Lerma03}).
\begin{proposition}[Strong-Feller property]		\label{Pr:StrongFellerProperty}
Both the unperturbed process $P$ ($\lambda=0$) and the perturbed process $P_{\lambda}$ ($\lambda>0$) satisfy the strong-Feller property.
\end{proposition}
\ifproofs
\begin{proof}
See Supplementary Material (Section~\ref{sec:SM:StrongFellerProperty}).
\end{proof}
\fi

The measure $\mu_{\lambda}\in\mathfrak{P}(\cZ)$ is called an \emph{invariant probability measure} (i.p.m.) for $P_{\lambda}$ if
\begin{equation*}
(\mu_{\lambda}P_{\lambda})(A) \df \int_{\cZ}\mu_{\lambda}(dz)P_{\lambda}(z,A) = \mu_{\lambda}(A), \qquad A\in\Bor(\cZ).
\end{equation*}
Since $\cZ$ defines a locally compact separable metric space and $P$, $P_{\lambda}$ satisfy the strong-Feller property, they both admit an i.p.m., denoted $\mu$ and $\mu_{\lambda}$, respectively \cite[Theorem~7.2.3]{Lerma03}.

We would like to characterize the \emph{stochastically stable states} $z\in\cZ$ of $P_{\lambda}$, that is any state $z\in\cZ$ for which any collection of i.p.m. $\{\mu_{\lambda}\in\mathfrak{P}(\cZ):\mu_{\lambda}P_{\lambda}=\mu_{\lambda},\lambda>0\}$ satisfies $\liminf_{\lambda\to{0}}\mu_{\lambda}(z)>0$. As the forthcoming analysis will show, the stochastically stable states will be a subset of the set of \emph{pure strategy states} (p.s.s.) defined as follows:
%


\begin{definition}[Pure Strategy State]	\label{def:PureStrategyState}
A pure strategy state is a state $s=(\alpha,x,\rho,\cdot)\in\cZ$ such that for all $i\in\mathcal{I}$, $x_i = e_{\alpha_i}$ and $\rho_i=u_i(\alpha)$, i.e., $x_i$ coincides with the vertex of the probability simplex $\Delta(\magn{\mathcal{A}_i})$ which assigns probability 1 to action $\alpha_i$, and $\rho_i$ coincides with the utility of agent $i$ under action profile $\alpha$.
\end{definition}

We will denote the set of pure strategy states by $\cS$. For any pure strategy state $s=(\alpha,x,\rho)$, define the $\delta$-neighborhood of $s$ as follows
\begin{eqnarray*}
\lefteqn{\Neigh{\delta}{s}  \df 
\left\{z'=(\alpha',x',\rho',\cdot)\in\cZ: \right. } \cr && \left. \alpha=\alpha'\,, |x-x'|<\delta\,, |\rho-\rho'|<\delta \right\}.
\end{eqnarray*}
Let also $\Neighx{\delta}{x}{s'} \df \mathcal{P}_{\cX}\left(\Neigh{\delta}{s'}\right),$ where $\mathcal{P}_{\cX}(O)$ is the \emph{canonical projection} of some set $O\in\Bor(\cZ)$ to $\cX$, defined by the product topology.

\subsection{Stochastic Stability}

The following theorem approximates the i.p.m. of the induced Markov chain for sufficiently small mutations parameter $\lambda>0$.
\begin{theorem}[Stochastic Stability]		\label{Th:StochasticStability:PureStrategyStates}
Let us consider sufficiently small $\epsilon>0$, $h>0$ and $\snoise>0$ such that $0<\epsilon\rewardpert{i}(\alpha)<1$ and $0<h<\rewardpert{i}(\alpha)$ almost surely\footnote{``Almost surely'' (a.s.) excludes paths of probability zero of the unperturbed process $\Prob_{z'}[\cdot]$ (due to the noise sequence $\noise$).} for all $\alpha\in\cA$ and $i\in\cI$. There exists a unique probability vector $\hat{\pi}=(\hat{\pi}_1,...,\hat{\pi}_{\magn{\cS}})$ such that, for any collection of i.p.m.'s $\{\mu_{\lambda}\in\mathfrak{P}(\cZ):\mu_{\lambda}P_{\lambda}=\mu_{\lambda}, \lambda>0\}$, 
\begin{itemize}
\item[(a)] $\lim_{\lambda\downarrow{0}}\mu_{\lambda}(\cdot) = \hat{\mu}(\cdot) \df \sum_{s\in\cS}\hat{\pi}_s\Dirac{s}(\cdot),$ where convergence is in the weak sense.
\item[(b)] The probability vector $\hat{\pi}=\hat{\pi}(\delta,\epsilon,\snoise)$ is an invariant distribution of the finite state Markov process $\hat{P}$, such that, for any $s,s'\in\cS$, and for any $\delta>\snoise>0$,
\begin{equation}	\label{eq:FiniteStateMarkovChain}
\hat{P}_{ss'} = \hat{P}_{ss'}(\delta,\epsilon,\snoise) \df \lim_{t\to\infty} QP^t(s,\Neigh{\delta}{s'}),
\end{equation}
where $Q$ is the t.p.f. corresponding to only one player trembling (i.e., following the uniform distribution of Equation~(\ref{eq:ActionUpdate})).
\end{itemize}
\end{theorem}
\begin{proof}
The proof follows similar reasoning with the proof of Theorem~3.1 in \cite{chasparis_stochastic_2019}. It builds upon the observation that the strategies under the unperturbed process ($\lambda=0$) will approach a vertex of the probability simplex w.p.1, i.e., the players will eventually play only one p.s.s. This observation is then exploited to demonstrate that the invariant probability measure (i.p.m.) of the induced Markov chain of the perturbed process will converge weakly as $\lambda\downarrow{0}$ to an i.p.m. whose support is on the set of p.s.s.'s. The detailed proof is presented in Section~\ref{sec:SM:ProofOfTheoremSSPureStrategyStates} of the Supplementary Material.
\end{proof}

Theorem~\ref{Th:StochasticStability:PureStrategyStates} establishes weak convergence of the i.p.m. of $P_\lambda$ with the invariant distribution of a finite Markov chain $\hat{P}$, whose support is on the set of p.s.s.'s, i.e., ${\rm supp}(\hat{\mu}) = \cS$.\footnote{The support of a probability measure $\mu$ is defined as ${\rm supp}(\mu)\df \{z\in\cZ:\,\forall O\in\Bor(\cZ) \mbox{ s.t. } z\in{O}\mbox{ then } \mu(O)>0 \}$.} Thus, from the ergodicity of $\mu_{\lambda}$, we have that the expected percentage of time that the process spends in any $O\in\Bor(\cZ)$ such that $\partial{O}\cap\cS\neq\varnothing$ is given by $\hat{\mu}(O)$ as $\lambda\downarrow{0}$ and time increases, i.e.,
\begin{equation*}
\lim_{\lambda\downarrow{0}}\left(\lim_{t\to\infty}\;
\frac{1}{t}\sum_{k=0}^{t-1}P_{\lambda}^{k}(x,O)\right) = \Hat{\mu}(O)\,.
\end{equation*}

We will refer to the support of the invariant distribution $ \hat{\mu}(\cdot)$, defined as ${\rm supp}(\hat{\mu})$, as the \emph{stochastically stable states} of $P_{\lambda}$ as $\lambda\downarrow{0}$. Often, we will denote this set by $\cS^*\subseteq\cS$.

The above weak convergence result is due to the ergodicity of the induced Markov chain resulted from $\lambda>0$. The noise in the observation signals does not influence the convergence properties (as compared to the noiseless case) when its size is sufficiently small. Furthermore, the size of $h>0$ plays also no role in this result, as long as the main hypotheses of the theorem are satisfied.

\section{Approximation of the Finite-State Markov Process}	\label{sec:ApproximationFiniteMarkovProcess}

\subsection{Background on finite Markov chains} \label{sec:FiniteMarkovChains}

In order to compute the invariant distribution of a finite-state, irreducible and aperiodic Markov chain, we are going to consider a characterization introduced by \cite{FreidlinWentzell84}. In particular, for finite Markov chains an invariant measure can be expressed as the ratio of sums of products consisting of transition probabilities. These products can be described conveniently by means of graphs on the set of states of the chain. In particular, let  $\cS$ be a finite set of states, whose elements will be denoted by $s_k$, $s_\ell$, etc., and let a subset $\mathcal{W}$ of $\cS$.

\begin{figure}[t!]
\centering
\includegraphics[scale=1]{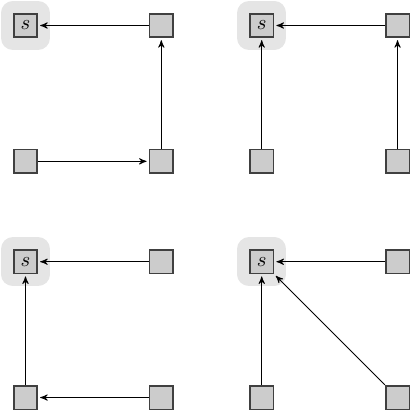}
\caption{Examples of $s$-graphs in case $\cS$ contains four states.}
\label{fig:Wgraphs}
\end{figure}

\begin{definition}  \label{def:W-graph_old}
($\mathcal{W}$-graph) A graph consisting of arrows
$s_k\rightarrow{s_\ell}$, such that 
$s_k\in{\cS\backslash{\mathcal{W}}},$ $s_\ell\in\cS$, $s_\ell\neq{s_k}$, is called a $\mathcal{W}$-graph if it satisfies the following
conditions:
\begin{enumerate}
\item{every point $s_k\in{\cS\backslash{\mathcal{W}}}$ is the initial point of
exactly one arrow;}
\item{there are no closed cycles in the graph; or, equivalently,
for any point $s_k\in{\cS\backslash{\mathcal{W}}}$ there
exists a sequence of arrows leading from it to some point
$s_\ell\in\mathcal{W}$.}
\end{enumerate}
\end{definition}

Figure~\ref{fig:Wgraphs} provides examples of $\mathcal{W}$-graphs for some state $s\in\cS$ when $\cS$ contains four states, where in this case $\mathcal{W}\equiv \{s\}$. We will denote by $\mathcal{G}\{\mathcal{W}\}$ the set of $\mathcal{W}$-graphs and we shall denote by  $g$ an element of this set.\footnote{Note that we introduced the notation $g$ for $\mathcal{W}$-graphs in order to distinguish it from the case of improvement graphs $G$.} Note further that a graph $g'\in\mathcal{G}\{s\}$ is a $s$-rooted graph, since $s$ has no links stemming from it. Often, we will use the notation ${\rm root}(g)\subseteq\cS$ to denote the subset of states that are the roots of $\mathcal{W}$-graph $g$. 

Let us consider nonnegative numbers $\hat{P}_{s_ks_\ell}$ representing transition probabilities from state $s_k$ to state $s_\ell$ in a finite Markov chain. The following Lemma holds:
\begin{lemma}[Lemma~6.3.1 in \cite{FreidlinWentzell84}]   \label{Lm:StationaryDistribution}
Let us consider a Markov chain with a finite set of states $\cS$ and transition probabilities $\{\hat{P}_{s_ks_\ell}\}$ and assume that every state can be reached from any other state in a finite number of steps. Then, the stationary distribution of the chain is $\hat{\pi} = [\hat{\pi}_{s}]$, where
\begin{equation}    \label{eq:StationaryDistribution}
\hat{\pi}_s = \frac{R_{s}}{\sum_{s_i\in\cS}R_{s_i}}\,, \quad
s\in\cS
\end{equation}
where $R_{s} \df \sum_{g\in{\mathcal{G}}\{s\}}\varpi(g)$, and $\varpi(g)$ denotes the transition probability along path $g$, i.e., $$\varpi(g) \df \prod_{(s_k\rightarrow{s_\ell})\in{g}}\hat{P}_{s_ks_\ell}.$$
\end{lemma}

In other words, in order to compute the weight that the stationary distribution assigns to a state $s\in\cS$, it suffices to compute the ratio of the sum of transition probabilities of all $\{s\}$-graphs over the corresponding sum of transition probabilities of all $\{s'\}$-graphs, $s'\in\cS$. 

\subsection{$\mathcal{W}$-graphs and improvement paths}	\label{sec:WgraphsImprovementPaths}

In this paper, we will be concerned with the evaluation of the transition probabilities $\hat{P}_{s_ks_\ell}$ defined in Equation~(\ref{eq:FiniteStateMarkovChain}) of Theorem~\ref{Th:StochasticStability:PureStrategyStates}, through which the stochastically stable states can be characterized as the support of the finite probability distribution in (\ref{eq:StationaryDistribution}). According to the definition of $\hat{P}_{s_ks_\ell}$, such transition probabilities are defined only for state transitions $s_k\to s_{\ell}$ that differ in the action of a single agent or briefly here \emph{one-step transitions} (due to the definition of the t.p.f. $Q$).  
\begin{lemma}[$\mathcal{W}$-graphs through one-step transitions] \label{Lm:OneStepTransitions}
Consider the hypotheses and the definition of the finite state Markov process $\hat{P}$ of Theorem~\ref{Th:StochasticStability:PureStrategyStates}. The conclusions of Lemma~\ref{Lm:StationaryDistribution} apply for $\hat{P}$ under the condition that the $\mathcal{W}$-graphs of Definition~\ref{def:W-graph_old} are defined within the set of one-step transitions.
\end{lemma}
\begin{proof}
According to the definition of the finite state Markov process $\hat{P}$ (cf.,~Equation~(\ref{eq:FiniteStateMarkovChain}) of Theorem~\ref{Th:StochasticStability:PureStrategyStates}), any transition $s_k\to s_\ell$ between two distinct p.s.s. $s_k\neq s_{\ell}$ is governed by the limiting t.p.f. $\lim_{t\to\infty}Q P^t(s_k,\Neigh{\delta}{s_\ell})$. Given that $P$ corresponds to the unperturbed t.p.f. ($\lambda=0$), and given the definition of a p.s.s., a change in action is only possible through the t.p.f. $Q$ where only a single player can tremble and change its action and therefore its strategy. Thus, under $\hat{P}$, only transitions to the same state or one-step transitions are possible. By the hypotheses of Theorem~\ref{Th:StochasticStability:PureStrategyStates} of $\epsilon,h>0$, we also have that any state can be reached from any other state through a sequence of one-step transitions, while the probability of remaining to the same state is strictly positive. Thus, the hypotheses of Lemma~\ref{Lm:OneStepTransitionProbabilityApproximation} are satisfied. 
\end{proof}

Hence, \textit{\textbf{for the forthcoming analysis}}, Definition~\ref{def:W-graph_old} and the set $\mathcal{G}(\mathcal{W})$ will be restricted within the set of one-step transitions. As a direct implication of  Lemma~\ref{Lm:OneStepTransitions}, a direct connection between $\mathcal{W}$-graphs and the improvement paths of Definition~\ref{def:ImprovementPath} can be established in weakly acyclic games. 
\begin{lemma}[$\cS_{\rm NE}$-graphs in weakly acyclic games]   \label{Lm:WeakAcyclicityAndWGraphsA}
In any weakly acyclic game, there exists a nonempty set of $\mathcal{W}$-graphs $\mathcal{G}_{\rm BR}\subseteq\mathcal{G}\{\cS_{\rm NE}\}$, such that for each graph $g\in\mathcal{G}_{\rm BR}$ the following property holds: for any p.s.s. $s\notin\cS_{\rm NE}$ there exists an improvement path $G_s \subseteq{g}$ starting from $s$ that leads to $\cS_{\rm NE}$. 
\end{lemma}
\ifproofs
\begin{proof}
By Definition~\ref{def:WeaklyAcyclicGame}  of weakly acyclic games, and for any p.s.s. $s\notin\cS_{\rm NE}$, there exists an improvement path $G_s$ that leads to a p.s.s. within $\cS_{\rm NE}$. By construction of an improvement path, any p.s.s. that belongs to this path is the starting point of exactly one arrow. Also, in an improvement path, each direct arrow corresponds to a better reply (i.e., one agent is performing a change in action that leads to a strict improvement on its utility), thus it is an one-step transition. We repeat the same process for all p.s.s. $s$ which do not belong to the set $\cS_{\rm NE}$ of Nash equilibria, thus creating a collection of improvement paths $\{G_{s}\}_{s\in\cS\backslash\cS_{\rm NE}}$. Define also the corresponding union of all improvement paths by $\overline{G}\df \bigcup_{s\in\cS\backslash\cS_{\rm NE}}G_s.$ By construction, $\overline{G}$ may include a p.s.s. which is the starting point of more than one directed edges (each of which corresponds to a better reply). Consider a graph $g^*\subseteq\overline{G}$ which is constructed as follows: for any $s\in\cS\backslash\cS_{\rm NE}$ which is the starting point of more than one edges in $\overline{G}$, we remove all but one of its edges starting from $s$. The set of edges to be removed is not important, since each of them corresponds to a better reply. Thus, in the resulting graph $g^*$, any $s\in\cS\backslash\cS_{\rm NE}$ is the starting point of a single link (as the definition of a $\mathcal{W}$-graph dictates) and there exists an improvement path starting from $s$ that leads to a Nash equilibrium in $\cS_{\rm NE}$. Then, by construction, $g^*\in\mathcal{G}\{\cS_{\rm NE}\}$ has the desired property. 
\end{proof}
\fi

Note that a graph within the $\cS_{\rm NE}$-graphs may contain one-step transitions that are not better replies. Lemma~\ref{Lm:WeakAcyclicityAndWGraphsA} simply states that among the family of $\cS_{\rm NE}$-graphs, there exists at least one graph consisting only of improvement paths that lead to a Nash equilibrium. We will exploit this observation in the forthcoming Section~\ref{sec:StochasticStabilityWeaklyAcyclicGames} to more explicitly characterize the set of stochastically stable states in weakly acyclic games. 

\begin{figure}[th!]
\centering\hspace{6pt}
\includegraphics[scale=1]{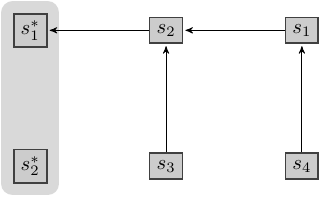}
\caption{Schematic example of an improvement $\mathcal{W}$-graph, where $\cS_{\rm NE}\df\{s_1^*,s_2^*\}$. For each p.s.s. $s\notin\cS_{\rm NE}$, namely $s_1,...,s_4$, there exists an improvement path (consisting only of better replies) that leads to a Nash equilibrium in $\cS_{\rm NE}$.}
\label{fig:ImprovementWGraphSketch}
\end{figure}

To simplify the forthcoming discussion, we will also introduce the term of improvement $\mathcal{W}$-graphs in weakly acyclic games as follows.
\begin{definition} [Improvement $\mathcal{W}$-graphs]
In any weakly acyclic game, we define the set of improvement $\mathcal{W}$-graphs as the non-empty set $\mathcal{G}_{\rm BR}\subseteq\mathcal{G}\{\cS_{\rm NE}\}$ that satisfy the property of Lemma~\ref{Lm:WeakAcyclicityAndWGraphsA}. 
\end{definition}
Note that according to Lemma~\ref{Lm:WeakAcyclicityAndWGraphsA}, improvement $\mathcal{W}$-graphs consist only of better replies (one-step) transitions, terminating at the set of Nash equilibria. Furthermore, according to the definition of $\mathcal{W}$-graphs, not all Nash equilibria need to be a terminal state of a $\{\cS_{\rm NE}\}$-graph. Furthermore, there are no links that stem from any Nash equilibrium, given the definition of a $\{\mathcal{S}_{\rm NE}\}$-graph. An example of such graph is depicted in Figure~\ref{fig:ImprovementWGraphSketch}.


\subsection{One-step transition probability approximation} \label{sec:OneStepTransitionProbabilityApproximation}

In order to characterize the stochastically stable states more explicitly based on Theorem~\ref{Th:StochasticStability:PureStrategyStates}, an approximation for the transition probabilities under $\hat{P}$ is necessary. Recall that according to the definition of the t.p.f. $Q\Pi$, and as $\lambda\downarrow{0}$, a transition from $s$ to $s'$ influences the stationary distribution only if $s$ differs from $s'$ in the action of a \emph{single} agent. We will often refer to such transitions as \emph{one-step transitions}. This observation will be capitalized by the forthcoming Lemma~\ref{Lm:OneStepTransitionProbabilityApproximation} to approximate the transition probability from $s$ to $s'$ under $\hat{P}$.

\begin{lemma}[One-step transition probability approximation]	\label{Lm:OneStepTransitionProbabilityApproximation}
Consider an one-step transition under t.p.f. $Q\Pi$ from action profile $\alpha$ to action profile $\alpha'$ which differ in the action of a single agent $j$, and let $s,s'\in\cS$ be the p.s.s.'s associated with $\alpha$ and $\alpha'$, respectively. Set $z'\df (\alpha',x',\rho',\cdot)$ to be the state of agent $j$ after $j$ perturbed once under t.p.f. $Q$ starting from p.s.s. $s$ and playing $\alpha_j'\neq\alpha_j$. For some $\delta>0$, define 
$$\OSA{s}{s'}{\delta,\epsilon,\snoise} \df \Prob_{z'}[\uptau(\Neigh{\delta}{s'}) \leq \infty]$$ 
which corresponds to the probability that the process eventually reaches $\Neigh{\delta}{s'}$ starting from $z'$. Consider further $\epsilon>0$ and $\snoise>0$ sufficiently small such that $0<\epsilon \tilde{u}_i(\alpha) <1$ for all $\alpha\in\cA$ and $i\in\cI$. Then, the one-step transition probability from $s$ to $s'$ under $Q\Pi$ satisfies, as $\snoise=\snoise(\delta)<\delta$ and $\delta\downarrow{0}$,
\begin{equation}	\label{eq:OneStepTransitionProbability_A}
\hat{P}_{ss'}(\delta,\epsilon,\snoise(\delta)) = \gamma_{j} \cdot \OSA{s}{s'}{\delta,\epsilon,\snoise(\delta)}\,,
\end{equation}
where $\hat{P}_{ss'}(\delta,\epsilon,\snoise(\delta))$ is defined in Equation~(\ref{eq:FiniteStateMarkovChain}) and $\gamma_{j}\df 1/(n\magn{\cA_j})$ corresponds to the probability that agent $j$ trembled and selected action $\alpha_j'\neq\alpha_j$ under $Q$.
\end{lemma}
\ifproofs
\begin{proof}
Under the better reply from $s$ to $s'$, agent $j$ experiences state $z'$ under t.p.f. $Q$, realized after agent $j$ trembled and played $\alpha_j'\neq\alpha_j$ starting from $s$. 
After $t$ time steps, the probability that the unperturbed process reaches a $\delta$-neighborhood of $s'$ is given by:
\begin{eqnarray*}
\lefteqn{QP^{t}(s,\Neigh{\delta}{s'}) } \cr && = \int_{\cZ}\gamma_j\Dirac{z'}(dy)P^{t}(y,\Neigh{\delta}{s'})=\gamma_jP^{t}(z',\Neigh{\delta}{s'}).
\end{eqnarray*}
Given that $\Neigh{\delta}{s'}$ is a continuity set of $Q\Pi(s,\cdot)$, from Equation~(\ref{eq:FiniteStateMarkovChain}) and  from Portmanteau theorem we have that, for any $\delta>\snoise>0$, $$\hat{P}_{ss'}=Q\Pi(s,\Neigh{\delta}{s'}) = \gamma_j \lim_{t\to\infty}P^t(z',\Neigh{\delta}{s'}).$$ Note also that 
\begin{equation}
P^{t}(z',\Neigh{\delta}{s'}) = \Prob_{z'}[Z_t\in\Neigh{\delta}{s'}] \leq \Prob_{z'}[\uptau(\Neigh{\delta}{s'})\leq {t}],
\end{equation}
given that the event of being in set $\Neigh{\delta}{s'}$ at time $t$ implies that the set $\Neigh{\delta}{s'}$ has been reached for the first time at time $t$ or earlier. Define the event $\Gamma_{\delta,t}\df\{\uptau(\Neigh{\delta}{s'})\leq{t}\}$, $t>0$, which is a non-decreasing event, i.e., $\Gamma_{\delta,t}\subseteq\Gamma_{\delta,t+1}$. Then, from continuity from below, we have that, for any $\delta>0$, 
\begin{equation}	\label{eq:OneStepTransitionProbability_A1}
\lim_{t\to\infty}P^{t}(z',\Neigh{\delta}{s'}) \leq \lim_{t\to\infty}\Prob_{z'}[\Gamma_{\delta,t}] = \Prob_{z'}[\Gamma_{\delta,\infty}] \equiv \breve{P}_{ss'}.
\end{equation}

Recall the shift operator $\theta_t$, defined as $\theta_t:\Omega\mapsto\Omega$ for some finite time step $t$, that satisfies $(Z_s\circ\theta_t)(\omega) = Z_s(\theta_{t}(\omega))=Z_{s+t}(\omega)$, i.e., it shifts the sequence by $t$ time steps backwards. Let also $$B_{t} \df \{\omega\in\Omega: \alpha(\tau) = \alpha(0)\,, \mbox{ for all } 0\leq \tau \leq t \},$$ which corresponds to the event that the same action has been played continuously until time $t$.

Note that one way of being in set $\Neigh{\delta}{s'}$ at time $t$ almost surely (with respect to the utility noise) is to reach this set before time $t$ and then continue playing $\alpha'$ thereafter (since by continuously playing $\alpha'$ will maintain the state within $\Neigh{\delta}{s'}$ almost surely when $\snoise=\snoise(\delta)<\delta$).\footnote{Note that when $\snoise=\snoise(\delta)<\delta$ the aspiration level will remain within $\Neigh{\delta}{s'}$ almost surely when starting within this set and playing $\alpha'$ thereafter. The strategy vector is not influenced by the utility noise, and will continue approaching $e_{\alpha_j'}$ as long as action profile $\alpha'$ is played.} Then, we can write
\begin{eqnarray}
\lefteqn{P^{t}(z',\Neigh{\delta}{s'})} \cr & \geq & \sum_{k=1}^{t}\Prob_{z'}[\Gamma_{\delta,k}\cap\Gamma_{\delta,k-1}^{c},Z\circ\theta_k\in{B}_{\infty}] \cr & = & 
\sum_{k=1}^{t}\Prob_{z'}[Z\circ\theta_k\in{B}_{\infty}|\Gamma_{\delta,k}\cap\Gamma_{\delta,k-1}^{c}]\cdot \Prob_{z'}[\Gamma_{\delta,k}\cap\Gamma_{\delta,k-1}^{c}] \cr & \geq & 
\Big(\inf_{z\in\Neigh{\delta}{s'}}\Prob_{z'}[{B}_{\infty}] \Big) \cdot \sum_{k=1}^{t} \Prob_{z'}[\Gamma_{\delta,k}\cap\Gamma_{\delta,k-1}^{c}] \label{eq:OneStepTransitionProbability_A1.1} \\
& \geq & \Big(\inf_{z\in\Neigh{\delta}{s'}}\Prob_{z}[B_\infty]\Big) \cdot \Prob_{z'}[\Gamma_{\delta,t}]. \label{eq:OneStepTransitionProbability_A1.2}
\end{eqnarray}
where in (\ref{eq:OneStepTransitionProbability_A1.1}) we have used the Markov property, while in (\ref{eq:OneStepTransitionProbability_A1.2}) we have used the fact that the family of sets $\{\Gamma_{\delta,k}\cap\Gamma_{\delta,k-1}^{c}\}_k$ are disjoint with one another, given that $\Gamma_{\delta,t}\subseteq\Gamma_{\delta,t+1}$, and therefore $$\sum_{k=1}^{t}\Prob_{z'}[\Gamma_{\delta,k}\cap\Gamma_{\delta,k-1}^{c}] = \Prob_{z'}[\Gamma_{\delta,t}]>0.$$ Given also that $\lim_{\delta\downarrow{0}}\inf_{z\in\Neigh{\delta}{s'}}\Prob_{z}[B_\infty]=1,$ we get that, as $\delta\downarrow{0}$,
\begin{equation}	\label{eq:OneStepTransitionProbability_A2}
\lim_{t\to\infty}P^{t}(z',\Neigh{\delta}{s'}) \geq \Prob_{z'}[\Gamma_{\delta,\infty}] \equiv \breve{P}_{ss'}\,.
\end{equation}
Thus, from Equations~(\ref{eq:OneStepTransitionProbability_A1})--(\ref{eq:OneStepTransitionProbability_A2}), we have that, as $\delta\downarrow{0}$, $$\lim_{t\to\infty}P^t(z',\Neigh{\delta}{s'}) = \breve{P}_{ss'}\,,$$ which concludes the proof.
\end{proof}
\fi

The above Lemma~\ref{Lm:OneStepTransitionProbabilityApproximation} provides an approximation of the one-step transition probability through the probability of the first hitting time to a small neighborhood of the destination state $s'$. Note that this approximation applies to any one-step transition, independently of whether the destination utility, $u_j(\alpha')$, is larger or smaller than the original utility, $u_j(\alpha)$. In the forthcoming subsection Section~\ref{sec:OneStepTransitionProbability:TwoTechnicalLemmas}, this approximation will further be specialized in the case of satisfactory and unsatisfactory transitions.

Given the approximation of the one-step transition probabilities of Lemma~\ref{Lm:OneStepTransitionProbabilityApproximation}, an approximation of the corresponding stationary distribution $\pi$ may also be derived, as defined in Theorem~\ref{Th:StochasticStability:PureStrategyStates}. Let us first define $\mathcal{G}^{(1)}\{s\}\subseteq\mathcal{G}\{s\}$ to be the set of $s$-graphs consisting solely of one-step transitions, i.e., for any $g\in\mathcal{G}^{(1)}\{s\}$ and any arrow $(s_k\to s_{\ell})\in{g}$, the associated action profiles, say $\alpha^{(k)},\alpha^{(\ell)}$, respectively, differ in a single action of a single agent. It is straightforward to check that $\mathcal{G}^{(1)}\{s\}\neq\varnothing$ for any $s\in\cS$.

\begin{lemma}[Approximation of stationary distribution]	\label{Lm:StationaryDistributionApproximation}
The stationary distribution of the finite Markov chain $\{\hat{P}_{s_ks_{\ell}}\}$, $\hat{\pi}=[\hat{\pi}_s]$, derived in Theorem~\ref{Th:StochasticStability:PureStrategyStates}, satisfies for $\snoise=\snoise(\delta)<\delta$,
\begin{equation}    \label{eq:StationaryDistributionSimplified}
\lim_{\delta\downarrow{0}}\hat{\pi}_s(\delta,\epsilon,\snoise(\delta)) = \lim_{\delta\downarrow{0}}\breve{\pi}_{s}(\delta,\epsilon,\snoise(\delta))\end{equation}
where $$\breve{\pi}_{s}(\delta,\epsilon,\snoise)\df \frac{\breve{R}_{s}(\delta,\epsilon,\snoise)}{\sum_{s_i\in\cS}\breve{R}_{s_i}(\delta,\epsilon,\snoise)}, \qquad
s\in\cS,$$ 
$\breve{R}_{s}(\delta,\epsilon,\snoise) \df \sum_{g\in{\mathcal{G}^{(1)}}\{s\}}\breve{\varpi}(g;\delta,\epsilon,\snoise),$ and 
\begin{equation}	\label{eq:TransitionProbabilityGraphApproximation}
\breve{\varpi}(g;\delta,\epsilon,\snoise) \df \bar{\gamma}_g \prod_{(s_k\to s_{\ell})\in{g}} \OSA{s_k}{s_\ell}{\delta,\epsilon,\snoise},
\end{equation}
for some constant $\bar{\gamma}_g \in (0,1)$.
\end{lemma}
\ifproofs
\begin{proof}
According to Lemma~\ref{Lm:StationaryDistribution}, for any $s\in\cS$, we have $\hat{\pi}_s=R_{s}/\sum_{s_i\in\cS}R_{s_i}$. Given the definition of the t.p.f. $Q$, where only one agent trembles, we should only consider one-step transition probabilities (as defined in Lemma~\ref{Lm:OneStepTransitionProbabilityApproximation}). Thus,
$$R_s = \sum_{g\in\mathcal{G}\{s\}}\prod_{(s_k\to s_{\ell})\in{g}}\hat{P}_{s_ks_{\ell}}.$$
According to Lemma~\ref{Lm:OneStepTransitionProbabilityApproximation} and Equation~(\ref{eq:OneStepTransitionProbability_A}), we have 
\begin{eqnarray*}
\lim_{\delta\downarrow{0}}R_s & = & \lim_{\delta\downarrow{0}}\sum_{g\in\mathcal{G}^{(1)}\{s\}}\prod_{(s_k\to s_{\ell})\in{g}} \gamma_{j(s_k,s_{\ell})}\OSA{s_k}{s_\ell}{\delta,\epsilon,\snoise}\cr 
& = & \lim_{\delta\downarrow{0}}\sum_{g\in\mathcal{G}^{(1)}\{s\}}\bar{\gamma}_{g} \prod_{(s_k\to s_{\ell})\in{g}} \OSA{s_k}{s_\ell}{\delta,\epsilon,\snoise}
\end{eqnarray*}
where $j(s_k,s_{\ell})$ denotes the single agent whose action changes from $s_k$ to $s_{\ell}$, and $$\bar{\gamma}_g \df \prod_{(s_k\to s_{\ell})\in{g}} \gamma_{j(s_k,s_{\ell})}\in(0,1).$$ Thus, the conclusion follows.
\end{proof}
\fi

Note that Lemma~\ref{Lm:StationaryDistributionApproximation} provides a simplification to Theorem~\ref{Th:StochasticStability:PureStrategyStates}, since it suffices to compute the transition probabilities of the $\mathcal{W}$-graphs consisting solely of one-step transitions. Furthermore, the one-step transitions can be approximated by Equation~(\ref{eq:OneStepTransitionProbability_A}), which can be the basis for a small explicit characterization in the following sections.

\subsection{Two technical lemmas}	\label{sec:OneStepTransitionProbability:TwoTechnicalLemmas}

Lemma~\ref{Lm:OneStepTransitionProbabilityApproximation} provides a general principle in the approximation of the one-step transition probability, which is a necessary step for characterizing the stationary distribution of Theorem~\ref{Th:StochasticStability:PureStrategyStates}. However, this approximation does not yet provide any details of how it can be computed. In this subsection, we provide a detailed derivation/approximation of the computation of the one-step transition probabilities in two cases, namely the \emph{satisfactory} and \emph{unsatisfactory} transitions. The proofs of the following lemmas are presented in detail in the supplementary material.

\begin{lemma}	[One-step satisfactory transition probability]	\label{Lm:OneStepTransitionProbabilitySatisfactory}
Consider the hypotheses of Lemma~\ref{Lm:OneStepTransitionProbabilityApproximation}. Let also the one-step transition from $s$ to $s'$ correspond to a better reply for agent $j$, i.e., the nominal utilities satisfy $u_j(\alpha') > u_j(\alpha)$ (i.e., this is a satisfactory transition). For given $\delta>0$, and for sufficiently small $\epsilon>0$ and $\snoise=\snoise(\delta)<\delta$ such that $H_j(\alpha)\df 1-\epsilon \rewardpert{j}(\alpha)\in(0,1)$ almost surely and $u_j(\alpha)>(1+2c)\snoise$ uniformly on $\cA$, the following hold:
\begin{itemize}
\item[(a)] $\OSA{s}{s'}{\delta,\epsilon,\snoise}$ satisfies as $\delta\downarrow{0}$,
\begin{equation}		\label{eq:OneStepTransitionProbabilitySatisfactory_MainA}
\OSA{s}{s'}{\delta,\epsilon,\snoise} \approx 
\prod_{t<{\uptau}_{\delta}^0(\epsilon)}\left(1-{H}_{j}(\alpha')^{t+1}\right), \quad \mbox{a.s.}\,,
\end{equation}
where $\uptau_{\delta}^0(\epsilon)$ is the first hitting time of the unperturbed process to the set $\Neighx{\delta}{x}{s'}$ when $\snoise\equiv{0}$, defined as follows:
\begin{equation}
\uptau_{\delta}^0(\epsilon) \df \left\lceil \frac{\log(\delta)}{\log({H}_{j}(\alpha'))} \right\rceil\,.
\end{equation}

\item[(b)] There exists positive constant $$\eta=\eta(\delta)\df \sum_{\ell=1}^{\infty}\frac{1}{\ell^2}(1-\delta^\ell)$$ such that, as $\epsilon> 0$ and $\delta\downarrow{0}$,
\begin{equation*}   \label{eq:OneStepTransitionProbabilitySatisfactory_MainB}
\OSA{s}{s'}{\delta,\epsilon, \snoise} \approx \exp\left(-\frac{\eta(\delta)}{\epsilon u_j(\alpha')}\right)\,,\quad \mbox{a.s.}\,.
\end{equation*}
\end{itemize}
\end{lemma}
\ifproofs
\begin{proof}
See Supplementary Material (Section~\ref{sec:SM:SatisfactoryTransition}).
\end{proof}
\fi

The following lemma provides an approximation of the transition probability under an one-step unsatisfactory transition.
\begin{lemma} [One-step unsatisfactory transition probability]	\label{Lm:OneStepTransitionProbabilityUnsatisfactory}
Consider the hypotheses of Lemma~\ref{Lm:OneStepTransitionProbabilityApproximation}. Let also $u_j(\alpha') < u_j(\alpha)$, i.e., this is an unsatisfactory one-step transition for agent $j$. For given $\delta,\epsilon>0$ and noise level $\snoise=\snoise(\delta)<\delta$ be sufficiently small such that $0<\epsilon \rewardpert{i}(\alpha)<1$ and $h<\rewardpert{i}(\alpha)$ almost surely for all $\alpha\in\cA$ and $i\in\cI$. As $\epsilon\downarrow{0}$ and $\delta\downarrow{0}$, we have:
\begin{equation} \label{eq:OneStepTransitionProbabilityUnsatisfactory_Eq0}
\OSA{s}{s'}{\delta,\epsilon,\snoise} \approx \exp\left(-\frac{\eta(\delta)}{\epsilon h}\right)\,, \mbox{a.s.}\,,
\end{equation}
for some positive constant $\eta(\delta)\df \sum_{\ell=1}^{\infty}(1-\delta^\ell)/\ell^2$.
\end{lemma}
\begin{proof}
See Supplementary Material (Section~\ref{sec:SM:UnsatisfactoryTransition}).
\end{proof}
Lemmas~\ref{Lm:OneStepTransitionProbabilitySatisfactory}--\ref{Lm:OneStepTransitionProbabilityUnsatisfactory} provide the tools for explicitly computing the one-step transition probabilities, as needed for the computation of the stationary distribution of Lemma~\ref{Lm:StationaryDistributionApproximation}. Note that both in the case of satisfactory and unsatisfactory transitions, the corresponding probabilities can be approximated by quantities associated with the destination utility.

\subsection{Resistance}		\label{sec:Resistance}

Given the explicit approximation of the one-step transition probabilities of the lemmas of Section~\ref{sec:OneStepTransitionProbability:TwoTechnicalLemmas}, we introduce the notion of \emph{resistance} of an one-step transition. It will allow for a clear characterization of stochastic stability. In this section, we introduce this notion and also characterize the stochastically stable states through it. 
\begin{definition}[Resistance]	\label{def:Resistance}
Consider an one-step transition from $s$ to $s'$, where $s,s'\in\cS$. Define the resistance of this transition as follows:
\begin{equation}	\label{eq:ResistanceOneStep}
\Res{s}{s'}{\delta,\epsilon,\snoise} \df -\log\left(\OSA{s}{s'}{\delta,\epsilon,\snoise}\right) \geq 0.
\end{equation}
\end{definition}
Intuitively, a smaller resistance of the one-step transition from $s$ to $s'$ will imply a larger transition probability along this transition. Note that the resistance depends on $\delta$, i.e., the size of the target neighborhood, since it is directly influencing the probability of reaching this neighborhood.

Based on this definition, we can define the resistance of a p.s.s. associated with a $\mathcal{W}$-graph $g\in\mathcal{G}^{(1)}\{s\}$ as the sum of the involved one-step resistances.
\begin{equation}	\label{eq:ResistanceGraph}
\ResGraph{s}{g}{\delta,\epsilon,\snoise} \df \sum_{(s_k\to s_{\ell})\in{g}}\Res{s_k}{s_\ell}{\delta,\epsilon,\snoise}\,, \quad g \in \mathcal{G}^{(1)}\{s\}\,.
\end{equation}
Given the definition of $\OSA{s}{s'}{\delta,\epsilon,\snoise}$ and Lemmas~\ref{Lm:OneStepTransitionProbabilitySatisfactory}--\ref{Lm:OneStepTransitionProbabilityUnsatisfactory},  we have the property that for any $s,s'\in\cS$,
\begin{equation*}
\lim_{\epsilon\downarrow{0}}\Res{s}{s'}{\delta,\epsilon,\snoise}=\infty \,, \mbox{ and }\, \lim_{\epsilon\downarrow{0}}\ResGraph{s}{g}{\delta,\epsilon,\snoise}=\infty\,,
\end{equation*}
which corresponds to the case that the corresponding transitions are not possible.

\subsection{Approximation of stochastically stable states}	\label{sec:ApproximationStochasticallyStableStates}

The approximation of the one-step transition probabilities, and the notion of resistance of such transitions, allow for approximating the stationary distribution of the finite Markov chain $\hat{P}$ of Theorem~\ref{Th:StochasticStability:PureStrategyStates}. 

For some $s\in\cS$, and considering the approximation of stationary distribution in Lemma~\ref{Lm:StationaryDistributionApproximation}, we denote $\MinResGraph{s}{\delta,\epsilon}$ as the \emph{minimum resistance among all possible $\{s\}$-graphs} in $\mathcal{G}^{(1)}\{s\}$ consisting solely of one-step transitions, i.e., 
\begin{equation}		\label{eq:DefinitionMinimumResistance}
\MinResGraph{s}{\delta,\epsilon,\snoise} \df \min_{g\in\mathcal{G}^{(1)}\{s\}} \ResGraph{s}{g}{\delta,\epsilon,\snoise}.
\end{equation}
The stochastically stable states can be identified as the states $s\in\cS$ of minimum resistance, as the following theorem demonstrates. We will denote the set of stochastically stable states by $\cS^*$. 
\begin{theorem}[Stochastically stable states]	\label{Th:StochasticallyStableStatesMinimumResistance}
For any given $\delta>0$, let the size of the noise satisfy $\snoise=\snoise(\delta)<\delta$. Then, the support of the stationary distribution of the finite Markov chain $\hat{P}$ (of Theorem~\ref{Th:StochasticStability:PureStrategyStates}) satisfies 
\begin{equation} \label{eq:StochasticallyStableStates}
\lim_{\epsilon\downarrow{0}}\lim_{\delta\downarrow{0}}\sum_{s\in\cS_r}\hat{\pi}_s(\delta,\epsilon,\snoise) = 1,
\end{equation}
i.e., $\cS^*\subseteq\cS_r$, where the set $\cS_r$ has the property that, for any $s\in\cS_r$
\begin{equation}	\label{eq:StochasticallyStableStatesMinimumResistance:Condition}
\lim_{\delta\downarrow{0}} \lim_{\epsilon\downarrow{0}} \left( \MinResGraph{s}{\delta,\epsilon,\snoise}  - \MinResGraph{s'}{\delta,\epsilon,\snoise} \right) < 0 \,, \quad \forall s'\in\cS\backslash\cS_r.
\end{equation}
\end{theorem}
\ifproofs
\begin{proof}
To simplify notation, within this proof we will skip stating explicitly the dependencies on $\delta,\epsilon,\snoise>0$ for the resistances and the one-step transition probabilities. By Lemma~4 
of the main part of the paper, and the definition of the resistance (Definition~8), for the subset $\cS_r\subset\cS$ and for any state $s\in \cS_r$ and graph $g\in\mathcal{G}^{(1)}\{s\}$, we have that,  
\begin{equation*}
\breve{\varpi}(g) = \bar{\gamma}_g \prod_{(s_k\to s_{\ell})\in{g}} \OSAs{s_k}{s_{\ell}} = \bar{\gamma}_g \cdot \exp\Big(-\ResGraphRed{s}{g} \Big),
\end{equation*}
and
\begin{equation*}
\breve{R}_s = \sum_{g\in\mathcal{G}^{(1)}\{s\}}\bar{\gamma}_g \cdot \exp\Big(-\ResGraphRed{s}{g}\Big).
\end{equation*}
Thus, for the states in $\cS\backslash\cS_r$, and for sufficiently small $\epsilon\downarrow{0}$, we have
\begin{eqnarray*}
\lefteqn{\sum_{s\in\cS\backslash\cS_r}\breve{R}_s = \exp\Big(-\min_{s\in\cS\backslash\cS_r}\MinResGraphRed{s}\Big) \cdot } \cr && \sum_{s\in\cS\backslash\cS_r}\sum_{g\in\mathcal{G}^{(1)}\{s\}} \bar{\gamma}_{g} \cdot \exp\Big(-\ResGraphRed{s}{g} + \min_{s\in\cS\backslash\cS_r}\MinResGraphRed{s} \Big),
\end{eqnarray*}
where we have simply factored out the exponential of the negative minimum resistance among all states in $\cS\backslash\cS_r$. Furthermore, for any $s\in\cS_r$, we can also write 
\begin{align}		\label{eq:StochasticallyStableStatesMinimumResistance:Eq1}
& \frac{\sum_{s'\in\cS\backslash\cS_r}\breve{R}_s}{\sum_{s\in\cS_r}\breve{R}_s} = \cr & \frac{\sum_{s'\in\cS\backslash\cS_r}\sum_{g\in\mathcal{G}^{(1)}\{s'\}} \bar{\gamma}_g \exp\Big(- \ResGraphRed{s'}{g} + \min_{s''\in\cS\backslash\cS_r}\MinResGraphRed{s''}  \Big)}{\sum_{s\in\cS_r}\sum_{g\in\mathcal{G}^{(1)}\{s\}} \bar{\gamma}_g \cdot \exp\Big(- \ResGraphRed{s}{g} + \min_{s''\in\cS\backslash\cS_r}\MinResGraphRed{s''}  \Big)}. \cr
\end{align}
By definition, we have $-r_{s'|g}+\min_{s''\in\cS\backslash\cS_r}\MinResGraphRed{s''} \leq 0$ for any $s\in\cS\backslash\cS_r$. Thus, as $\epsilon\downarrow{0}$ the numerator approaches $1$ as $\epsilon\downarrow{0}$.  If the set $\cS_r\subset\cS$ satisfies condition \begin{equation}	\label{eq:StochasticallyStableStatesMinimumResistance:Condition}
\lim_{\delta\downarrow{0}} \lim_{\epsilon\downarrow{0}} \left( \MinResGraph{s}{\delta,\epsilon,\snoise}  - \MinResGraph{s''}{\delta,\epsilon,\snoise} \right) < 0 \,, \quad \forall s''\in\cS\backslash\cS_r\,,
\end{equation} 
then for any $s\in\cS_r$ and for sufficiently small $\epsilon>0$ and $\delta>0$, $\MinResGraphRed{s} < \min_{s''\in\cS\backslash\cS_r} \MinResGraphRed{s''}$. This further implies that for any $s\in\cS_r$, there is a $\{s\}$-graph, say $g^*=g^*(s)$, such that $\MinResGraphRed{s}=\ResGraphRed{s}{g^*} < \min_{s''\in\cS\backslash\cS_r} \MinResGraphRed{s''}$. In this case, each term of the denominator in the above ratio could be written as follows:
\begin{eqnarray*}
\lefteqn{\sum_{g\in\mathcal{G}^{(1)}\{s\}} \bar{\gamma}_g \cdot \exp\Big(- \ResGraphRed{s}{g} + \min_{s''\in\cS\backslash\cS_r}\MinResGraphRed{s''}  \Big) }\cr && = \bar{\gamma}_{g^*}\exp\Big( - \ResGraphRed{s}{g^*} + \min_{s''\in\cS\backslash\cS_r}\MinResGraphRed{s''}  \Big) + \cr && \sum_{g\in\mathcal{G}^{(1)}\{s\}\backslash g^*} \bar{\gamma}_g \cdot \exp\Big(- \ResGraphRed{s}{g} + \min_{s''\in\cS\backslash \cS_r}\MinResGraphRed{s''}  \Big).
\end{eqnarray*}
A similar decomposition could be performed for each $s\in\cS_r$. Note that the first term of the r.h.s. approaches $\infty$ as $\epsilon\downarrow{0}$, given that $-r_{s|g^*}+\min_{s''\in\cS\backslash\cS_r}r_{s''}^*>0$. On the other hand, for any $s'\in\cS\backslash\cS_r$, $-r_{s'|g}+\min_{s''\in\cS\backslash\cS_r}r_{s''}^*<0$ for sufficiently small $\epsilon>0$ and $\delta>0$, which implies that the exponential terms of the numerator approach $0$ as $\epsilon\downarrow{0}$. In this case, the overall ratio in (\ref{eq:StochasticallyStableStatesMinimumResistance:Eq1}) approaches $0$ as $\epsilon\downarrow{0}$. Thus, we conclude that
\begin{equation} 	\label{eq:StochasticallyStableStatesMinimumResistance:Eq2}
\frac{\sum_{s'\in\cS\backslash\cS_r}\breve{R}_{s'}(\delta,\epsilon,\snoise)}{\sum_{s\in\cS_r}\breve{R}_s(\delta,\epsilon,\snoise)}\xrightarrow{\epsilon\downarrow{0}}0\,, \mbox{ for all } s\in\cS_r.
\end{equation}

Denote by $\hat{\pi}_{s}=\hat{\pi}_{s}(\delta,\epsilon,\snoise)$ the probability assigned by the stationary distribution $\hat{\pi}$ to a p.s.s. $s\in\cS_r\subset\cS$. Then, according to Lemma~\ref{Lm:StationaryDistributionApproximation}, 
we have:
\begin{eqnarray*}
\lefteqn{\lim_{\epsilon\downarrow{0}}\lim_{\delta\downarrow{0}}\sum_{s\in\cS_r}\hat{\pi}_{s}(\delta,\epsilon,\snoise) } \cr & = & \lim_{\epsilon\downarrow{0}}\lim_{\delta\downarrow{0}}\frac{\sum_{s\in\cS_r}\breve{R}_{s}(\delta,\epsilon,\snoise)}{\sum_{s'\in\cS}\breve{R}_{s'}(\delta,\epsilon,\snoise)} \cr & = & \lim_{\delta\downarrow{0}}\lim_{\epsilon\downarrow{0}}\frac{\sum_{s\in\cS_r}\breve{R}_{s}(\delta,\epsilon,\snoise)}{\sum_{s\in\cS_r}\breve{R}_{s}(\delta,\epsilon,\snoise) + \sum_{s'\in\cS\backslash {s}}\breve{R}_{s'}(\delta,\epsilon,\snoise)} \cr
& = & \lim_{\delta\downarrow{0}}\lim_{\epsilon\downarrow{0}}\frac{1}{1+\sum_{s'\in\cS\backslash\cS_r}\breve{R}_{s'}(\delta,\epsilon,\snoise)/\sum_{s\in\cS_r}\breve{R}_{s}(\delta,\epsilon,\snoise)}.
\end{eqnarray*}
Note that the interchange of limits in the second equality is valid due to the finiteness of the limits of the transition probabilities. Given Equation~(\ref{eq:StochasticallyStableStatesMinimumResistance:Eq2}), we conclude that $$\lim_{\epsilon\downarrow{0}}\lim_{\delta\downarrow{0}}\sum_{s\in\cS_r}\hat{\pi}_{s}(\epsilon)=1.$$ Conversely, $\lim_{\epsilon\downarrow{0}}\lim_{\delta\downarrow{0}}\sum_{s\in\cS\backslash\cS_r}\pi_{s}(\epsilon)=0$.
Thus, the stochastically stable states may only be contained within the set $\cS_r\subseteq\cS$ where $\cS_r$ satisfies condition (\ref{eq:StochasticallyStableStatesMinimumResistance:Condition}).
\end{proof}
\fi

In words, Theorem~\ref{Th:StochasticallyStableStatesMinimumResistance} states that if the states in a set $\cS_r\subset\cS$ exhibit minimum resistance (as a group of states) that is smaller than the minimum resistance of each other state in $\cS\backslash \cS_r$, then the group of states $\cS_r$ defines the support of the distribution $\hat{\pi}$ for sufficiently small $\epsilon>0$ and $\delta>0$. Furthermore, if we also take sufficiently small $\lambda>0$, then according to Theorem~\ref{Th:StochasticStability:PureStrategyStates}, the set of states $\cS_r$ defines the set of stochastically stable states of the perturbed process $P_{\lambda}$, i.e., $\cS_r\equiv \cS^*$. Note that this theorem applies to any game that satisfies the positive-utility property with no additional structural conditions.  

Note further that Theorem~\ref{Th:StochasticallyStableStatesMinimumResistance} does not in fact really on any specifics of how the resistance is being calculated/approximated (e.g., specifics of the utility algorithm, or specifics of the aspiration factor). In fact, this theorem applies to the standard PLA case as well (although in this case the approximation of the resistance may be a bit different \cite{chasparis_stochastic_2019}). It provides a significant simplication in the calculation of the stochastically stable states, that is instead of considering all possible $\mathcal{W}$-graphs as Lemma~\ref{Lm:StationaryDistribution} indicates, it suffices to consider the minimum resistance $\mathcal{W}$-graph(s).  

\section{Stochastic Stability in Weakly Acyclic Games and Aspiration Action-Functional}	\label{sec:StochasticStabilityWeaklyAcyclicGames}

In the following theorem, we provide a specialization of Theorem~\ref{Th:StochasticallyStableStatesMinimumResistance} to the case of weakly acyclic games, where we show that the stochastically stable states are within the set of pure Nash equilibria.

\begin{theorem}[Stochastically Stable States in Weakly Acyclic Games] \label{Th:StationaryDistributionInWeaklyAcyclicGames}
For any weakly acyclic game that satisfy strict local stability, and for any $\delta>0$, let the size of the noise satisfy $\snoise=\snoise(\delta)<\delta.$ Let also $h>0$ be such that $$h < \min_{i\in\cI}\min_{\alpha\in\cA}u_i(\alpha).$$ Then, as $\epsilon,\delta,h\downarrow{0}$, the stochastically stable states $\cS^*$ satisfy condition (\ref{eq:StochasticallyStableStates}) of Theorem~\ref{Th:StochasticallyStableStatesMinimumResistance}   with $\cS^*\subseteq\cS_{r}\subseteq\cS_{\rm NE}$.
\end{theorem}
\ifproofs
\begin{proof}
Let us consider the $\mathcal{W}$-graphs of the pure Nash equilibria (consisting only of one-step transitions) and denoted by $\mathcal{G}\{\cS_{\rm NE}\}$. By Lemma~\ref{Lm:WeakAcyclicityAndWGraphsA}, there exists at least one graph $g^*\in\mathcal{G}\{\cS_{\rm NE}\}$, with the property that: for any p.s.s. $s'\notin\cS_{\rm NE}$, there exists an improvement path $G_s\subseteq g^*$ starting from $s'$ that leads to $\cS_{\rm NE}$. In other words, all transitions within $g^*$ are characterized by the approximation of Lemma~\ref{Lm:OneStepTransitionProbabilitySatisfactory} (of satisfactory transitions). This implies that, as $\epsilon,\delta\downarrow{0}$, the minimum resistance among all $\mathcal{G}\{\cS_{\rm NE}\}$ will satisfy: 
\begin{equation}	\label{eq:StochasticallyStableStates:1}
r^{*}_{\cS_{\rm NE}} \approx \sum_{(s_k\to s_\ell)\in g^*} \frac{\eta(\delta)}{\epsilon u_{j(s_k,s_\ell)}(\alpha)}
\end{equation}
where $j(s_k,s_\ell)\in\cI$ denotes the agent that performed a change in action under the one-step transition $s_k\to s_{\ell}$.
Let us consider instead the $\mathcal{W}$-graphs of any state $s'\notin\cS_{\rm NE}$. All such graphs will contain at least one transition of the form $s_k\to s_\ell$, such that $s_k\in\cS_{\rm NE}$ and $s_\ell\notin\cS_{\rm NE}$. If strict local stability (cf.,~\ref{def:StrictLocalStability}) is satisfied for the set of pure Nash equilibria, then any $\{s'\}$-graph will contain at least one unsatisfactory transition (as defined in Lemma~\ref{Lm:OneStepTransitionProbabilityUnsatisfactory}). Thus, the corresponding resistance will satisfy as $\epsilon,\delta\downarrow{0}$,
\begin{equation*}		\label{eq:StochasticallyStableStates:2}
r_{s'|g} \geq \frac{\eta(\delta)}{\epsilon h}. 
\end{equation*} 
From (\ref{eq:StochasticallyStableStates:1}--(\ref{eq:StochasticallyStableStates:2}), we conclude that as $h\downarrow{0}$, $$r_{\cS_{\rm NE}}^* < r_{s'|g}\,, \quad \mbox{ for all } g\in\mathcal{G}\{s'\}.$$ Then, from Theorem~\ref{Th:StochasticallyStableStatesMinimumResistance}, we conclude that the set of stochastically stable states is ncessarily contained within the set of Nash equilibria, $\cS_{\rm NE}$.
\end{proof}
\fi

It is evident from the proof of Theorem~\ref{Th:StationaryDistributionInWeaklyAcyclicGames} that the stochastically stable states are determined by paths (sequences of one-step transitions) that minimize the accumulated resistance. Thus, convergence is not a property of a static (state-dependent) function, as it is the case in potential games (cf.,~\cite{MondererShapley96}), rather it is a dynamic or path property dependent on action trajectories. Let us introduce the functional: for any $g\in\mathcal{G}_{\rm BR}$, let 
\begin{equation}	\label{eq:AspirationActionFunctional}
\Psi(g) \df \frac{\eta(\delta)}{\epsilon} \sum_{(s_k\to s_{\ell})\in{g}}\frac{1}{u_{j(s_k,s_\ell)}(\alpha_\ell)} > 0,
\end{equation}
which assigns an aspiration cost to every improvement path. We will refer to this functional as the \emph{aspiration action-functional}. This is analogous to the concept of the action functional of Freidlin-Wentzell (cf.,~\cite[Chapter~3]{FreidlinWentzell84}) for perturbed continuous-time dynamic systems, which represents the cost of deviating from the deterministic (unperturbed) dynamics. Here, instead, the functional (\ref{eq:AspirationActionFunctional}) represents the cost of following an improvement trajectory. This cost functional is used to rank all admissible improvement paths according to the cumulative aspiration cost.

\begin{theorem}[Aspiration action-functional minimization]	\label{Th:AspirationActionFunctionalAndStochasticStability}
For any weakly-acyclic game and under the hypotheses of Theorem~\ref{Th:StationaryDistributionInWeaklyAcyclicGames}, let $\mathcal{G}_{\rm BR}^*\subseteq\mathcal{G}_{\rm BR}\subseteq\mathcal{G}\{\cS_{\rm NE}\}$ denote the improvement $\mathcal{W}$-graphs that attain minimum resistance among all improvement $\mathcal{W}$-graphs, i.e., as $\epsilon,\delta\downarrow{0}$, we have 
\begin{equation}
\mathcal{G}_{\rm BR}^* \df \arg\min_{g\in\mathcal{G}_{\rm BR}} r_{\cS_{\rm NE}|g} \approx \arg\min_{g\in\mathcal{G}_{\rm BR}}\Psi(g).
\end{equation}
Then, as $h\downarrow{0}$, the stochastically stable states $\cS^*$ satisfy 
\begin{equation} 	\label{eq:RootsImprovementPaths}
\cS^* \subseteq \cS_{\rm BR}^* \df \bigcup_{g^*\in\mathcal{G}^*_{\rm BR}}{\rm root}(g^*).
\end{equation}
\end{theorem}
\ifproofs
\begin{proof}
Let us denote by $\cS_{\rm BR}^*$ as the roots of the minimizers of the action functional, i.e., for any $s^*\in\cS_{\rm BR}^*$, there exists $g^*\in\mathcal{G}^{*}_{\rm BR}$ such that $s^*\in{\rm root}(g^*)$. Let us know consider the $\mathcal{W}$-graphs of each one of the states in $\cS_{\rm BR}^*$ independently in order to compare their resistances. For any $s^*\in\cS_{\rm BR}^*$, and as $h\downarrow{0}$, the minimum resistance among all $\mathcal{G}\{s^*\}$ graphs will be given by:
\begin{equation}
r_{s^*}^* \approx \min_{\{g\in\mathcal{G}_{\rm BR}: s^*\in{\rm root}(g)\}} \Psi(g) + \left(\magn{\cS_{\rm NE}}-1\right)\frac{\eta(\delta)}{\epsilon{h}}.  
\end{equation}
The first part of the r.h.s. captures the minimum resistance among all improvement $\mathcal{W}$-graphs that are rooted to $s^*$, while the second part captures the resistances of all links initiated from the other Nash equilibria. Note that by strict local stability, all links initiated from the other Nash equilibria will necessarily be unsatisfactory transitions. Note further that since $s^*\in\cS^*_{\rm BR}$, we should also have: $$\min_{\{g\in\mathcal{G}_{\rm BR}: s^*\in{\rm root}(g)\}} \Psi(g) \equiv \min_{\{g\in\mathcal{G}_{\rm BR}\}} \Psi(g) \df \Psi^*,$$ given that $s^*$ is among the roots of the improvement graphs that minimizes resistance.

Let us now consider a Nash equilibrium state $s\in\cS_{\rm NE}$ such that $s\notin \cS_{\rm BR}^*$. Given that $s$ satisfies strict local stability, there will be an improvement $\mathcal{W}$-graph that has one of its roots in $s$. Note that the minimum resistance of all $\mathcal{W}$-graphs rooted to $s$ will satisfy, as $h\downarrow{0}$,
\begin{equation}
r_{s}^* \approx \min_{\{g\in\mathcal{G}_{\rm BR}: s\in{\rm root}(g)\}} \Psi(g) + \left(\magn{\cS_{\rm NE}}-1\right)\frac{\eta(\delta)}{\epsilon{h}}.  
\end{equation}
Given that $s\notin \cS_{\rm BR}^*$, we will have that
$$\min_{\{g\in\mathcal{G}_{\rm BR}: s\in{\rm root}(g)\}} \Psi(g) > \Psi^*,$$ which implies that for sufficiently small $h>0$, $r^*_{s} > r^*_{s^*}$, which concludes the proof.
\end{proof}
\fi

In other words, Theorem~\ref{Th:StationaryDistributionInWeaklyAcyclicGames} states that the roots of the minimum resistance improvement paths $\mathcal{G}_{\rm BR}^*$ will be the stochastically stable equilibria and they belong to the set of Nash equilibria. This convergence concept corresponds to a trajectory (or graph) property of the game, and it is not dependent on the payoff of the individual equilibrium action profiles. Two Nash equilibria may have exactly the same payoff profiles, however stochastic stability will favor the one that exhibits a minimum cost improvement graph leading to it. The aspiration action functional of Equation~(\ref{eq:AspirationActionFunctional}) determines the measure of this (trajectory) cost.

This behavior (of the stochastic stability being determined by a trajectory cost) is attributed to the fact that the aspiration-based reinforcement term of APLA is a local processing concept that is driven by local aspiration costs. The ``easier'' it is for the players to follow a path that leads to a Nash equilibrium, the smaller the resistance of this equilibrium.  

The minimum resistance Nash equilibria may not easily be characterized by statis properties of the game (e.g., the Pareto efficiency). However, if the players' interests are sufficiently aligned with each other, we should expect that the minimizers of the aspiration action functional will be action profiles that provide a large payoff for all players in the stochastically stable states. If, for example, the Pareto efficient equilibria are among the roots of these paths, then they will be among the stochastically stable states. This is for example the case in the so-called Stag-Hunt coordination games (cf.,~\cite{Skyrms_2014}), as we will discuss in more detail in the next section. 

Note further that the aspiration action-functional minimization in APLA dynamics simplifies the computation of stochastically stable states compared to the standard PLA dynamics. In particular, standard stochastic stability derivations would require a minimization over all possible $\mathcal{W}$-graphs (as it is required in the PLA dynamics). Instead, for the APLA dynamics, the minimization in (\ref{eq:AspirationActionFunctional}) is computationally more efficient since it is restricted to the improvement paths only. This reduces significantly the domain of paths over which the resistance minimization should be performed.

\section{Simulation study} \label{sec:StagHuntSimulationStudy}

In this subsection, we provide the results of a simulation study performed on $2$-player $2$-action coordination games of Table~\ref{Tb:CoordinationGame}. According to \cite{Lewis02}, coordination games are characterized by a) \emph{a level of similarity of interest among agents} (i.e., the payoff differences between agents are small), and b) \emph{the existence of multiple coordination equilibria} (i.e., Nash equilibria where no agent would have been better off had it alone acted otherwise). For example, when we select $a-c\geq d-b$, then the coordination game of Table~\ref{Tb:CoordinationGame} corresponds to the so-called Stag-Hunt game (cf.,~\cite{Skyrms_2014}).

\begin{table}[h]
\centering
\begin{game}{2}{2}
      & A    & B\\
A   &$a,a$   &$b,c$\\
B   &$c,b$   &$d,d$
\end{game}
\caption{Coordination game with $a>c>0$, $d>b>0$, and $a>d$.} \label{Tb:CoordinationGame}
\end{table}

Specific convergence guarantees can be derived for this class of games, by utilizing the results of this paper and in the context of either PLA or APLA dynamics. The following result describes stochastic stability under the PLA dynamics.
\begin{proposition}[Stochastic stability of PLA in Coordination Games]		\label{Pr:PLAStochasticStability2x2game}
Consider the 2-player, 2-action game of Table~\ref{Tb:CoordinationGame} with $a>c>0$, $d>b>0$, and $a>d$. Let us also consider the standard PLA dynamics defined in \cite{chasparis_stochastic_2019}, which can be derived directly by the APLA dynamics of Algorithm~\ref{Al:APLA} by setting $h=\zeta={0}$ and assume the hypotheses of Theorem~\ref{Th:StochasticStability:PureStrategyStates}, with the size of the noise satisfying $\overline{v}=\overline{v}(\delta)<\delta$. Denote by $s_{(A,A)}$ and $s_{(B,B)}$ to be the p.s.s.'s corresponding to action profiles $(A,A)$ and $(B,B)$, respectively. Then, as $\epsilon,\delta\downarrow{0}$, the stationary distribution of the finite Markov chain $\hat{P}$ of Theorem~\ref{Th:StochasticStability:PureStrategyStates} satisfies: 
\begin{itemize}
\item[(a)] if $a-c < d-b$, then $$\lim_{\epsilon\downarrow{0}}\lim_{\delta\downarrow{0}} \hat{\pi}_{s_{(B,B)}}(\delta,\epsilon,\overline{v}(\delta))=1,$$ i.e., $(B,B)$ corresponds to the unique stochastically stable state;
\item[(b)] if $a-c\geq{d-b}$ and $c\leq{b}$, then $$\lim_{\epsilon\downarrow{0}} \lim_{\delta\downarrow{0}} \hat{\pi}_{s_{(A,A)}}(\delta,\epsilon,\overline{v}(\delta))=1,$$ i.e., $(A,A)$ corresponds to the unique stochastically stable state.
\end{itemize}
\end{proposition}
\begin{proof}
See Supplementary Material (Section~\ref{sec:SM:PLAStochasticStability2x2game}).
\end{proof}

An immediate conclusion of Proposition~\ref{Pr:PLAStochasticStability2x2game} is the fact that in 2-player, 2-action games, \emph{risk-dominance} (cf.,~\cite{samuelson_evolutionary_1998}) is a necessary condition for a Nash equilibrium to be a stochastically stable state. Note that this conclusion agrees with the generally observed behavior in evolutionary dynamics (cf.,~\cite[Chapter~7]{samuelson_evolutionary_1998}) in the presence of noise. Selectively, we could mention here the stochastic stability analysis presented by \cite{kandori_learning_1993}, based on the concept of \emph{stochastically stable equilibrium} of \cite{FosterYoung90} and in the context of replicator dynamics with finite population. Therein it was shown that the equilibrium with the largest basin of attraction (e.g., the risk-dominant equilibrium in the Stag-Hunt game) will always be selected in the long run. As also shown by \cite{kandori_learning_1993}, the result still applies for any adjustment process as long as a weak monotonicity property is satisfied (which is the case for replicator dynamics). Similar conclusions have also been derived by the \emph{adaptive learning} with finite memory of Young \cite[Theorem~4.1]{Young04}, where it is shown that risk dominant equilibria are the only stochastically stable states for sufficiently large memory.

In the case of APLA dynamics, we can directly apply Theorems~\ref{Th:StationaryDistributionInWeaklyAcyclicGames}--\ref{Th:AspirationActionFunctionalAndStochasticStability} to assess stochastic stability for the game of Table~\ref{Tb:CoordinationGame}.
\begin{proposition}[Stochastic stability of APLA in Coordination Games] \label{Pr:APLAStochasticStability2x2game}
Consider the $2$-player, $2$-action game of Table~\ref{Tb:CoordinationGame} with $a>c>0$, $d>b>0$, and $a>d$. Let us also consider the APLA dynamics of Algorithm~\ref{Al:APLA} for some  $h,\zeta>{0}$ and assume the hypotheses of Theorem~\ref{Th:StochasticStability:PureStrategyStates}, with the size of the noise satisfying $\overline{v}=\overline{v}(\delta)<\delta$. Denote by $s_{(A,A)}$ the p.s.s. corresponding to action profile $(A,A)$, respectively. Then, the stationary distribution of the finite Markov chain $\hat{P}$ of Theorem~\ref{Th:StochasticStability:PureStrategyStates} (approximated by Theorem~\ref{Th:StochasticallyStableStatesMinimumResistance} as $\epsilon,\delta\downarrow{0}$) satisfies $$\lim_{h\downarrow{0}}\lim_{\epsilon\downarrow{0}}\lim_{\delta\downarrow{0}}\hat{\pi}_{s_{(A,A)}}(\delta,\epsilon,\overline{v}(\delta))=1,$$ i.e., $(A,A)$ corresponds to the unique stochastically stable state. 
\end{proposition}
\begin{proof}
The proof is a direct implication of Theorem~\ref{Th:AspirationActionFunctionalAndStochasticStability} and the fact that, starting from any p.s.s. that does not correspond to a Nash equilibrium (e.g., $(A,B)$ or $(B,A)$), the improvement path (or just improvement link in this case) with the minimum resistance is the one leading to $(A,A)$ as $h\downarrow{0}$.
\end{proof}

Note that the conclusion of Proposition~\ref{Pr:APLAStochasticStability2x2game} does not depend on which state is risk-dominant, as it was the case for the standard PLA dynamics of Proposition~\ref{Pr:PLAStochasticStability2x2game}. The Pareto Nash equilibrium (A,A) will be the unique stochastically stable outcome independently.
 
We conducted a simulation study for $a=5, c=3, b=1, d=4$, which corresponds to a Stag-Hunt game satisfying $a-c<d-b$. In this case, the pure Nash equilibrium (B,B) corresponds to the risk-dominant equilibrium. According to Propositions~\ref{Pr:PLAStochasticStability2x2game}--\ref{Pr:APLAStochasticStability2x2game}, the action profile (B,B) should be the unique stochastically stable state for the PLA dynamics, while the action profile (A,A) should be the unique stochastically stable state for the APLA dynamics. 

\begin{figure}[th!]
\centering
\includegraphics[scale=1]{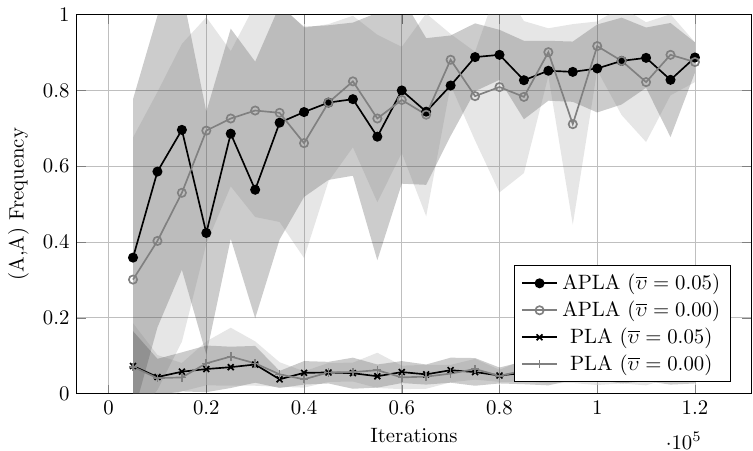}
\caption{Mean frequency of occurrence of the (A,A) action profile of Table~\ref{Tb:CoordinationGame} ($a=5, c=3, b=1, d=4$) over the whole simulation time and over 10 simulation runs. The following configuration parameters have been used: $\epsilon=\nu=0.06$, $h=\lambda=0.04$, $\zeta=30$.}
\label{fig:SimulationStagHuntResultsTime}
\end{figure}

\begin{figure}[th!]
\centering
\includegraphics[scale=1]{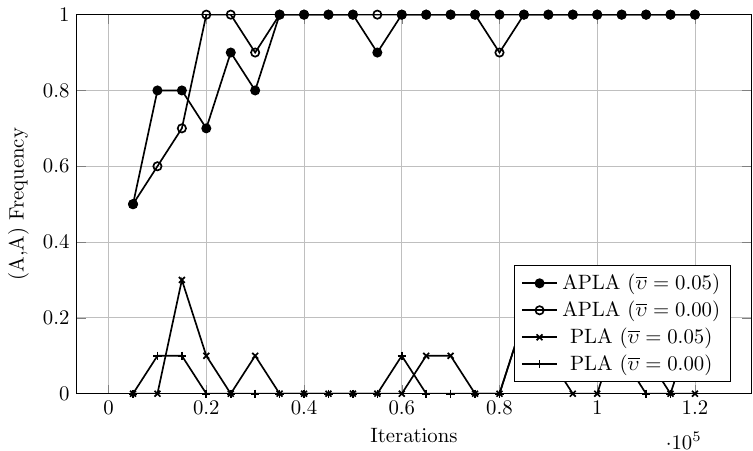}
\caption{Mean frequency of occurrence of the (A,A) action profile of Table~\ref{Tb:CoordinationGame} ($a=5, c=3, b=1, d=4$) at the end of the simulation time and over 10 simulation runs. The following configuration parameters have been used: $\epsilon=\nu=0.06$, $h=\lambda=0.04$, $\zeta=30$.}
\label{fig:SimulationStagHuntResultsFinal}
\end{figure}

The following figures demonstrate the statistical (Monte Carlo) analysis of the behavior of APLA under different configuration parameters and in comparison with the performance of PLA dynamics. The simulations have been performed under noisy ($\snoise>0$) and noiseless rewards ($\snoise=0$). The PLA/APLA algorithms are run over a simulation length that varies, and we record the frequency of observing the payoff-dominant (A,A) action profiles. In particular, Figure~\ref{fig:SimulationStagHuntResultsTime} provides the mean and standard deviation of the frequency of observing the (A,A) action profile, when for each simulation length 10 simulations are executed. Instead, Figure~\ref{fig:SimulationStagHuntResultsFinal} provides the mean and standard deviation of the frequency of observing the (A,A) action profile at the end of the simulation run, over 10 simulation runs. All experiments have been conducted using the publicly available simulator at \cite{chasparis2025apla}.

It is evident that under the standard PLA dynamics, the frequency of occurrence of the payoff-dominant equilibrium is quite small, and independently of the size of the noise $\snoise$. We verify here that in the case of PLA, the pure Nash equilibrium (B,B) is the unique stochastically stable equilibrium. Conversely, in the case of APLA, the frequency of occurrence of the payoff-dominant equilibrium (A,A) goes to 1 as the simulation length increases. These observations validate Theorem~\ref{Th:StochasticallyStableStatesMinimumResistance}, Theorem~\ref{Th:AspirationActionFunctionalAndStochasticStability} and Propositions~\ref{Pr:PLAStochasticStability2x2game}--\ref{Pr:APLAStochasticStability2x2game}. However, note that the simulation time at which the stochastically stable equilibrium is reached/played for first time is often rather early, as Figure~\ref{fig:SimulationStagHuntResultsFinal} verifies. 

The frequency with which the stochastically stable action profile will be observed will definitely depend on the configuration parameters, especially the step sizes $\epsilon$ and $\nu$. As we decrease the step sizes, we may reduce the escape probabilities, thus minimizing variation in the selection of action profiles, however it will take longer simulation time for the frequency to reach 1. For engineering systems this might be prohibitive. However, alternative configuration setups may accommodate different objectives. For example, we may accept an adaptive selection of step sizes, e.g., larger at the beginning to accommodate fast exploration and smaller at later simulation times to reduce variation in the selection of action profiles. The interested reader may experiment with such adaptive setups by using directly the simulation framework at \cite{chasparis2025apla}.

\section{Conclusions}	\label{sec:Conclusions}

This paper introduced aspiration-based perturbed learning automata (APLA), that is a reinforcement-based learning scheme designed to overcome some known limitations of classical payoff-based dynamics in multi-player/action strategic-form games. By coupling action reinforcement with an aspiration factor that reflects agents' satisfaction levels, APLA provides an equilibrium selection mechanism in distributed environments subject to noisy observations. We established stochastic stability results for APLA in multi-player positive-utility and weakly-acyclic games, showing that the dynamics converge in a weak sense to the set of pure Nash equilibria. The methodology presented also simplifies the computation of stochastically stable states, since it translates the problem into a minimization of the aspiration action-functional, and corresponds to a graph property of the game. This analysis can be utilized for deriving conditions under which convergence to Pareto efficient Nash equilibria can be attained.

These results broaden the class of games for which reinforcement-based learning admits convergence guarantees, beyond the potential and coordination game structures traditionally considered in the literature. To the best of our knowledge, APLA constitutes the first reinforcement-based or learning-automaton scheme to ensure convergence in weakly-acyclic multi-player games. 

A specialization to the Stag-Hunt game and accompanying simulations validated the theoretical findings and illustrated the influence of aspiration levels and noise on equilibrium selection. Overall, the proposed framework advances the scope of distributed learning in games and suggests several directions for future work, including adaptive aspiration rules, continuous-action extensions, and applications in large-scale networked systems.

%
%

%

\newpage

\appendix

This part of the report collects supplementary material of the paper, including detailed proofs and also additional simulation tests. Throughout this document, we may cross-reference equations and results of the main document. In order to distinguish between the two documents we will often refer to the Supplementary Material as ``SM'' and the Main Document as ``MD''.

\section{Sample Strategy Evolution under APLA} \label{sec:SM:SampleStrategyEvolution}

\begin{figure}[h!]
\centering
\includegraphics[scale=0.9,keepaspectratio]{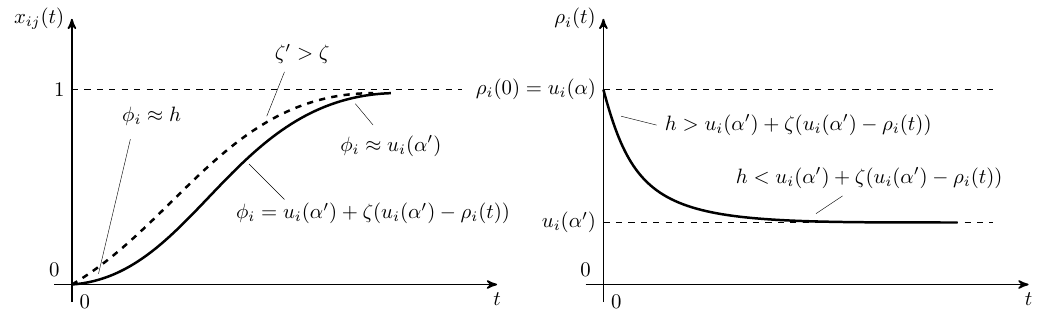}\\(a)\\
\includegraphics[scale=0.9,keepaspectratio]{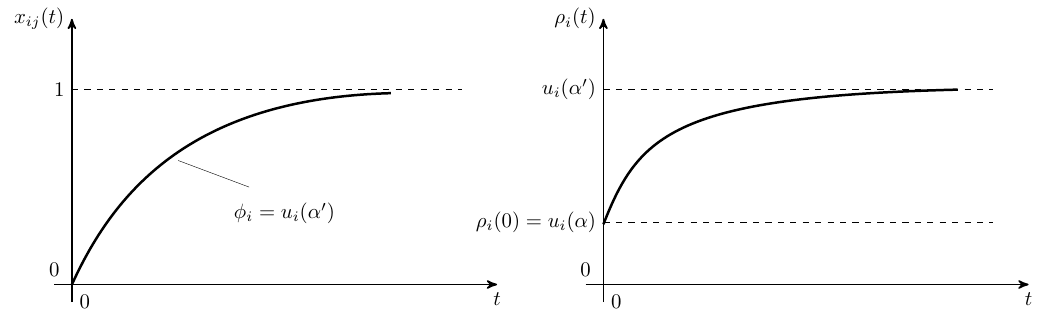}\\ (b)
\caption{A sample evolution of strategy $x_{ij}$ of player $i$ when starting from $x_{ij}(0)=0$ and $\rho_i(0)=u_i(\alpha)$ and action profile $\alpha^+\equiv\alpha'$ is played repeatedly, such that $\alpha_i'=j$ and $\alpha_{-i}'=\alpha_{-i}$. We demonstrate two cases: (a) $u_i(\alpha')<\rho_i(0)\equiv u_i(\alpha)$, i.e., agent $i$ experiences an \emph{unsatisfactory} reward, and b) $u_i(\alpha')>\rho_i(0)$, i.e., agent $i$ experiences a \emph{satisfactory} reward.}
\label{fig:ExamplesOfStrategyEvolution}
\end{figure}

Figure~\ref{fig:ExamplesOfStrategyEvolution} of SM provides a schematic evolution of the strategy of player $i$ with respect to the selected action $\alpha_i'=j$. In part (a), we show the case of how the strategy evolves when $\alpha^+\equiv\alpha'=(\alpha_i',\alpha_{-i}')$ is an unsatisfactory action for player $i$ compared to the initial action $\alpha$, while in part (b), we show the case where $\alpha^+\equiv\alpha'$ is a satisfactory action for player $i$.

\section{Non-Zero-Sum Games} \label{sec:SM:NonZeroSumGames}

The following table provides examples of non-zero-sum games in the case of two players and two actions under the positive-utility property assumption. 


\begin{table}[!ht]
\centering
\begin{minipage}{0.3\textwidth}
\centering
\begin{game}{2}{2}
& A & B\\
A &$5,5$ &$1,3$\\
B &$3,1$ &$4,4$
\end{game}\\[4pt] (a) Stag-hunt game
\end{minipage}\quad 
\begin{minipage}{0.3\textwidth}
\centering
\begin{game}{2}{2}
& A & B\\
A &$3,3$ &$1,1$\\
B &$1,1$ &$2,2$
\end{game}\\[4pt] (b) Typewriter game
\end{minipage}\quad
\begin{minipage}{0.3\textwidth}
\centering
\begin{game}{2}{2}
& A & B\\
A &$3,3$ &$1,4$\\
B &$4,1$ &$2,2$
\end{game}\\[4pt] (c) Prisoner's Dilemma
\end{minipage}
\caption{Types of non-zero-sum games}
\label{Tb:SHG}
\end{table}

Let us consider the Stag-hunt game to discuss some challenges in terms of learning to play desirable outcomes by repeatedly playing a game. This is a type of non-zero-sum game that also belongs to the family of coordination games. Although Nash equilibria are stochastically stable in coordination games under perturbed learning automata, not all Nash equilibria may be desirable. An example may be drawn from the classical Stag-Hunt coordination game of Table~\ref{Tb:SHG}. 
%
%
In this game, the first player selects the row of the payoff matrix and the second player selects the column. The first element of the selected entry determines the reward of the row player, and the second element determines the reward of the column player. This game has two pure Nash equilibria, which correspond to the symmetric plays $(A,A)$ and $(B,B)$. Ideally, we would prefer that agents eventually learn to play $(A,A)$ which corresponds to the payoff-dominant (or Pareto-efficient) equilibrium. However, existing results in perturbed learning automata \cite{ChasparisShamma11_DGA,ChasparisShammaRantzer15,chasparis_stochastic_2019} demonstrate that $(B,B)$ may prevail asymptotically with positive probability. The reason lies in the cost that an agent experiences when the other agent deviates from a Nash equilibrium, which captures the notion of \emph{risk dominance} (cf.,~\cite{HarsanyiSelten88}). In fact, $(B,B)$ is the \emph{risk-dominant} equilibrium in the Stag-Hunt game of Table~\ref{Tb:SHG}.

\section{Load-Balancing Games} \label{sec:SM:LoadBalancingGames}

A class of resource allocation problems in computing systems, usually referred to as \emph{load-balancing games}, can be formulated as \emph{weakly acyclic games}. In such problems, we are given a set of \emph{tasks} (or computing \emph{threads}) that need to be executed in a multi-core computing system (comprising multiple CPU cores). An objective may correspond to the minimization of the \emph{makespan}, that is the maximum \emph{load} over all available CPU cores. In this case, the computing \emph{load} of a CPU core corresponds to the total computing bandwidth requested by all tasks assigned to this core, that is the frequency with which the CPU core is reserved by all tasks.

\begin{figure}[h!]
\centering
\includegraphics[scale=1]{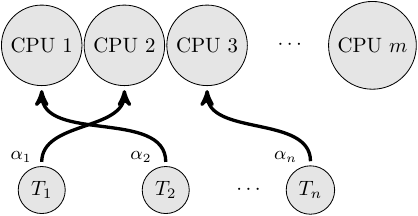}
\caption{A sketch of a load-balancing allocation problem in the context of a multi-core computing system. Each running task independently pins itself to a single CPU-core. Multiple tasks may run on the same CPU-core.}
\label{fig:LoadBalancing}
\end{figure}

More formally, there exist $m$ machines with \emph{speeds} $s_1, s_2, ..., s_m$ and $n$ tasks with \emph{weights} $w_1, w_2, ..., w_n$, where the weight of a task $i$ characterizes its operation level (e.g., the computing bandwidth requested). In order to make the discussion more specific, we will focus on the case of computing systems, where machine $j$ will be associated with CPU-core $j$ and task $i$ will be associated with the computing thread $i$. Under such framework, the speed $s_j$ of a machine $j$ can be represented as the number of instructions per sec (IPS), and the weight $w_i$ of task $i$ could correspond to the \emph{number of completed instructions per sec per unit of available bandwidth}. However, the speed $s_j$ may not necessarily be known in advance (usually average over many different types of tasks). Also, the weight $w_i$ may also not be available in advance, while it may change throughout the execution time of a task. For now, let us assume that these quantities are constant, but not necessarily available in advance. \emph{We would like that threads independently learn to allocate themselves into the available CPU-cores in order to maximize their total average processing speed.} A schematic of this assignment problem is depicted in Figure~\ref{fig:LoadBalancing}.

We introduce the relevant notation of a strategic-form game. We denote $\cI\df\{1,...,n\}$ to be the set of threads requesting resources, and $\mathcal{J}\df\{1,...,m\}$ to be the set of CPU-cores available. In this setting, each thread may be thought of as a player that can decide independently with respect to which one of the available machines to run at. In other words, $\alpha_i\in\mathcal{J}$ corresponds to the action of task $i$, which may be any one of the available machines $\mathcal{J}$, and $\alpha\df(\alpha_1,...,\alpha_n)$ corresponds to the action profile over all tasks (or \emph{assignment}). 

As it is the case in standard operating systems, each task may run in either one of the available machines under no constraints, e.g., all tasks may run on the same machine. However, the number of tasks running on the same machine influences the speed with which these tasks will run (a high number of tasks on the same machine will lead to a low processing speed for these tasks and vice versa). In particular, the \emph{load} of machine $j\in\mathcal{J}$ under assignment $\alpha$ is defined as 
\begin{equation}    \label{eq:LoadBalancingLoad}
\ell_j(\alpha) \df \frac{\sum_{\{k\in\mathcal{I}:\alpha_k = j\}}w_k}{s_{j}} > 0.
\end{equation}
We will also denote the maximum load under assignment profile $\alpha$ as $L(\alpha)\df\max_{j\in\mathcal{J}}\ell_j(\alpha_j)$. In other words, $L(\alpha)$ corresponds to the \emph{makespan}, cf.,~\cite[Chapter~20]{nisan_selfish_2007}. 

Although the speed $s_j$ of CPU-core $j$ and the weight $w_i$ of thread $i$ may not be known in advance, the \emph{running speed} (number of instructions completed per sec) of a thread on a given core can be measured in real-time quite accurately. We define the utility of thread $i$ as its running speed on core $j$, which can be expressed as follows:
\begin{equation}	\label{eq:LoadBalancingUtility}
u_i(\alpha_i=j,\alpha_{-i}) \df \frac{w_i}{\sum_{\{k\in\mathcal{I}:\alpha_k=j\}}w_k}s_j  = \frac{w_i}{\ell_j(\alpha)},
\end{equation}
where we have assumed that the operating system allocates fairly the available bandwidth in CPU core $j$ over all threads and proportionally to their weights. It is important to note that $w_i$ and $\ell_j(\alpha)$ may not be known or easily measured. However, the utility $u_i$ can directly be measured on regular time intervals and per thread. Thus, it can directly be integrated into the implementation of PLA and APLA algorithms. Such game-theoretic design is slightly different from the classical treatment of load-balancing games (see, e.g., \cite{nisan_selfish_2007}), where the cost function of a thread is defined as the load of the machine. 

The strategic-form game, characterized by the tuple $\langle\mathcal{I},\mathcal{A},\{u_i\}_i\rangle$ will be referred to as a \emph{load-balancing game}. We are specifically interested in allocations that correspond to \emph{(pure) Nash equilibria}, that is allocations $\alpha^*$ at which no thread would have the incentive to switch to a different CPU core. In particular, an allocation $\alpha^*$ is a \emph{Nash equilibrium} if $u_i(\alpha_i',\alpha_{-i}^*) \leq u_i(\alpha_i^*,\alpha_{-i}^*)$ for all $\alpha_i'\neq\alpha_i^*$.

\begin{proposition}[Existence of Nash equilibria] 
Consider the load-balancing game characterized by the tuple $\langle\mathcal{I},\mathcal{A},\{u_i\}_i\rangle$ with a utility function defined by (\ref{eq:LoadBalancingUtility}). Then, the set of pure Nash equilibria is non-empty, i.e., $\cA_{\rm NE}\neq\varnothing$. Furthermore, the load-balancing game is a weakly acyclic game.
\end{proposition}
\ifproofs
\begin{proof}
Let us consider any allocation profile $\alpha$ which is not a pure Nash equilibrium. In other words, there exist a thread $i$, currently running on CPU core $j$, and another core $j$, such that, switching from core $j$ to another core $l$ \emph{strictly} increases the utility of thread $i$ (i.e., its processing speed). In particular, given that:
\begin{align}
u_i(\alpha_i=j,\alpha_{-i}) - u_i(\alpha_i'=l,\alpha_{-i}) & = w_i\frac{\ell_l(\alpha') - \ell_j(\alpha)}{\ell_l(\alpha')\ell_j(\alpha)}
\end{align} 
we conclude that, if $u_i(\alpha')>u_i(\alpha)$ (i.e., $\alpha'$ is a better reply to $\alpha$) then $\ell_j(\alpha) > \ell_l(\alpha')$. In other words, if thread $i$ strictly improves its speed by switching from core $j$ to core $l$, it implies that the load of core $j$ (when $i$ runs on core $j$) is strictly larger than the load of core $l$ (when $i$ runs on core $l$). We conclude that $L(\alpha')\leq L(\alpha)$, i.e., under any better reply, the makespan reduces or remains the same. Furthermore, the number of threads that have a load which is equal or higher than $\ell_j(\alpha)$ has now strictly been decreased. We conclude that this process may only terminate at a state where no thread can improve its speed any further, i.e., at a Nash equilibrium, which implies that the game is weakly acyclic.
\end{proof}
\fi

The importance of this proposition lies on the fact that there exists a set of Nash equilibria at which all threads perform well at least locally. There are certain performance guarantees that can be derived with respect to the performance of the tasks at a Nash equilibrium which go a bit away from the scope of this paper, and it may be of independent interest. However, in the context of this work, performing an equilibrium selection and establishing convergence of a more ``preferable'' (in a way to be defined) Nash equilibrium is of interest. This will be discussed in detail in the forthcoming section \ref{sec:SM:SimulationsLoadBalancingGames} where we discuss the simulations of APLA in load-balancing games.

Note that the set of Nash equilibria may not necessarily coincide with the set of optimal allocations $\cA^*$ (e.g., minimum malespan, or maximum total execution speed for all tasks). In fact, the set of optimal allocations may or may not be part of the set of Nash equilibria. However, certain guarantees can be established with respect to the utility achieved at the worst Nash equilibrium as compared to the utility received at an optimal allocation. The following proposition provides a lower bound on the performance of any Nash equilibrium as compared to the performance of an optimal assignment. We only investigate the case of identical CPU cores, since this condition is satisfied by our experimental setup.

\begin{proposition}[Performance of Nash equilibria]     \label{Pr:SM:PerformanceNashEquilibria}
For the case of identical CPU cores and for any pure Nash equilibrium assignment $\alpha\in\cA_{\rm NE}$, the makespan satisfies
\begin{equation}    \label{eq:UpperBoundMakespan}
    L(\alpha) \leq \frac{2\magn{\mathcal{J}}}{\magn{\mathcal{J}}+1} \cdot L^*
\end{equation}
where $\magn{\mathcal{J}}$ denotes the number of available CPU cores. Furthermore, the utility of any thread $i\in\cI$ at any pure Nash equilibrium assignment $\alpha\in\cA_{\rm NE}$ satisfies
\begin{equation}    \label{eq:UpperBoundUtility}
    u_i(\alpha) \geq \frac{\left(\magn{\mathcal{J}}+1\right)}{2\magn{\mathcal{J}}} \cdot \frac{w_i}{L^*}.
\end{equation}
\end{proposition}
\ifproofs
\begin{proof}
The proof of the first statement (\ref{eq:UpperBoundMakespan}) follows the exact same reasoning with Theorem~20.5 in \cite{nisan_selfish_2007}. The proof of the second statement (\ref{eq:UpperBoundUtility}) follows directly from the definition of the utility (\ref{eq:LoadBalancingUtility}) and the first statement (\ref{eq:UpperBoundMakespan}). In particular, let us consider any thread $i$ with weight $w_i$. Its speed will satisfy:
\begin{equation*}
    u_i \geq \frac{w_i}{L(\alpha)} \geq \frac{\left(\magn{\mathcal{J}}+1\right)}{2\magn{\mathcal{J}}} \cdot \frac{w_i}{L^*}.
\end{equation*}
which concludes the proof.
\end{proof}
\fi

The above proposition provides a lower-bound for the utility that can be achieved at a Nash equilibrium assignment. In particular, note that the ratio $u_i^*\df\nicefrac{w_i}{L^*}$ corresponds to the least maximum speed that a thread can achieve under an optimal assignment. Thus, in a 10 CPU-core architecture, condition (\ref{eq:UpperBoundUtility}) implies that $u_i(\alpha)\geq \nicefrac{11}{20}\cdot u_i^*$. Such lower bound is a bit conservative, however it provides a significant guarantee. 

From Equation~(\ref{eq:UpperBoundUtility}), we may also conclude that:
\begin{equation*}
    \frac{1}{\magn{\mathcal{J}}}\sum_{i\in\cI}u_i \geq \frac{\left(\magn{\mathcal{J}}+1\right)}{2\magn{\mathcal{J}}} \cdot \left(\frac{1}{\magn{\mathcal{J}}} \sum_{i\in\mathcal{I}} \frac{w_i}{L^*}\right),
\end{equation*}
which also establishes a similar lower bound with respect to our original (desirable) objective of maximizing the average speed over all threads. We conclude that if threads settle on a Nash equilibrium assignment, then there is a certain guarantee with respect to their average running speed.

\section{Proof of Proposition~\ref{Pr:StrongFellerProperty} (Strong-Feller Property)}		\label{sec:SM:StrongFellerProperty}

Recall that agent $i$ revises its strategy vector $x_i\in\Delta(\magn{\mathcal{A}_i})$ to a new strategy $x_i^+\df x_i(t+1)$ as follows: 
\begin{eqnarray}	\label{eq:ReinforcementLearningModel}
x_i^+ & = & x_i + \epsilon \cdot (e_{\alpha_i^+} - x_i ) \cdot \phi_i\left(\rewardpert{i},\rewardpert{i}-\rho_i\right) \df \Sigma_{i}(\alpha^+,x_i,\rho_i,v_i),
\end{eqnarray}
for some constant step-size $\epsilon>0$. Furthermore, agent $i$ revises its aspiration level $\rho_i\in[\underline{\rho},\overline{\rho}]$ to a new aspiration-level $\rho_i^+\df\rho_i(t+1)$ as follows:
\begin{eqnarray}	\label{eq:AspirationUpdate}
\rho_i^+ & = & \rho_i + \epsilon \nu(\epsilon) \cdot \left( \rewardpert{i}-\rho_i \right) 
\df {K}_i(\alpha^+,\rho_i,v_i).
\end{eqnarray}
for some constant $\nu=\nu(\epsilon)>0$.

Let us consider the perturbed process $P_{\lambda}$. The proof for the unperturbed process will directly follow by setting $\lambda=0$. Let us also consider any sequence $\{z^{(k)}=(\alpha^{(k)},x^{(k)},\rho^{(k)},
\noise^{(k)})\}$ such that $z^{(k)}\to{z}=(\alpha,x,\rho,\noise)\in\cZ$. 

For any open set $O\in\Bor(\cZ)$, let $\mathbb{I}_{O}:\cZ\to\{0,1\}$ denote the index function, i.e.,
\begin{eqnarray*}
\mathbb{I}_{O}(x) \df \begin{cases}
1 & \mbox{ if } x\in{A}, \\
0 & \mbox{ else.}
\end{cases}
\end{eqnarray*} 
The transition probability function of the perturbed process satisfies:
\begin{eqnarray*}
\lefteqn{P_{\lambda}(z^{(k)}=(\alpha^{(k)},x^{(k)},\rho^{(k)},\noise^{(k)}),O) }\cr 
& = & \sum_{\alpha\in\mathcal{P}_{\cA}(O)}\Big\{\prod_{i\in\cI}\tilde{x}_{i\alpha_i}^{(k)} \cdot 
\mathbb{I}_{\mathcal{P}_{\cX_i}(O)} \left(\Sigma_i\left(\alpha,x_i^{(k)},\rho_i^{(k)},\noise_i^{(k)}\right)\right)  \cdot  \mathbb{I}_{\mathcal{P}_{\mathcal{R}_i}(O)}\left(K_i\left(\alpha,\rho_i^{(k)},\noise_i^{(k)}\right)\right)\Big\} \cdot \mathbb{I}_{\mathcal{P}_{\dnoise}(O)}\left(\noise_i^{(k+1)}\right)\Big\}  \cr
\end{eqnarray*}
where $\mathcal{P}_{\cX_i}(O)$, $\mathcal{P}_{\mathcal{A}}(O)$, $\mathcal{P}_{\mathcal{W}_i}(O)$, and $\mathcal{P}_{\dnoise}(O)$ are the \emph{canonical projections} defined by the product topology, and  
$$\tilde{x}_{i\alpha_i}^{(k)}\df (1-\lambda)x_{i\alpha_i}^{(k)} + {\lambda}/{\magn{\cA_i}}$$ is the perturbed strategy of agent $i$ based on which the action profile is selected at the next iteration. Given that the random variable $\noise_i$, which represents the noise in the utility measurements of each agent $i$, is continuously distributed on a closed and bounded interval $\dnoise$, then the same holds for $\Sigma_i(\alpha,\cdot,\cdot,\cdot)$ and ${K}_i(\alpha,\cdot,\cdot)$ since they are both linear functions of their arguments. 
%
Thus, the conclusion follows directly by the definition of the strong Feller property. 

\section{Proof of Theorem \ref{Th:StochasticStability:PureStrategyStates} (Stochastic Stability)} \label{sec:SM:ProofOfTheoremSSPureStrategyStates}

In this section, we provide the main steps for the proof of  Theorem~\ref{Th:StochasticStability:PureStrategyStates}, which are similar to the corresponding ones under the PLA dynamics \cite{chasparis_stochastic_2019} with small adaptations. We begin by investigating the asymptotic behavior of the unperturbed process $P$, and then we characterize the i.p.m. of the perturbed process with respect to the pure strategy states $\cS$. The proof is going to be presented through a series of propositions.

\subsection{Unperturbed Process}	\label{Sc:UnperturbedProcess}

Recall that the \emph{unperturbed process} with t.p.f. $P$ has been defined so that \emph{no agent trembles}, i.e., $\lambda=0$. We first present two technical propositions that will help us characterize the behavior of the unperturbed process. 

\begin{proposition}[Constant action selection]		\label{Pr:SM:ConstantActionSelection}
Let us consider an initial strategy $x(0)$, action profile $\alpha(0)=\alpha$, and aspiration level $\rho(0)$. If action profile $\alpha$ is played continuously for all future times, then the following holds:
\begin{enumerate}
\item[(a)] the strategy of agent $i$, $x_{ij}$, where $j$ corresponds to the selected action $\alpha_i(0)$, evolves with time as follows:
\begin{equation}		\label{eq:SteadyActionSelection_Strategy}
x_{ij}(t) = 1 - (1-x_{ij}(0))\prod_{k=0}^{t-1}H_i(\alpha,\rho_i(k))\,, \quad t=1,2,...,
\end{equation}
where $H_i(\alpha,\rho_i(k)) \df 1 - \epsilon \cdot \phi_i(u_i(\alpha),u_i(\alpha)-\rho_i(k))$. 

\item[(b)] the aspiration level evolves with time as follows
\begin{equation} \label{eq:SteadyActionSelection_AspirationLevel}
\snoise - X^{t}\Delta{u}_i(\alpha) - Y^{t}\snoise \leq \rho_i(t) - u_i(\alpha) \leq -\snoise - X^{t}\Delta{u}_i(\alpha) + Y^{t}\snoise,
\end{equation}
for $t>0$ almost surely (with respect to the measurement noise), where $X \df 1-\epsilon \nu(\epsilon)$, $Y\df 1+\epsilon\nu(\epsilon)$ and $\Delta{u}_i(\alpha) = u_i(\alpha)-\rho_i(0)$.

\end{enumerate}
\end{proposition}
\begin{proof}
(a) To simplify notation, let us denote $\phi_i(k)\equiv \phi_i(u_i(\alpha),u_i(\alpha)-\rho_i(k))$ and $H_i(k)\equiv H_i(\alpha,\rho_i(k))\df 1-\epsilon\phi_i(k)$. Recursively, the strategy of agent $i$ evolves as follows:
\begin{align*}
x_{ij}(1) & = x_{ij}(0) + \epsilon  (1-x_{ij}(0))\phi_i(0) \cr
& = 1 - (1 - x_{ij}(0)) + \epsilon  (1-x_{ij}(0))\phi_i(0) \cr
& = 1 - (1-x_{ij}(0)) H_{i}(0) \cr
x_{ij}(2) & = x_{ij}(1) + \epsilon (1-x_{ij}(1))\phi_i(1) \cr 
& = 1-(1-x_{ij}(0))H_i(0) + \epsilon  (1-x_{ij}(0))H_{i}(0) \phi_i(1)  \cr
& = 1 - (1-x_{ij}(0))H_i(0) (1 - \epsilon \phi_i(1)) \cr
& = 1 - (1-x_{ij}(0))H_i(0)H_i(1)
\end{align*} 
Let $x_{ij}(t) = 1-(1-x_{ij}(0))\prod_{k=0}^{t-1}H_i(k).$ Then,
\begin{align*}
x_{ij}(t+1) = & 1-(1-x_{ij}(0))\prod_{k=0}^{t-1}H_i(k)+\epsilon (1-x_{ij}(0))\prod_{k=0}^{t-1}H_i(k)\phi_i(t) \cr 
= & 1-(1-x_{ij}(0))\prod_{k=0}^{t-1}H_i(k)(1-\epsilon \phi_i(t)) \cr
= & 1-(1-x_{ij}(0))\prod_{k=0}^{t}H_i(k)\,.
\end{align*}
Thus, the conclusion follows by induction.

(b) Recursively, the aspiration level of agent $i$ evolves as follows:
\begin{eqnarray*}
\rho_i(1) & = & \rho_i(0) + \epsilon \nu(\epsilon) \left( \rewardpert{i}(\alpha) - \rho_i(0)\right) \cr
& \geq &  \rho_i(0) + \epsilon\nu(\epsilon) \Delta{u}_i(\alpha) - \epsilon\nu(\epsilon)\snoise \cr
& = & u_i(\alpha) - X \Delta{u}_i(\alpha) + \snoise - Y\snoise \cr 
\rho_i(1) & \leq & \rho_i(0) + \epsilon\nu(\epsilon)\Delta{u}_i(\alpha) + \epsilon\nu(\epsilon)\snoise \cr 
& = & u_i(\alpha) - X \Delta{u}_i(\alpha) - \snoise + Y \snoise,
\end{eqnarray*}
almost surely. Let $\rho_i(t)$ satisfy the desired recursion. Then, we have
\begin{eqnarray*}
\rho_i(t+1) & = & \rho_i(t) + \epsilon\nu(\epsilon) (\rewardpert{i}(\alpha) - \rho_i(t)) \cr
& \leq & u_i(\alpha) -\snoise  - X^{t}\Delta{u}_i(\alpha) + Y^{t}\snoise + \cr && \epsilon\nu(\epsilon) \left( u_i(\alpha) + \snoise - u_i(\alpha) - \snoise + X^t\Delta{u}_i(\alpha) + Y^{t}\snoise\right) \cr
& = & u_i(\alpha) - \snoise - (1-\epsilon\nu(\epsilon))X^{t}\Delta{u}_i(\alpha) + (1+\epsilon\nu(\epsilon))Y^{t}\snoise \cr
& = & u_i(\alpha) - \snoise - X^{t+1}\Delta{u}_i(\alpha) + Y^{t+1}\snoise
\end{eqnarray*}
and
\begin{eqnarray*}
\rho_i(t+1) & \geq & u_i(\alpha) +\snoise - X^{t}\Delta{u}_i(\alpha) - Y^{t}\snoise + \cr && \epsilon\nu(\epsilon)\left(u_i(\alpha)-\snoise - u_i(\alpha) + \snoise + X^{t}\Delta{u}_i(\alpha) - Y^{t-1}(1-Y)\snoise\right) \cr & = & 
u_i(\alpha) + \snoise - (1-\epsilon\nu(\epsilon))X^{t}\Delta{u}_i(\alpha) - (1+\epsilon\nu(\epsilon)) Y^{t} \snoise \cr 
& = & u_i(\alpha) + \snoise - X^{t+1}\Delta{u}_i(\alpha) - Y^{t+1}\snoise
\end{eqnarray*}
almost surely. Thus, the conclusion follows by induction.
\end{proof}

In the above proposition, we do not impose any constraint in the aspiration level and aspiration factor. As long as the initial conditions of Proposition~\ref{Pr:SM:ConstantActionSelection} are satisfied, the strategy associated with action $j$ will evolve as Equations~(\ref{eq:SteadyActionSelection_Strategy})--(\ref{eq:SteadyActionSelection_AspirationLevel}) dictate.

We will utilize the above properties to assess the behavior of the unperturbed process as time increases. For $t\geq{0}$ define the sets
\begin{eqnarray*}
A_{t} & \df & \{\omega\in\Omega: \alpha(\tau) = \alpha(t)\,, \mbox{ for all } \tau\geq{t} \}\,, \cr
B_{t} & \df & \{\omega\in\Omega: \alpha(\tau) = \alpha(0)\,, \mbox{ for all } 0\leq \tau \leq t \}\,.
\end{eqnarray*}
In other words, $A_t$ corresponds to the event that the same action is selected for all times after $t$, while $B_t$ corresponds to the event that the same action has been played until time $t$. Note that $\{B_{t}:t\geq{0}\}$ is a non-increasing sequence, i.e., $B_{t+1}\subseteq B_{t}$, while $\{A_{t}:t\geq{0}\}$ is non-decreasing, i.e., $A_{t}\subseteq A_{t+1}$. Let also
\begin{eqnarray*}
A_{\infty} \df \bigcup_{t=0}^{\infty}A_t \mbox{ and } B_{\infty} \df \bigcap_{t=1}^{\infty}B_t.
\end{eqnarray*}
In other words, $A_{\infty}$ corresponds to the event that agents eventually play the same action profile, while $B_{\infty}$ corresponds to the event that agents never change their actions.

Let us define also the event 
\begin{eqnarray*}
\Gamma_{\delta,t} \df \{\omega\in\Omega: \rho_i(\tau)\geq u_i(\alpha)-\delta\,,\mbox{ for some } \alpha\in\cA\,, \forall i\in\cI\,,\forall\tau\geq{t}\}.
\end{eqnarray*}
i.e., the event $\Gamma_{\delta,t}$ corresponds to the aspiration level being at least $\delta$-close (from below) to the nominal utility of some action profile or larger.

Let $\uptau(D)$ denote the first hitting time of the unperturbed process to a set $D\in\Bor(\cZ)$. Recall the shift operator $\theta_t$, defined as $\theta_t:\Omega\mapsto\Omega$ for some finite time step $t$, that satisfies $(Z_s\circ\theta_t)(\omega) = Z_s(\theta_{t}(\omega))=Z_{s+t}(\omega)$, i.e., it shifts the sequence by $t$ time steps backwards. Note also that $A_t\circ\theta_t=B_{\infty}$ or equivalently $A_t = \theta_t^{-1}(B_{\infty})$, where $\theta_t^{-1}$ defines the inverse shift operator. 

The first proposition states that, for any initial state $z$, the unperturbed process will end up playing the same action for all future times.
\begin{proposition}[Convergence to p.s.s.]	\label{Pr:SM:ConvergenceToPureStrategyStates}
Let us assume that $\epsilon>0$, $h>0$ and $\snoise>0$ are sufficiently small such that $0<\epsilon \rewardpert{i}(\alpha)<1$ and $h<\rewardpert{i}(\alpha)$ almost surely for all $\alpha\in\cA$ and $i\in\cI$. Then, the following hold:
\begin{itemize}
\item[(a)] $\inf_{z\in{\cZ}} \Prob_{z}[B_{\infty}] > 0\,,$
\item[(b)] for any $\delta>0$ and noise of size $\snoise<\delta$, $\inf_{z\in{\cZ}} \Prob_{z}[A_{\infty}\cap\Gamma_{\delta,\infty}] = 1.$
\end{itemize}
\end{proposition}
\ifproofs
\begin{proof}
(a) To simplify notation, let us denote $\phi_i(k)\equiv \phi_i(\rewardpert{i}(\alpha),\rewardpert{i}(\alpha)-\rho_i(k))$ and $H_i(k)\equiv H_i(\alpha,\rho_i(k))\df 1-\epsilon\phi_i(k)$. Let us consider an action profile $\alpha=(\alpha_1,...,\alpha_n)\in\cA$ and an initial strategy profile $x(0)=(x_1(0),...,x_n(0))$ such that $x_{i\alpha_i}(0)>0$ for all $i\in\cI$. Note that if the same action profile $\alpha$ is selected consecutively up to time $t$, then, according to Proposition~\ref{Pr:SM:ConstantActionSelection}, the strategy of agent $i$ evolves as follows: 
\begin{eqnarray}    \label{eq:ConvergenceToPSS:AccumulatedStrategy}
    x_{i\alpha_i}(t) & = & 1-(1-x_{i\alpha_i}(0))\prod_{k=0}^{t-1}H_i(k)\,, \quad t>0\,, \cr
    & = & 1 - (1-x_{i\alpha_i}(0))\prod_{k=0}^{t-1}\left(1-\epsilon \phi_i(k)\right)\,, \quad t>0\,, \cr
    & \geq & 1 - (1-x_{i\alpha_i}(0))\prod_{k=0}^{t-1}\left(1-\epsilon h\right)\,, \quad t>0\,,
\end{eqnarray}
where we have used the property that $H_i(t)\df 1-\epsilon \phi_i(t) \leq 1-\epsilon{h}$ for all $t\geq{0}$, under the assumption that $h>0$ is sufficiently small such that $h<\rewardpert{i}(\alpha)$, almost surely, for all $i\in\cI$ and $\alpha\in\cA$. Given also that $1-\epsilon{h}<1$, we have that $\lim_{t\to\infty}\prod_{k=0}^{t-1}(1-\epsilon h) = 0$ which further implies that $\lim_{t\to\infty}x_{i\alpha_i}(t)=1$. 

The event $B_t$ is non-increasing, thus from continuity from above we have
\begin{equation}	\label{eq:ConvergenceToPSS:Binfty}
\Prob_{z}[B_{\infty}] = \lim_{t\to\infty}\Prob_{z}[B_t] = \lim_{t\to\infty}\prod_{k=0}^{t}\prod_{i=1}^{n}x_{i\alpha_i}(k).
\end{equation}
Note that $\Prob_{z}[B_{\infty}] > 0$ if and only if 
\begin{equation}	\label{eq:ConvergenceToPSS:Condition1}
\sum_{t=0}^{\infty}\log(x_{i\alpha_i}(t)) > -\infty, \mbox{ for all } i\in\mathcal{I}.
\end{equation}
Let us introduce the variable $y_i(t) \df 1-x_{i\alpha_i}(t),$ which corresponds to the probability of agent $i$ selecting any action other than $\alpha_i$. Hence, $\lim_{t\to\infty}y_i(t) = 0$. Condition (\ref{eq:ConvergenceToPSS:Condition1}) is equivalent to
\begin{equation}	\label{eq:ConvergenceToPSS:Condition2}
-\sum_{t=0}^{\infty}\log(1-y_i(t)) < \infty,	\mbox{ for all } i\in\mathcal{I}.
\end{equation}
Note that $y_{i}(t+1)/y_i(t) = 1-\epsilon \phi_i < 1$, due to the fact that $h\leq \phi_i\leq \rewardpert{i}(\alpha)$. Then, the Ratio test, cf.,~\cite[Theorem~6.2.4]{Reed98} implies that the series of positive terms $\sum_{t=1}^{\infty}y_i(t)$ is convergent. Thus, from L'Hospital's rule (cf.,~\cite[Theorem~5.13]{Rudin64}),
\begin{equation}	\label{eq:LHopital}
\lim_{t\to\infty}\frac{-\log(1-y_i(t))}{y_i(t)} = \lim_{t\to\infty}\frac{1}{1-y_i(t)} = 1 > 0. 
\end{equation}
From the Limit Comparison Test (cf.,~\cite[Theorem~6.2.2]{Reed98}), we conclude that condition (\ref{eq:ConvergenceToPSS:Condition2}) holds, which equivalently implies that $\Prob_z[B_{\infty}]>0$. 
Lastly, due to (\ref{eq:ConvergenceToPSS:AccumulatedStrategy}), $\Prob_z[B_{\infty}]$ is continuous with respect to the initial strategy $x(0)$ which takes values in a bounded and closed set $\cX$. Thus, by \cite[Theorem~3.2.2]{Reed98}, we conclude that $\inf_{z\in\cZ}\Prob_{z}[B_{\infty}] > 0$.

(b) Define the event $$E_{\delta,\ell} \df \{z=(\alpha,x,\rho,\noise)\in\cZ: x_{i\alpha_i} > 1-\epsilon^{\ell}\,, \rho_i \geq u_i(\alpha) - \delta\,, \forall i\in\cI \},$$ i.e., $E_{\ell}$ corresponds to the event of a strategy being $\epsilon^{\ell}$ close to a vertex of $\cX$, and the aspiration level being at least $\delta$-close (from below) to the corresponding utility or larger. Then, for $\ell>0$, we have:
\begin{eqnarray}    
\Prob_z[A_t\cap\Gamma_{\delta,t}] & \geq & \sum_{k=1}^{t}\Prob_{z}\left[ \uptau(E_{\delta,\ell}) = k\,, Z \circ\theta_k\in B_{\infty} \right] \quad \mbox {(a.s.)} \label{eq:A_infty_proof_eq1.1} \\
& = & \sum_{k=1}^{t}\Prob_{z}\left[Z \circ\theta_k\in B_{\infty} | \uptau(E_{\delta,\ell}) = k \right] \cdot \Prob_{z}\left[  \uptau(E_{\delta,\ell}) = k \right] \label{eq:A_infty_proof_eq1.2} \\
& \geq & \inf_{z\in E_{\delta,\ell}}\Prob_{z}[B_{\infty}]\cdot \sum_{k=1}^{t}\Prob_{z}\left[ \uptau(E_{\delta,\ell})=k \right] \cr 
& \geq & \inf_{z\in E_{\delta,\ell}}\Prob_{z}[B_{\infty}]\cdot \inf_{z\in E_{\delta,\ell}^{c}}\Prob_{z}\left[\uptau(E_{\delta,\ell})\leq t\right]. \label{eq:A_infty_proof_eq1.3}
\end{eqnarray}
Note that (\ref{eq:A_infty_proof_eq1.1}) results from the fact that when the unperturbed process has reached $E_{\delta,\ell}$ and the same action is played thereafter, the stategy will continue to satisfy $x_{i\alpha_i} > 1-\epsilon^{\ell}$ and the aspiration level $\rho_i \geq u_i(\alpha) - \delta$ almost surely (a.s.), given that $\snoise<\delta$. Furthermore, in (\ref{eq:A_infty_proof_eq1.2}) we have used the properties of the conditional probability and in (\ref{eq:A_infty_proof_eq1.3}) we have used the Markov property. 

In the remainder of the proof, we will investigate the terms of the right-hand side of the above inequality. Consider the subsequence $t_k=k\ell^m$, for some $m=m(\ell)>0$ such that the time block of $\ell^m$ iterations is sufficiently large so that $E_{\delta,\ell}$ can be reachable from any state in $E_{\delta,\ell}^c$ (a.s). Then,
\begin{equation*}
\Prob_{z}\left[\uptau(E_{\delta,\ell})\leq t_k|\uptau(E_{\delta,\ell})>t_{k-1}\right] \geq \inf_{z\in E_{\delta,\ell}^{c}}\Prob_{z}[B_{\ell^{m}}] \geq \inf_{z\in E_{\delta,\ell}^{c}}\Prob_{z}[B_{\infty}],
\end{equation*}
where the last inequality is due to the fact that the process $\{B_t:t\geq 0\}$ is non-increasing, i.e., $B_{t+1}\subseteq B_{t}$. Furthermore, from part (a) we have that $\inf_{z\in E_{\ell}^{c}}\Prob_{z}[B_{\infty}] > 0$. Hence, from the counterpart of the Borel-Cantelli Lemma (cf.,~\cite[Lemma~1]{bruss_counterpart_1980}) and the fact that $\{\uptau(E_{\delta,\ell})\leq t_k\}\subseteq \{\uptau(E_{\delta,\ell})\leq t_{k+1}\}$ (i.e., it is a non-decreasing sequence of events), we have that, for any $\ell>0$, 
\begin{equation}    \label{eq:A_infty_proof_eq2}
    \lim_{k\to\infty}\inf_{z\in E_{\delta,\ell}^{c}}\Prob_{z}[\uptau(E_{\delta,\ell})\leq t_k] = 1.
\end{equation}
Finally, set $k=\ell$. Then, $t_k=t_{\ell}=\ell^{m+1}$. Given (\ref{eq:A_infty_proof_eq1.3})--(\ref{eq:A_infty_proof_eq2}) and the fact $(A_t\cap\Gamma_{\delta,t})\subseteq(A_{t+1}\cap\Gamma_{\delta,t+1})$, we have from continuity from below
\begin{equation*}
    \Prob_{z}[A_{\infty}\cap\Gamma_{\delta,\infty}] = \lim_{\ell\to\infty}\Prob_{z}[A_{t_\ell}\cap\Gamma_{\delta,t_\ell}] \geq \lim_{\ell\to\infty}\inf_{z\in E_{\delta,\ell}} \Prob_z[B_{\infty}] = 1,
\end{equation*}
where in the last equality we have used the definition of $E_{\delta,\ell}$. Thus, the conclusion follows.
\end{proof}
\fi

Note that in the proof of Proposition~\ref{Pr:SM:ConvergenceToPureStrategyStates}, the exact way the aspiration level is updated plays no role in approaching a p.s.s. The property that  is needed is that the aspiration factor admits only positive values, which is directly implied by Property~\ref{P:PositiveUtilityProperty}, i.e., the positive-utility property. Furthermore the noise in the utility function also plays no role, as long as the utility values do not violate the positive utility property almost surely. This is satisfied as long as the size of the noise is strictly smaller than the smallest possible (positive) reward received.

Under these conditions, the convergence properties of the strategy vector of the unperturbed process seems rather unaffected of even possibly noisy utilities. The reason for this is that the noise or the aspiration factor influence the level/amount of reinforcement, but not the sign of reinforcement (which is always positive). In a way, the strategy vector acts as a filter in absorbing the variations in the utility or the aspiration factor. 

The following proposition provides asymptotic properties of the unperturbed process, also in connection with its invariant probability measure (i.p.m.) and its support. We let $C_b(\cZ)$ denote the Banach space of real-valued continuous functions on $\cZ$ under the sup-norm (denoted by $\|\cdot\|_{\infty}$) topology. For $f\in\Cc(\cZ)$, define
\begin{equation*}
P_{\lambda}f(z) \df \int_{\cZ}P_{\lambda}(z,dy)f(y),
\end{equation*}
and 
\begin{equation*}
\mu[f] \df \int_{\cZ}\mu(dz)f(z), \mbox{ for } \mu\in\mathfrak{P}(\cZ).
\end{equation*}

\begin{proposition}[Limiting t.p.f. of unperturbed process]	\label{Pr:SM:LimitingUnperturbedTPF}
Let $\mu$ denote an i.p.m. of $P$. Then, there exists a t.p.f. $\Pi$ on $\cZ\times\Bor(\cZ)$ with the following properties:
\begin{itemize}
\item[(a)] for $\mu$-a.e. $z\in\cZ$, $\Pi(z,\cdot)$ is an i.p.m. for $P$;
\item[(b)] for all $f\in\Cc(\cZ)$, $\lim_{t\to\infty}\|P^tf-\Pi f\|_{\infty}=0$;
\item[(c)] $\mu$ is an i.p.m. for $\Pi$;
\item[(d)] the support\footnote{The \emph{support} of a measure $\mu$ on $\cZ$ is the unique closed set $F\subset\Bor(\cZ)$ such that $\mu(\cZ\backslash{F})=0$ and $\mu(F\cap{O})>0$ for every open set $O\subset\cZ$ such that $F\cap{O}\neq\varnothing$.} of $\Pi$ is on $\cS$ for all $z\in\cZ$.
\end{itemize}
\end{proposition}
\ifproofs
\begin{proof}
The state space $\cZ$ is a locally compact separable metric space\footnote{A topological metric space $\cZ$ is called locally compact if every point $z$ of $\cZ$ has a compact neighborhood. A topological space is called separable if it contains a countable, dense subset.} and the t.p.f. of the unperturbed process $P$ admits an i.p.m. due to the strong-Feller property as shown in Proposition~\ref{Pr:StrongFellerProperty}. 
Thus, statements (a), (b) and (c) follow directly from \cite[Theorem~5.2.2 (a), (b), (e)]{Lerma03}. 

(d) Let us assume that the support of $\Pi$ includes points in $\cZ$ other than the p.s.s. in $\cS$. Then, there exists an open set $O\in\Bor(\cZ)$ such that $O\cap\cS=\varnothing$ and $\Pi(z^*,O)>0$ for some $z^*\in\cZ$. According to (b), $P^{t}$ converges weakly to $\Pi$. Thus, from the Portmanteau theorem (cf.,~\cite[Theorem~1.4.16]{Lerma03}), we have that $\liminf_{t\to\infty} P^{t}(z^*,O) \geq \Pi(z^*,O)>0.$ However, this contradicts Proposition~\ref{Pr:SM:ConvergenceToPureStrategyStates} of the Supplementary Material. 
\end{proof}
\fi

Proposition~\ref{Pr:SM:LimitingUnperturbedTPF} states that the limiting unperturbed t.p.f. converges weakly to a t.p.f. $\Pi$ which accepts the same i.p.m. as $P$. Furthermore, \emph{the support of $\Pi$ is the set of p.s.s. in $\cS$}. This is a rather handy property, since the limiting perturbed process $P_{\lambda}$ can also be ``related'' (in a weak-convergence sense) to the t.p.f. $\Pi$, as it will be shown in the following section.

\subsection{Invariant probability measure (i.p.m.) of the perturbed process}

Note that the t.p.f. of the perturbed process can be decomposed as follows:
\begin{subequations}
\begin{equation}	\label{eq:tpf_decomposition_step1}
P_{\lambda} = (1-\varphi(\lambda))P + \varphi(\lambda)Q_{\lambda}
\end{equation}
where $$\varphi(\lambda) \df 1-(1-\lambda)^{n}$$ is the probability that at least one agent trembles (since $(1-\lambda)^{n}$ is the probability that no agent trembles), and $Q_{\lambda}$ corresponds to the t.p.f. when at least one agent trembles. Note that $\varphi(\lambda)\to{0}$ as $\lambda\downarrow{0}$. Define also $Q$ as the t.p.f. of the one-step process where \emph{exactly one agent trembles}, i.e., it plays an action uniformly at random, and $Q^*$ as the t.p.f. where \emph{at least two agents tremble}. Then, we may write:
\begin{equation}	\label{eq:tpf_decomposition_step2}
Q_{\lambda} = (1-\psi(\lambda))Q+\psi(\lambda)Q^*,
\end{equation}
\end{subequations}
where
$$\psi(\lambda) \df 1-\frac{n\lambda(1-\lambda)^{n-1}}{1-(1-\lambda)^{n}}$$
corresponds to the probability that at least two agents tremble given that at least one agent trembles. It also satisfies $\psi(\lambda)\to{0}$ as $\lambda\downarrow{0}$, which establishes an approximation of $Q_{\lambda}$ by $Q$ as the perturbation factor $\lambda\downarrow{0}$. 

Define also the infinite-step t.p.f. when trembling only at the first step (briefly, \emph{lifted} t.p.f.) as follows: 
\begin{equation}	\label{eq:tpf_lifted}
P_{\lambda}^{L} \df \varphi(\lambda)\sum_{t=0}^{\infty}(1-\varphi(\lambda))^{t}Q_{\lambda} P^{t} = Q_{\lambda} R_{\lambda}
\end{equation}
where
\begin{equation}	\label{eq:tpf_resolvent}
R_{\lambda} \df \varphi(\lambda)\sum_{t=0}^{\infty}(1-\varphi(\lambda))^{t}P^{t},
\end{equation}
i.e., $R_{\lambda}$ corresponds to the \emph{resolvent} t.p.f. 

In the following proposition, we establish weak-convergence of the lifted t.p.f. $P_{\lambda}^{L}$ to $Q\Pi$ as $\lambda\downarrow{0}$, which will further allow for an explicit characterization of the weak limit points of the i.p.m. of $P_{\lambda}$. The proof follows the exact same reasoning with \cite[Proposition~4.3]{chasparis_stochastic_2019} under the PLA dynamics. Again both the aspiration factor and the noise term have no explicit influence in this derivation. Their influence is entailed within the unperturbed process t.p.f. $P$. 

\begin{proposition}[i.p.m. of perturbed process]		\label{Pr:SM:WeakLimitPointsOfPerturbedInvariantMeasures}
Consider the decomposition of the perturbed t.p.f. as defined by Equations~(\ref{eq:tpf_decomposition_step1})--(\ref{eq:tpf_decomposition_step2}). Consider also the lifted t.p.f. $P_{\lambda}^{L}$ and the resolvent t.p.f. $R_{\lambda}$, as defined by Equations~(\ref{eq:tpf_lifted})--(\ref{eq:tpf_resolvent}), respectively. The following hold:
\begin{itemize}
\item[(a)] For $f\in\Cc(\cZ)$, $\lim_{\lambda\to{0}}\|R_{\lambda}f-\Pi{f}\|_{\infty} = 0.$
\item[(b)] For $f\in\Cc(\cZ)$, $\lim_{\lambda\to{0}}\|P_{\lambda}^{L}f-Q\Pi{f}\|_{\infty} = 0$.
\item[(c)] Any invariant distribution $\mu_{\lambda}$ of $P_\lambda$ is also an invariant distribution of $P_{\lambda}^{L}$.
\item[(d)] Any weak limit point in $\mathfrak{P}(\cZ)$ of $\mu_{\lambda}$, as $\lambda\to{0}$, is an i.p.m. of $Q\Pi$.
\end{itemize}
\end{proposition}
\ifproofs
\begin{proof}
(a) For any $f\in C_b(\cZ)$, we have
\begin{eqnarray*}
\|R_{\lambda}f - \Pi{f}\|_{\infty} & = & \supnorm{\varphi(\lambda)\sum_{t=0}^{\infty}(1-\varphi(\lambda))^tP^tf - \Pi f} \cr & = & \supnorm{\varphi(\lambda)\sum_{t=0}^{\infty}(1-\varphi(\lambda))^t(P^t f - \Pi f)} \cr
\end{eqnarray*}
where we have used the property $\varphi(\lambda)\sum_{t=0}^{\infty}(1-\varphi(\lambda))^t=1$. Note that
\begin{eqnarray*}
\varphi(\lambda)\sum_{t=T}^{\infty}(1-\varphi(\lambda))^t\supnorm{P^tf-\Pi f} \leq (1-\varphi(\lambda))^{T}\sup_{t\geq{T}}\supnorm{P^tf - \Pi f}. 
\end{eqnarray*}
From Proposition~\ref{Pr:SM:LimitingUnperturbedTPF}(b), we have that for any $\delta>0$, there exists $T=T(\delta)>0$ such that the r.h.s. is uniformly bounded by $\delta$ for all $t\geq T$. Thus, the sequence $$A_T\df\varphi(\lambda)\sum_{t=0}^{T}(1-\varphi(\lambda))^t(P^tf - \Pi f)$$ is Cauchy and therefore convergent (under the sup-norm). In other words, there exists $A\in\mathbb{R}$ such that $\lim_{T\to\infty}\supnorm{A_{T}-A}=0.$ 
For every $T>0$, we have
\begin{eqnarray*}
\supnorm{R_{\lambda}f-\Pi{f}} \leq \supnorm{A_{T}} + \supnorm{A - A_T}.
\end{eqnarray*}
Note that 
\begin{eqnarray*}
\supnorm{A_T} \leq \varphi(\lambda) \sum_{t=0}^{T} (1-\varphi(\lambda))^{t}  \supnorm{P^t f-\Pi f}.
\end{eqnarray*}
If we take $\lambda\downarrow{0}$, then the r.h.s. converges to zero. Thus,
\begin{equation*}
\|R_{\lambda}f-\Pi{f}\|_{\infty} \leq \supnorm{A-A_T}, \mbox{ for all } T>0,
\end{equation*}
which concludes the proof.

(b) For any $f\in C_{b}(\cZ)$, we have
\begin{eqnarray*}
\lefteqn{ \|P_{\lambda}^{L}f-Q\Pi{f}\|_{\infty} } \cr
& \leq & \|Q_{\lambda}(R_{\lambda}f-\Pi{f})\|_{\infty} + \|Q_{\lambda}\Pi{f} - Q\Pi{f}\|_{\infty} \cr
& \leq & \|R_{\lambda}f-\Pi{f}\|_{\infty} + \|Q_{\lambda}\Pi{f}-Q\Pi{f}\|_{\infty}.
\end{eqnarray*}
The first term of the r.h.s. approaches 0 as $\lambda\downarrow{0}$ according to (a). The second term of the r.h.s. also approaches 0 as $\lambda\downarrow{0}$ since $Q_{\lambda}\rightarrow{Q}$ as $\lambda\downarrow{0}$.

(c) By definition of the perturbed t.p.f. $P_{\lambda}$, we have
\begin{equation*}
P_{\lambda}R_{\lambda} = (1-\varphi(\lambda))PR_{\lambda} + \varphi(\lambda)Q_{\lambda}R_{\lambda}.
\end{equation*}
Note that $Q_{\lambda}R_{\lambda}=P_{\lambda}^{L}$ and $(1-\varphi(\lambda))PR_{\lambda} = R_{\lambda} - \varphi(\lambda)I,$ where $I$ corresponds to the identity operator. Thus, 
\begin{equation*}
P_{\lambda}R_{\lambda} = R_{\lambda}-\varphi(\lambda)I+\varphi(\lambda)P_{\lambda}^{L}.
\end{equation*}
For any i.p.m. of $P_{\lambda}$, $\mu_{\lambda}$, we have
\begin{equation*}
\mu_{\lambda}P_{\lambda}R_{\lambda} = \mu_{\lambda}R_{\lambda}-\varphi(\lambda)\mu_{\lambda}+\varphi(\lambda)\mu_{\lambda}P_{\lambda}^{L},
\end{equation*}
which equivalently implies that $\mu_{\lambda} = \mu_{\lambda}P_{\lambda}^{L},$ since $\mu_{\lambda}P_{\lambda} = \mu_{\lambda}$. We conclude that $\mu_{\lambda}$ is also an i.p.m. of $P_{\lambda}^{L}$.

(d) Let $\hat{\mu}$ denote a weak limit point of $\mu_{\lambda}$ as $\lambda\downarrow{0}$. To see that such a limit exists, take $\hat{\mu}$ to be an i.p.m. of $P$. Then, 
\begin{eqnarray*}
\lefteqn{\|P_{\lambda}f-P{f}\|_{\infty}} \cr & \geq & \|\mu_{\lambda}(P_{\lambda}f-P{f})\|_{\infty} \cr 
& = & \|\mu_\lambda P_{\lambda} f - Pf - \hat{\mu}Pf + \hat{\mu}Pf \|_{\infty} \cr 
& = & \|\mu_{\lambda}(I-P)[f] - \hat{\mu} (I - P) [f]\|_{\infty} \cr
& = & \|(\mu_{\lambda}-\hat{\mu})(I-P)[f]\|_{\infty}.
\end{eqnarray*} 
Note that the weak convergence of $P_{\lambda}$ to $P$, it necessarily implies that $\mu_{\lambda}\Rightarrow\hat{\mu}$. Note further that
\begin{eqnarray*}
\hat{\mu}[f] - \hat{\mu}Q\Pi{f}= (\hat{\mu}[f]-\mu_{\lambda}[f]) + \mu_{\lambda}[P_{\lambda}^{L}f-Q\Pi{f}]+ (\mu_{\lambda}[Q\Pi{f}]-\hat{\mu}[Q\Pi{f}]),
\end{eqnarray*}
where we have used the fact that $\mu_{\lambda}=\mu_{\lambda}P_{\lambda}^{L}$ from (c). The first and the third term of the r.h.s. approaches 0 as $\lambda\downarrow{0}$ due to the fact that $\mu_{\lambda}\Rightarrow\hat{\mu}$. The same holds for the second term of the r.h.s. due to part (b). Thus, we conclude that any weak limit point of $\mu_{\lambda}$ as $\lambda\downarrow{0}$ is an i.p.m. of $Q\Pi$.
\end{proof}
\fi

Proposition~\ref{Pr:SM:WeakLimitPointsOfPerturbedInvariantMeasures} establishes convergence (in a weak sense) of the i.p.m. $\mu_{\lambda}$ of the perturbed process to an i.p.m. of $Q\Pi$. In the following section, this convergence result will allow for a more explicit characterization of $\mu_{\lambda}$ as $\lambda\downarrow{0}$.

\subsection{Equivalent finite-state Markov process}	\label{sec:FiniteStateMarkovChainEquivalence_old}

Define the finite-state Markov process $\hat{P}$  such that, for any $s,s'\in\mathcal{S}$, $$\hat{P}_{ss'}\df\lim_{t\to\infty}QP^{t}(s,\Neigh{\delta}{s'}),$$ for any $\delta$ sufficiently small such that $\delta>\snoise$. Note that $\hat{P}_{ss'} = \hat{P}_{ss'}(\delta,\epsilon,\snoise)$, i.e., it depends on the selection of $\delta$, the step size $\epsilon$, and the size of the noise $\snoise<\delta$. In the absence of noise, $\delta$ could be taken arbitrarily small, however when $\snoise>0$, we need to ensure the hypotheses of Proposition~\ref{Pr:SM:ConvergenceToPureStrategyStates}(b).

\begin{proposition} [Unique i.p.m. of $Q\Pi$]		\label{Pr:SM:UniqueInvariantPMofQPi}
Let us consider the hypotheses of Proposition~\ref{Pr:SM:ConvergenceToPureStrategyStates}(a) of sufficiently small $\epsilon>0$, $h>0$ and $\snoise>0$ such that $0<\epsilon\rewardpert{i}(\alpha)<1$ and $0<h<\rewardpert{i}(\alpha)$ almost surely for all $\alpha\in\cA$ and $i\in\cI$. Then, the following hold:
\begin{itemize}
\item[(a)] There exists a unique i.p.m. $\hat{\mu}$ of $Q\Pi$ that satisfies 
\begin{equation}	\label{eq:InvariantMeasureQP_derivation}
\hat{\mu}(\cdot) = \sum_{s\in\mathcal{S}}\hat{\pi}_s\Dirac{s}(\cdot)
\end{equation}
for some constants $\hat{\pi}_s\geq{0}$, $s\in\mathcal{S}$. 
\item[(b)] For any sufficiently small $\delta>\snoise$, define the limiting transition probabilities 
\begin{equation}		\label{eq:DefinitionTransitions}
\hat{P}_{ss'} = \hat{P}_{ss'}(\delta,\epsilon,\snoise) \df \lim_{t\to\infty}QP^t(s,\Neigh{\delta}{s'}).
\end{equation}
Then, 
$\hat{\pi}=\hat{\pi}(\delta,\epsilon,\snoise) \df \hat{\mu}(\Neigh{\delta}{s})$ with $\hat{\pi}=(\hat{\pi}_1,...,\hat{\pi}_{\magn{\mathcal{S}}})$ is the unique invariant distribution of $\hat{P}$, i.e., $\hat{\pi}=\hat{\pi}\hat{P}$.
\end{itemize}
\end{proposition}
\ifproofs
\begin{proof}
(a) From Proposition~\ref{Pr:SM:LimitingUnperturbedTPF}(d), we have that the support of $\Pi$ is the set of p.s.s. $\mathcal{S}$. Thus, the support of $Q\Pi$ is also on $\mathcal{S}$. From Proposition~\ref{Pr:SM:WeakLimitPointsOfPerturbedInvariantMeasures}, we also know that $Q\Pi$ admits an i.p.m., say $\hat{\mu}$, whose support is also $\mathcal{S}$. Thus, $\hat{\mu}$ admits the form of (\ref{eq:InvariantMeasureQP_derivation}), for some constants $\hat{\pi}_{s}\geq{0}$, $s\in\mathcal{S}$.

(b) From Proposition~\ref{Pr:SM:ConvergenceToPureStrategyStates},   for any two distinct $s,s'\in\cS$ and for any sufficiently small $\delta>\snoise>0$, note that $\Neigh{\delta}{s'}$ is a continuity set of $Q\Pi(s,\cdot)$, i.e., $$Q\Pi(s,\partial\Neigh{\delta}{s'})=0\,.$$ Thus, from Portmanteau theorem, given that $QP^{t}\Rightarrow Q\Pi$, $$Q\Pi(s,\Neigh{\delta}{s'}) = \lim_{t\to\infty}QP^{t}(s,\Neigh{\delta}{s'}) = \hat{P}_{ss'}.$$ 

Given that $\hat{\mu}$ is an i.p.m. of $Q\Pi$, then for any $A\in\Bor(\cZ)$, we have
\begin{eqnarray*}
\hat{\mu}(A) & = & (\hat{\mu}Q\Pi)(A) \cr 
& = & \int_{\cZ}\hat{\mu}(dz)Q\Pi(z,Z) \cr
& = & \int_{\cZ}\left(\sum_{s\in\cS}\hat{\pi}_s \Dirac{s}(dz)\right) Q\Pi(z,A) \cr
& = & \sum_{s\in\cS} \hat{\pi}_s \int_{\cZ} \Dirac{s}(dz)Q\Pi(z,A) \cr 
& = & \sum_{s\in\cS} \hat{\pi}_s Q\Pi(s,A).
\end{eqnarray*}
For any $\delta>\snoise$, define $\hat{\pi}_s = \hat{\pi}_s(\delta,\epsilon,\snoise) \df \hat{\mu}(\Neigh{\delta}{s})$. Then,
$$\hat{\pi}_{s'} = \hat{\mu}(\Neigh{\delta}{s'}) = \sum_{s\in\mathcal{S}}\hat{\pi}_s Q\Pi(s,\Neigh{\delta}{s'}) = \sum_{s\in\mathcal{S}}\hat{\pi}_s\hat{P}_{ss'},$$ which shows that $\hat{\pi}$ is an invariant distribution of $\hat{P}$, i.e., $\hat{\pi}=\hat{\pi}\hat{P}$.

It remains to establish uniqueness of the invariant distribution of $Q\Pi$. Note that the set $\cS$ of p.s.s. states is isomorphic with the set $\cA$ of action profiles. Let $\overline{Q\Pi}\in[0,1]^{\magn{\cA}\times\magn{\cA}}$ indicate a transition probability function, capturing the transition probabilities among any two action profiles under the $Q\Pi$ t.p.f., i.e., $\overline{Q\Pi}(\alpha,\alpha')$ corresponds to the probability that action profile $\alpha'$ is played continuously starting from the p.s.s. corresponding to $\alpha$. Let us consider the assumptions of a uniformly bounded noise, such that $0<\epsilon\rewardpert{i}(\alpha)<1$ and $h<\rewardpert{i}(\alpha)$ almost surely for all $\alpha\in\cA$ and $i\in\cI$. Then, it has been shown in Proposition~\ref{Pr:SM:ConvergenceToPureStrategyStates}(a) that $\overline{Q\Pi}(\alpha,\alpha')>0$ for all $\alpha'\in\cA$ and $i\in\cI$. Finite induction then shows that $(\overline{Q\Pi})^{n}(\alpha,\alpha')>0$ for all $\alpha,\alpha'\in\cA$. It follows that if we restrict the domain of $Q\Pi$ to $\mathcal{S}$, it defines an irreducible stochastic matrix. Therefore, $Q\Pi$ may only admit a unique i.p.m.
\end{proof}
\fi

Note that the above Proposition~\ref{Pr:SM:UniqueInvariantPMofQPi} provides the form and uniqueness of the invariant distribution of the limiting finite Markov chain governed by $Q\Pi$. The above result is independent of the size of the noise, which simply has to be sufficiently small so that the hypotheses of Proposition~\ref{Pr:SM:UniqueInvariantPMofQPi} are satisfied. The transition probabilities between two p.s.s. $s$ and $s'$, $\hat{P}_{ss'}$, will depend on the noise level $\snoise$ but we do not necessarily need to take the limit as the noise approaches zero. This is due to the fact that the aspiration factor $\phi_i$ filters out the noisy observations, and the strategy may always approach the p.s.s. $s'$ as long as the action profile $\alpha'$ is selected. 

\subsection{Proof of Theorem~\ref{Th:StochasticStability:PureStrategyStates}}	

Theorem~\ref{Th:StochasticStability:PureStrategyStates}(a)--(b) is a direct implication of Propositions~\ref{Pr:SM:WeakLimitPointsOfPerturbedInvariantMeasures}--\ref{Pr:SM:UniqueInvariantPMofQPi}.

\section{Bounds on the first hitting time of one-step satisfactory transitions} \label{sec:SM:ApproximationFirstHittingTime}

In this section, we will be concerned with the evolution of the strategy vector $x$ and the derivation of bounds on the first hitting time to a small neighborhood of a p.s.s. Define $$\Neighx{\delta}{x}{s'} \df \mathcal{P}_{\cX}\left(\Neigh{\delta}{s'}\right),$$ where $\mathcal{P}_{\cX}(O)$ is the \emph{canonical projection} of the set $O$ to $\cX$, defined by the product topology. We will derive (almost surely) bounds of the first hitting time of the strategy $x$ as the size of the noise approaches zero ($\snoise\downarrow{0}$). In particular, we are interested on the first hitting time of the strategy vector to the set $\Neighx{\delta}{x}{s'}$ for some p.s.s. $s'$, when players start from a p.s.s. $s$, that differs only in the action of a single player, and employ the $Q\Pi$ dynamics. The first proposition establishes such bounds when during the transition from $s$ to $s'$, only action profile $\alpha'$ is selected (that is the action corresponding to p.s.s. $s'$).

\begin{proposition}[Approximation of the first hitting time]		\label{Pr:SM:ApproximationFirstHittingTime}
Consider an one-step transition under t.p.f. $Q\Pi$ from action profile $\alpha$ to action profile $\alpha'$ which differ in the action of a single agent $j$, and let $s,s'\in\cS$ be the p.s.s.'s associated with $\alpha$ and $\alpha'$, respectively. Let $\alpha'$ be a satisfactory transition for agent $j$, i.e., $u_j(\alpha')>u_j(\alpha)$. Set $z'\df (\alpha',x',\rho',\cdot)$ to be the state of agent $j$ after $j$ perturbed once under t.p.f. $Q$ starting from p.s.s. $s$ and playing $\alpha_j'\neq\alpha_j$. Consider further $\epsilon>0$ and $\snoise>0$ sufficiently small such that $0<\epsilon \tilde{u}_i(\alpha) <1$ for all $\alpha\in\cA$ and $i\in\cI$, almost surely. Starting from p.s.s. $s$, consider the event of selecting action profile $\alpha'$ continuously over time under the $Q\Pi$ process. Under this condition (of continuously selecting action profile $\alpha'$), denote the first hitting time of the strategy vector to the set $\Neighx{\delta}{x}{s'}$, under the  $Q\Pi$ dynamics, where we take $\delta=\delta(\snoise)>\snoise$, as $$\uptau^*_{\delta}(\epsilon,\snoise) \df \inf_{t\geq{0}}\left\{x(t)\in\Neighx{\delta}{x}{s'}\right\},$$ which is a random variable driven by both the action selection process $\{\alpha(t)\}$ as well as the utility noise $\{\noise(t)\}$. Define\footnote{We use the notation $\overline{\uptau}_{\delta}^*$ to denote the essential infimum of the first hitting time, because it approximates the shortest path to $\Neighx{\delta}{x}{s'}$ when receiving the largest possible realization of the utility (excluding sequences of probability zero). Similar is also the reasoning for the notation of the essential supremum $\underline{\uptau}_\delta^*$ of the first hitting time, because it approximates the shortest path when receiving the smallest possible realization of the utility.} 
\begin{equation}
\overline{\uptau}^*_{\delta}(\epsilon,\snoise)\df \ess\inf_{\noise}[\uptau^*_{\delta}(\epsilon)]\,, \quad \underline{\uptau}^*_{\delta}(\epsilon,\snoise) \df \ess\sup_{\noise}{\uptau}^*_{\delta}(\epsilon)\,, 
\end{equation}
where 
\begin{equation*}
\overline{\uptau}_{\delta}^*(\epsilon,\snoise) = \ess\inf_{\noise}[\uptau^*_{\delta}(\epsilon)] \df \sup\left\{c\geq{0}: \Prob_{z'}[\uptau^*_{\delta}(\epsilon) \geq c ] =1\right\}
\end{equation*}
and 
\begin{equation*}
\underline{\uptau}_{\delta}^*(\epsilon,\snoise) = \ess\sup_{\noise}[\uptau^*_{\delta}(\epsilon)] \df \inf\left\{c\geq{0}: \Prob_{z'}[\uptau^*_{\delta}(\epsilon) \leq c ] =1\right\}
\end{equation*}
which define the essential infimum and essential supremum of the first hitting time $\uptau_{\delta}^*(\epsilon)$, respectively, i.e., when we exclude sequences of noise $\noise$ of probability zero. By definition of these bounds we have:
\begin{equation}\label{eq:SM:ApproximationFirstHittingTime_000}
\overline{\uptau}_{\delta}^*(\epsilon,\snoise) \leq \uptau_{\delta}^*(\epsilon,\snoise) \leq \underline{\uptau}_{\delta}^*(\epsilon,\snoise)\,, \quad \mbox{a.s.}\,.
\end{equation}
The following also hold:
\begin{itemize}
\item[(a)] The lower bound of the first hitting time satisfies
\begin{equation} \label{eq:SM:ApproximationFirstHittingTime_001}
\overline{\uptau}_{\delta}^*(\epsilon,\snoise) = \sup\{k\geq{0}:1-(1-\epsilon u_j(\alpha') - \epsilon\snoise)^k > 1-\delta\},
\end{equation}
\item[(b)] The upper bound of the first hitting time satisfies
\begin{equation} \label{eq:SM:ApproximationFirstHittingTime_002}
\underline{\uptau}_{\delta}^*(\epsilon,\snoise)=\inf\{k\geq{0}:1-(1-\epsilon u_j(\alpha') +\epsilon\kappa(\snoise))^k > 1-\delta\},
\end{equation}
where $\kappa(\snoise)\df (1+2c)\snoise$.
\item[(c)] Define ${\uptau}^0_{\delta}(\epsilon)$ as the first hitting time of the noiseless unperturbed process (i.e., when we take $\snoise\equiv{0}$). Then, the following approximation holds:
\begin{equation}	\label{eq:SM:ApproximationFirstHittingTime_01}
\lim_{\snoise\downarrow{0}}\overline{\uptau}_{\delta}^*(\epsilon,\snoise) = \lim_{\snoise\downarrow{0}}\underline{\uptau}_{\delta}^*(\epsilon,\snoise) = \lim_{\snoise\downarrow{0}}\uptau^*_{\delta}(\epsilon,\snoise) \equiv \uptau_{\delta}^0(\epsilon)\,,\quad \mbox{a.s.}\,.
\end{equation}
\end{itemize} 
\end{proposition}
\begin{proof}
Let us assume that the $Q\Pi$ process starts from some p.s.s. $s\in\cS$ (which assigns zero probability to playing action $\alpha'$) and agent $j$ plays action $\alpha'$ continuously thereafter (due to an initial perturbation under $Q$). Recall that there is positive probability of such path to occur according to Proposition~\ref{Pr:SM:ConvergenceToPureStrategyStates} of the Supplementary Material. Under such path, the strategy of agent $j$ evolves as follows (as shown in Proposition~\ref{Pr:SM:ConstantActionSelection} of the Supplementary Material):
\begin{equation}  \label{eq:SM:ApproximationFirstHittingTime_02}
x_{j\alpha_j'}(t) = 1 - (1-x_{j\alpha'_j}(0))\prod_{k=0}^{t-1}H_j(\alpha(k),\rho_j(k)),$$ where $x_{j\alpha_j'}(0)=0$, and $$H_j(\alpha(k),\rho_j(k))\df 1-\epsilon\phi_j(\rewardpert{j}(\alpha(k)),\rewardpert{j}(\alpha(k))-\rho_j(k))\,, \quad k\geq{0}.
\end{equation}

Given that agent $j$ starts from p.s.s. $s$ (with nominal utility $u_j(\alpha)$) and selects action profile $\alpha'$ continuously thereafter (with nominal utility $u_j(\alpha')>u_j(\alpha)$), then $\rho_j(t) - \reward{j}(\alpha') \leq \snoise \leq \delta$ almost surely for all $t>0$.\footnote{Starting from any $\rho_j(t)$ such that $u_j(\alpha')-\rho_j(t)\geq -\snoise$, note that, given $\epsilon\nu(\epsilon)<1$, 
\begin{eqnarray*}
\rho_j(t+1) & = & \rho_j(t) + \epsilon\nu(\epsilon) \left( \rewardpert{j}(\alpha') - \rho_j(t) \right) \cr & \leq & 
\rho_j(t) + \epsilon\nu(\epsilon) \left(\reward{j}(\alpha') +\snoise - \rho_j(t) \right) \cr 
& = & \reward{j}(\alpha') + (1-\epsilon\nu(\epsilon))(\rho_j(t) - \reward{j}(\alpha')) + \epsilon\nu(\epsilon)\snoise \cr 
& \leq & \reward{j}(\alpha') + (1-\epsilon\nu(\epsilon))\snoise + \epsilon\nu(\epsilon) \snoise \cr
& = & \reward{j}(\alpha') + \snoise,
\end{eqnarray*}
which implies that $u_j(\alpha')-\rho_j(t+1) \geq -\snoise$ a.s..
}

%

Thus, given the definition of the aspiration factor, there exists a positive constant $\kappa=\kappa(\snoise)\df (1+2c)\snoise>0$, such that $\kappa(\snoise)\to{0}$ as $\delta\downarrow{0}$ and\footnote{Under a path of continuously selecting action profile $\alpha'$ under the $Q\Pi$ dynamics, and if $\reward{j}(\alpha') - \rho_j(t) \geq -\snoise$, note that $\rewardpert{j}(\alpha')-\rho_j(t)\geq u_j(\alpha')-\snoise-\rho_j(t)\geq -2\snoise$ a.s.. This further implies that for sufficiently small $\delta>\snoise>0$, the aspiration factor satisfies 
\begin{eqnarray*}
\phi_j & \geq & \rewardpert{j}(\alpha')+\zeta (\rewardpert{j}(\alpha')-\rho_j(t)) \cr 
& \geq & \reward{j}(\alpha')-\snoise+\zeta(-2\snoise) \cr
& = & \reward{j}(\alpha') - (1+2c)\snoise.
\end{eqnarray*}
This sets a lower bound for the aspiration factor independently on the value of the aspiration level. 
} $$u_j(\alpha') - \kappa(\snoise) \leq \phi_j(\rewardpert{j}(\alpha'),\rewardpert{j}(\alpha')-\rho_j(t)) \leq u_j(\alpha')+\snoise,$$ a.s.. Note that these bounds apply independently of the aspiration level update. Thus, from Proposition~\ref{Pr:SM:ConstantActionSelection}, we can derive a lower and upper bound for the strategy vector of agent $j$ as follows:
\begin{equation}	\label{eq:SM:ApproximationFirstHittingTime_03}
1 - \left(1 - \epsilon(u_j(\alpha')-\kappa(\snoise)\right)^t \leq x_{j\alpha_j'}(t) \leq 1 - \left( 1 - \epsilon(u_j(\alpha')+\snoise)\right)^t
\end{equation}

(a) Let us define $$k^*\df \sup\left\{k\geq{0}:1-(1-\epsilon u_j(\alpha') -\epsilon\snoise)^k \leq 1-\delta\right\}.$$ In other words, $k^*$ corresponds to the maximum iteration under which the strategy has not reached $1-\delta$ when continuously receiving the maximum utility. Then, for every $k\leq{k^*}$, we have that $x_{j\alpha_j'}(k)\leq 1-\delta$ a.s., which further implies that $$\Prob_{z'}[\uptau_{\delta}^*(\epsilon,\snoise) \geq k^*+1]=1.$$ By definition of $k^*$, equivalently we have\footnote{Let us consider $k>{k^*+1}$. In order for the process to reach $\Neighx{\delta}{x}{s'}$ for the first time at a time $k>k^*+1$, it implies that there was a time instance $0<t<k$ and a constant $a\in\mathbb{R}$ such that $\rewardpert{j}(\alpha(t+1))<u_j(\alpha')+a<u_j(\alpha') + \snoise$ (i.e., the perturbed reward took a value smaller than its possible maximum value). If not, then the process would have reached $\Neighx{\delta}(s')$ at time $k\leq k^*+1$. Given that $\Prob_{z'}[\noise(t)\in[-\snoise,a)\,, a<\snoise]<1$, then $\Prob_{z'}[\uptau_{\delta}^*(\epsilon,\snoise) \geq k > k^*+1]<1$.} $$k^*\df \sup\left\{k\geq{0}:\Prob_{z'}[\uptau_{\delta}^*(\epsilon,\snoise)\geq k+1]=1\right\}.$$

(b) Let us define $$k^*\df \inf\left\{k\geq{0}:1-(1-\epsilon u_j(\alpha')+\epsilon\kappa(\snoise))^k > 1-\delta\right\}.$$ In other words, $k^*$ corresponds to the minimum possible iteration of the strategy reaching $1-\delta$ under the minimum possible utility. Under the condition that action $\alpha'$ is played continuously, and given that $$x_{j\alpha_j'}(k) \geq 1 - (1-\epsilon u_j(\alpha') + \epsilon\kappa(\snoise))^k\, \quad \mbox{a.s.}\,,$$ then $x_{j\alpha_j'}(k)>1-\delta$ for all $k\geq k^*$ a.s.\,. This further implies that $$\Prob_{z'}\left[\uptau^*_{\delta}(\epsilon,\snoise) \leq k^*\right]=1.$$ By definition of $k^*$, equivalently we have\footnote{Note that for any $k<k^*$, we have $$\Prob_{z'}[\uptau_{\delta}^*(\epsilon,\snoise)\leq{k}<k^*]<1.$$ This is due to the fact that if the strategy $x_{j\alpha_j'}(k)>1-\delta$, then there exists $t<k$ for which the measurement noise took value that is strictly larger than its lowest extreme, i.e., there is $a>0$ such that $\noise(t)>-\snoise+a$. The probability of the noise accepting such values is non-zero, which would imply $\Prob_{z'}[\uptau_{\delta}^*(\epsilon,\snoise)\leq{k}<k^*]<1$.} $$k^*\equiv\inf\left\{k\geq{0}:\Prob_{z'}\left[\uptau^*_{\delta}(\epsilon,\snoise) \leq k\right]=1\right\}$$ based on which we conclude that $$\underline{\uptau}_{\delta}^*(\epsilon,\snoise) = k^*.$$

(c) Consider the notation $x_j^0(t)$, $t\geq{0}$, to indicate the strategy evolution when there is no noise (i.e., $\snoise\equiv{0}$) and $x_j(t)$ otherwise.  

Using the Binomial Theorem \cite[Chapter 5]{graham_concrete_2017}, note that 
\begin{equation} \label{eq:SM:ApproximationFirstHittingTime_04}
(1-\epsilon u_j(\alpha') + \epsilon\kappa(\snoise))^t = \sum_{k=0}^{t}\left(\begin{array}{c} t\\k \end{array}\right)(1-\epsilon \reward{j}(\alpha'))^{t-k}\left(\epsilon\kappa(\snoise)\right)^{k}.
\end{equation}
Let us define the following quantity: $$\varepsilon_1 = \varepsilon_1(\snoise,t)\df \sum_{k=1}^{t}\left(\begin{array}{c} t\\k \end{array}\right)(1-\epsilon \reward{j}(\alpha'))^{t-k}\left(\epsilon\kappa(\snoise)\right)^{k}.$$ Then, we have: 
\begin{equation}  \label{eq:SM:ApproximationFirstHittingTime_05}
(1-\epsilon u_j(\alpha')+\epsilon\kappa(\snoise))^t = (1-\epsilon u_j(\alpha'))^t + \varepsilon_1(\snoise,t).
\end{equation}

Let us assume that $x^0(t)\in\Neighx{\delta-\varepsilon_1}{x}{s'}$ which implies that $${x}_{j\alpha'_j}^0(t)\equiv 1 - (1-\epsilon u_j(\alpha'))^t > 1 - (\delta - \varepsilon_1) = 1 - \delta + \varepsilon_1.$$ Then, from Equations~(\ref{eq:SM:ApproximationFirstHittingTime_03})--(\ref{eq:SM:ApproximationFirstHittingTime_05}), we have
\begin{eqnarray}		\label{eq:Pr:SM:ApproximationFirstHittingTime_01}
1-x_{j\alpha_j'}(t) & \leq &  1-1+(1-\epsilon u_j(\alpha') + \epsilon \kappa(\snoise))^t \cr & = & 1 - 1 + (1-\epsilon u_j(\alpha'))^{t} + \varepsilon_1 \cr 
& = & 1 - {x}^0_{j\alpha_j'}(t) + \varepsilon_1 \cr 
& < & 1 - (1-\delta+\varepsilon_1) + \varepsilon_1 \cr 
& = & \delta
\end{eqnarray}
which equivalently implies that $x(t)\in\Neighx{\delta}{x}{s'}$ and hence $\uptau^*_{\delta} \leq {\uptau}^0_{\delta-\varepsilon_1}$ almost surely. From the derivation in (\ref{eq:Pr:SM:ApproximationFirstHittingTime_01}), and the first inequality, note that we also have 
\begin{equation}  \label{eq:Pr:SM:ApproximationFirstHittingTime_UB}
\uptau^*_{\delta}(\epsilon,\snoise) \leq \underline{\uptau}_{\delta}^*(\epsilon,\snoise) \leq {\uptau}^0_{\delta-\varepsilon_1}(\epsilon)\,, \quad \mbox{a.s..}
\end{equation}

Likewise, we will derive a lower bound for $\uptau_\delta^*(\epsilon)$. Define the constant $$\varepsilon_2 = \varepsilon_2(\snoise,t)\df \sum_{k=1}^{t}\left(\begin{array}{c} t\\k \end{array}\right)(1-\epsilon \reward{j}(\alpha(k)))^{t-k}\left(-\epsilon\snoise\right)^{k} > 0.$$ Then, by using again the Binomial Theorem, we have
\begin{equation*}
(1-\epsilon u_j(\alpha')-\epsilon\snoise)^t = (1-\epsilon u_j(\alpha'))^t + \varepsilon_2(\snoise,t).
\end{equation*}
Let $x(t)\in\Neighx{\delta}{x}{s'}$ which implies that $x_{j\alpha_j'}(t) > 1-\delta$. Then we also have that  
\begin{eqnarray}		\label{eq:Pr:SM:ApproximationFirstHittingTime_02}
& 1-1+(1-\epsilon u_j(\alpha') - \epsilon\snoise)^t & \leq 1-{x}_{j\alpha'}(t) < \delta \cr  \Rightarrow &
1-1+(1-\epsilon u_j(\alpha'))^t + \varepsilon_2 & < \delta \cr
\Rightarrow & 1-{x}^0_{j\alpha_j'}(t) & < \delta - \varepsilon_2
\end{eqnarray}
which equivalently implies that $x_{j}^0(t)\in\Neighx{\delta-\varepsilon_2}{x}{s'}$. This further implies that ${\uptau}_{\delta-\varepsilon_2}^0(\epsilon) \leq \uptau^*_{\delta}(\epsilon)$ almost surely. In the above derivation (\ref{eq:Pr:SM:ApproximationFirstHittingTime_02}), and given the first inequality, note that we can also deduce that 
\begin{equation}		\label{eq:Pr:SM:ApproximationFirstHittingTime_LB}
{\uptau}_{\delta-\varepsilon_2}^0\leq \overline{\uptau}^*_{\delta}(\epsilon,\snoise) \leq \uptau^*_{\delta}(\epsilon,\snoise)\,, \quad \mbox{ a.s.}\,.
\end{equation}
From Equations~(\ref{eq:Pr:SM:ApproximationFirstHittingTime_UB}) and (\ref{eq:Pr:SM:ApproximationFirstHittingTime_LB}), we conclude that $${\uptau}^0_{\delta-\varepsilon_2}(\epsilon) \leq \overline{\uptau}^*_{\delta}(\epsilon,\snoise) \leq \uptau_{\delta}^*(\epsilon,\snoise) \leq \underline{\uptau}_{\delta}^*(\epsilon,\snoise) \leq {\uptau}^0_{\delta-\varepsilon_1}(\epsilon)\,, \quad \mbox{a.s.}$$ Given the monotonicity of the hitting times in monotonically increasing/decreasing sets and given also that $\lim_{\snoise\downarrow{0}}\varepsilon_1(\snoise,t)=\lim_{\snoise\downarrow{0}}\varepsilon_2(\snoise,t)=0$, we conclude that 
\begin{equation}
\label{eq:SM:OneStepTransitionProbabilitySatisfactory_A01}
\lim_{\snoise\downarrow{0}}{\uptau}^0_{\delta-\varepsilon_1(\snoise)}(\epsilon)=\lim_{\snoise\downarrow{0}}{\uptau}^0_{\delta-\varepsilon_2(\snoise)}(\epsilon) = \uptau_{\delta}^{0}(\epsilon)\,, \quad \mbox{a.s.}\,.
\end{equation} 
As a direct consequence of the Sandwich Theorem \cite[Proposition~2.2.2]{Reed98}, we further conclude that:
\begin{equation*}	
\lim_{\snoise\downarrow{0}}\overline{\uptau}_{\delta}^*(\epsilon,\snoise) = \lim_{\snoise\downarrow{0}}\underline{\uptau}_{\delta}^*(\epsilon,\snoise) = \lim_{\snoise\downarrow{0}}\uptau^*_{\delta}(\epsilon,\snoise) \equiv \uptau_{\delta}^0(\epsilon)\,, \quad \mbox{a.s.},
\end{equation*} 
which relates the first hitting time of the noisy process with the noiseless process as the size of the noise approaches zero. This concludes the proof.
\end{proof}

\section{Lower bound on the first hitting time of one-step transitions}		\label{sec:SM:MinimumFirstHittingTime}

Proposition~\ref{Pr:SM:ApproximationFirstHittingTime} relates the first hitting times of the process with and without noise under the condition that the action profile $\alpha'$ is played continuously. Considering now the same transition from p.s.s. $s$ to $s'$, we would like to also find a lower bound for the first hitting of the process when action profile $\alpha'$ is not necessarily played continuously. This lower bound is provided by the following proposition.

\begin{proposition}[Lower bound of first hitting time] \label{Pr:SM:MinimumFirstHittingTime}
Consider an one-step satisfactory transition under t.p.f. $Q\Pi$ from action profile $\alpha$ to action profile $\alpha'$ which differ in the action of a single agent $j$, with $u_j(\alpha')>u_j(\alpha)$. Let $s,s'\in\cS$ be the p.s.s.'s associated with $\alpha$ and $\alpha'$, respectively. For any $\delta>0$ such that $\delta>\snoise$, let us define the first hitting time of the process $Q\Pi$ to the set $\Neighx{\delta}{x}{s'}$ as 
\begin{equation*}
\uptau_{\delta}(\epsilon,\snoise) \df \inf_{t\geq{0}}\left\{x(t)\in\Neighx{\delta}{x}{s'}\right\}
\end{equation*}
which is a random variable driven by both the action selection as well as the utility noise. Then, for sufficiently small $\snoise$ such that $u_j(\alpha)>(1+2c)\snoise$ uniformly on $\alpha\in\cA$, the following holds:
\begin{equation}		\label{eq:SM:OneStepTransitionProbabilitySatisfactory_SM2_01}
\uptau_{\delta}(\epsilon,\snoise)\geq \overline{\uptau}^*_{\delta}(\epsilon,\snoise)\,, \quad \mbox{a.s.}\,.
\end{equation}
\end{proposition}
\begin{proof}
Note that on any sample path from $s$ to $\Neighx{\delta}{x}{s'}$, under the $Q\Pi$ t.p.f., only two action profiles can be played, namely $\alpha$ and $\alpha'$. These two action profiles differ in the action of a single agent $j\in\cI$, i.e., $\alpha_j'\neq\alpha_j$ and $\alpha_i'=\alpha_i$ for all $i\neq{j}$. Thus, independently of the effect of the noise, any deviation from playing $\alpha'$ involves playing action $\alpha$, since only action profiles $\alpha$ and $\alpha'$ are assigned positive probability along such path. As a consequence, along any path from $s$ to $s'$, the strategy vector of agent $j$ is such that $x_{j\alpha_j}>0,$ and $ x_{j\alpha_j'}>0$ while all other actions of agent $j$ are assigned probability zero. 

Let us consider any strategy vector of agent $j$ along this path at time step $t$, $x_{j}(t)$. We need to show that the shortest path to $\Neighx{\delta}{x}{s'}$ starting from $x_{j}(t)$ is the one where action profile $\alpha'$ is played continuously. To this end, let us consider an alternative path, where at time $t+1$ action profile $\alpha$ is selected, while at time $t+2$ action profile $\alpha'$ is selected instead. We will investigate the strategy evolution along these two steps, denoted by $\tilde{x}$. That is,
\begin{eqnarray*}
\tilde{x}_{j\alpha_j'}(t+1) & = & \tilde{x}_{j\alpha_j'}(t) + \epsilon\phi_{j}(\rewardpert{j}(\alpha),\rewardpert{j}(\alpha) - \rho_j(t))\cdot (0 - \tilde{x}_{j\alpha_j'}(t)) \cr
\tilde{x}_{j\alpha_j'}(t+2) & = & \tilde{x}_{i\alpha_j'}(t+1) + \epsilon\phi_{j}(\rewardpert{j}(\alpha'),\rewardpert{j}(\alpha') - \rho_j(t+1))\cdot (1 - \tilde{x}_{j\alpha_j'}(t+1))
\end{eqnarray*}
given that agent $j$ selected action $\alpha_j$ at time $t+1$ and action $\alpha_j'$ at time $t+2$. Note that independently of the evolution of the aspiration level, the following lower and upper bounds apply to the aspiration factor:
\begin{equation}  \label{eq:Pr:SM:MinimumFirstHittingTime_01}
h\leq \phi_{j}(\rewardpert{j}(\alpha),\rewardpert{j}(\alpha) - \rho_j(t)) \leq u_j(\alpha)+\snoise\,, \quad \mbox{a.s.}\,.
\end{equation}
When $\alpha'$ is selected at time $t+2$ and given that the nominal utilities satisfy $u_j(\alpha')>u_j(\alpha)$, note that (as we showed in the proof of Proposition~\ref{Pr:SM:ApproximationFirstHittingTime}) there exists constant $\kappa=\kappa(\snoise)\df (1+2c)\snoise>0$, such that $\kappa(\snoise)\to{0}$ as $\delta\downarrow{0}$ and 
\begin{equation} \label{eq:Pr:SM:MinimumFirstHittingTime_02}
u_j(\alpha') - \kappa(\snoise) \leq \phi_j(\rewardpert{j}(\alpha'),\rewardpert{j}(\alpha')-\rho_j(t+1)) \leq u_j(\alpha')+\snoise\,, \quad \mbox{a.s.}\,.
\end{equation}
Note that these bounds apply independently of the aspiration level update. From the bounds of Equations~(\ref{eq:Pr:SM:MinimumFirstHittingTime_01})--(\ref{eq:Pr:SM:MinimumFirstHittingTime_02}), we have that 
\begin{equation*}
(1-\epsilon h)x_{j\alpha_j'} \leq \tilde{x}_{j\alpha_j'}(t+1) \leq (1-\epsilon u_j(\alpha) - \epsilon\snoise) \cdot x_{j\alpha_j'}(t),
\end{equation*}
and 
\begin{eqnarray}	\label{eq:Pr:SM:MinimumFirstHittingTime_03}
\lefteqn{\tilde{x}_{j\alpha_j'}(t+2)} \cr  & \leq & (1-\epsilon u_{j}(\alpha)-\epsilon\snoise) x_{j\alpha_j'}(t) + \epsilon (u_j(\alpha')+\snoise) (1-x_{j\alpha_j'}(t)) + \mbox{h.o.t.}(\epsilon),
\end{eqnarray}
where $\mbox{h.o.t.}(\epsilon)$ denotes \emph{higher order terms of $\epsilon$}. Likewise, for the case that action profile $\alpha'$ has been played at times $t+1$ and $t+2$ (i.e., there is no deviation), we have
\begin{eqnarray} \label{eq:Pr:SM:MinimumFirstHittingTime_04}
\lefteqn{x_{j\alpha_j'}(t+2)}\cr & \geq & x_{j\alpha_j'}(t) + \epsilon(2u_j(\alpha')-\kappa(\snoise)+\snoise)(1-x_{j\alpha_j'}(t)) + \mbox{h.o.t.}(\epsilon).
\end{eqnarray}
From Equations~(\ref{eq:Pr:SM:MinimumFirstHittingTime_03})--(\ref{eq:Pr:SM:MinimumFirstHittingTime_04}), we conclude that 
\begin{eqnarray}
\lefteqn{x_{j\alpha_j'}(t+2) - \tilde{x}_{j\alpha_j'}(t+2)} \cr 
& \geq & \epsilon(u_j(\alpha)+\snoise)x_{j\alpha_j'}(t) + \epsilon(u_j(\alpha')-\kappa(\snoise))(1-x_{j\alpha_j'}(t)).
\end{eqnarray}
Note that as long as $\snoise>0$ is sufficiently small such that $$u_j(\alpha')-\kappa(\snoise) = u_j(\alpha') - (1+2c)\snoise > 0,$$ the above lower bound is uniformly bounded away from zero for any $x_{j\alpha_j'}(t)\in[0,1]$. In conclusion, any deviation from playing action profile $\alpha'$ may only reduce the probability of playing action $\alpha_j'$,  $x_{j\alpha_j'}$, almost surely. This further implies that the shortest path from $s$ to $s'$ may only be attained by selecting $\alpha'$ continuously, i.e., $\uptau_{\delta}(\epsilon,\snoise)\geq \uptau_{\delta}^*(\epsilon,\snoise)$ a.s.\,. Given Proposition~\ref{Pr:SM:ApproximationFirstHittingTime} and Equation~(\ref{eq:SM:ApproximationFirstHittingTime_000}), this further implies that $\uptau_{\delta}(\epsilon,\snoise)\geq \overline{\uptau}_{\delta}^*(\epsilon,\snoise)$, a.s.\,.
\end{proof}

\section{Proof of Lemma~\ref{Lm:OneStepTransitionProbabilitySatisfactory} (Satisfactory Transition)
}	\label{sec:SM:SatisfactoryTransition}

In this proof, we will be using the notation and definitions of Sections~\ref{sec:SM:ApproximationFirstHittingTime}--\ref{sec:SM:MinimumFirstHittingTime}. We first need to introduce some additional notation. Consider the unperturbed process initiated at state $z'$, i.e., $Z_0=z'$. Let us also define the set
$$\underline{D}_{j,\ell}(\alpha') \df \left\{(\alpha,x,\rho,\cdot)\in\cZ: x_{j\alpha_j'} > 1 - \underline{H}_{j}(\snoise)^\ell\right\},$$ 
where $\underline{H}_{j}(\snoise) \df 1-\epsilon (u_j(\alpha')-\kappa(\snoise))$, $\kappa(\snoise)\df (1+2c)\snoise$ (as derived in the proof of Proposition~\ref{Pr:SM:ApproximationFirstHittingTime}). Furthermore, let 
$$\overline{D}_{j,\ell}(\alpha') \df \left\{(\alpha,x,\rho,\cdot)\in\cZ: x_{j\alpha_j'} > 1 - \overline{H}_{j}(\snoise)^\ell\right\},$$ 
where $\overline{H}_{j}(\snoise) \df 1-\epsilon (u_j(\alpha')+\snoise)$. Note that the set 
${E}_{j,\ell}(\alpha') \df \overline{D}_{j,\ell}(\alpha')^c \cap \underline{D}_{j,\ell}(\alpha')$ define the space over which the strategy vector lies in at iteration $\ell$ almost surely, independently of the aspiration level when action $\alpha'$ is played continuously (under a satisfactory transition).

Note that in the case of a satisfactory transition the aspiration level plays no role in $H_j(\alpha(k),\rho_j(k))$. To simplify notation, in this section, we will denote this quantity by $H_j(\alpha')\df H_j(\alpha(k),\rho_j(k))\equiv 1-\epsilon u_j(\alpha')$.

%

(a) One event under which the strategy vector $x_j(t)$ starts from $s$ and reaches $\Neighx{\delta}{x}{s'}$, under the $Q\Pi$ dynamics, can be realized almost surely when action profile $\alpha'$ is played continuously for $\underline{\uptau}_{\delta}^*$ number of times starting from $s$, and under the condition that $\snoise=\snoise(\delta)<\delta$ (cf.,~Proposition~\ref{Pr:SM:ApproximationFirstHittingTime}).\footnote{Note that $\underline{\uptau}_{\delta}^*$ corresponds to the essential supremum of the first hitting time when the satisfactory action profile $\alpha'$ is played continuously (i.e., it corresponds to the number of steps required to reach $\Neighx{\delta}{x}{s'}$ when the noise takes its smallest possible value). Thus, if action profile $\alpha'$ is selected for at least $\underline{\uptau}_{\delta}^*$ number of times, then $\Neighx{\delta}{x}{s'}$ has been reached a.s.\,.} Using the Markov property, we have:
\begin{equation}		\label{eq:SM:OneStepTransitionProbabilitySatisfactory_A1.0}
\breve{P}_{ss'}(\delta,\epsilon,\snoise) \geq \Prob_{z'}\left[\alpha(t+1)=\alpha', \forall  t < \underline{\uptau}^*_\delta(\epsilon,\snoise) \right] \cdot \inf_{z\in \Neighx{\delta}{x}{s'}}\Prob_{z}[B_\infty], 
\end{equation}
where recall that $$B_t \df \{\omega\in\Omega:\alpha(\tau)=\alpha(0)\,, \forall 0\leq\tau\leq{t}\}.$$
Given that actions are selected only based on the strategy vector, and independently of the aspiration level, note that $$\lim_{\delta\downarrow{0}} \inf_{z\in \Neighx{\delta}{x}{s'}}\Prob_{z}[B_\infty] = 1.$$
Furthermore, note that:
\begin{eqnarray}		
\lefteqn{\Prob_{z'}\left[\alpha(t+1)=\alpha', \forall  t < \underline{\uptau}^*_\delta(\epsilon,\snoise) \right]} \cr 
& \geq & \prod_{t < \underline{\uptau}_{\delta}^*}\inf_{z\in \underline{D}_{j,t}(\alpha')}\Prob_{z'}\left[\alpha(t+1)=\alpha' | Z_{t}=z\right] \label{eq:SM:OneStepTransitionProbabilitySatisfactory_A1.1} \\
& \geq & \prod_{t < \underline{\uptau}_{\delta}^*} \left(1-\underline{H}_{j}(\snoise)^{t+1} \right) \cr 
& = & \prod_{t < {\uptau}_{\delta}^*} \left(1-\underline{H}_{j}(\snoise)^{t+1} \right) \cdot \prod_{t={\uptau}_{\delta}^*}^{\underline{\uptau}_{\delta}^*-1} \left(1-\underline{H}_{j}(\snoise)^{t+1} \right)\,. \label{eq:SM:OneStepTransitionProbabilitySatisfactory_A1.2}
\end{eqnarray}
Note that in (\ref{eq:SM:OneStepTransitionProbabilitySatisfactory_A1.1}), we have used the Markov property, and in (\ref{eq:SM:OneStepTransitionProbabilitySatisfactory_A1.2}), we have used Proposition~\ref{Pr:SM:ApproximationFirstHittingTime} and the property that $\uptau_{\delta}^* \leq \underline{\uptau}_{\delta}^*$ a.s.. Using the properties of the logarithmic function, we have that
\begin{eqnarray*}
\lefteqn{\log\Prob_{z'}\left[\alpha(t+1)=\alpha', \forall  t < \underline{\uptau}^*_\delta \right]} \cr 
& \geq & \sum_{t < \uptau_{\delta}^*}\log\left(1-\underline{H}_{j}(\snoise)^{t+1}\right) + \sum_{t = \uptau_{\delta}^*}^{\underline{\uptau}_{\delta}^*-1}\log\left(1-\underline{H}_{j}(\snoise)^{t+1}\right).
\end{eqnarray*}
Furthermore, note that the tail of this summation is also bounded as follows:
\begin{equation*}
\sum_{t=\uptau_{\delta}^*}^{\underline{\uptau}_{\delta}^*-1}\log\left(1-\underline{H}_{j}(\snoise)^{t+1} \right) \geq (\underline{\uptau}_{\delta}^*-\uptau_{\delta}^*)\log\left(1-\underline{H}_j(\snoise)\right) \xrightarrow{\snoise\downarrow{0}} 0,\quad \mbox{a.s.}\,,
\end{equation*}
where we have used the property of Equation (\ref{eq:SM:OneStepTransitionProbabilitySatisfactory_A1.2}) and the fact that $\underline{H}_j(\snoise)<1$ for sufficiently small step size $\epsilon>0$ and noise size $\snoise>0$, which implies that $\log(1-\underline{H}_j(\snoise)^{t+1}) \geq \log(1-\underline{H}_j(\snoise))$.
Thus, according to Proposition~\ref{Pr:SM:ApproximationFirstHittingTime}, we have
\begin{eqnarray*}
\lefteqn{\lim_{\snoise\downarrow{0}}\log \Prob_{z'}\left[\alpha(t+1)=\alpha', \forall  t < \underline{\uptau}^*_\delta(\epsilon,\snoise) \right] \geq  \lim_{\snoise\downarrow{0}}\sum_{t<\uptau_{\delta}^*(\epsilon,\snoise)}\log\left(1-\underline{H}_j(\snoise)^{t+1}\right) } \cr & = & \sum_{t<\lim_{\snoise\downarrow{0}}\uptau_{\delta}^*(\epsilon,\snoise)}\lim_{\snoise\downarrow{0}}\log(1-\underline{H}_{j}(\snoise)^{t+1}) = \sum_{t<\uptau_{\delta}^0(\epsilon)}\log(1-H_j(\alpha')^{t+1}),
\end{eqnarray*}
where we have also used the fact that the limit of each of the terms of the series converges as $\snoise\downarrow{0}$ and the number of terms is finite and stabilizes as $\snoise\downarrow{0}$.
From the above inequality, and given that the natural logarithm is a monotonically increasing function, we conclude that $$\lim_{\snoise\downarrow{0}} \Prob_{z'}\left[\alpha(t+1)=\alpha', \forall  t < \underline{\uptau}^*_\delta (\epsilon,\snoise) \right] \geq \prod_{t<{\uptau}_{\delta}^0(\epsilon)}(1-{H}_j(\alpha')^{t+1})\,, \quad \mbox{a.s.}\,.$$ Furthermore, from (\ref{eq:SM:OneStepTransitionProbabilitySatisfactory_A1.0}), we also have that as $\delta\downarrow{0}$ and $\snoise(\delta)<\delta$, 
\begin{equation}	\label{eq:SM:OneStepTransitionProbabilitySatisfactory_LowerBound}
\breve{P}_{ss'}(\delta,\epsilon,\snoise) \geq \prod_{t<{\uptau}_{\delta}^0(\epsilon)}(1-{H}_j(\alpha')^{t+1})\,, \quad \mbox{a.s.}\,.
\end{equation}

\textit{In the remainder of the proof, we will derive an upper bound for the above limit.} Note that according to Proposition~\ref{Pr:SM:MinimumFirstHittingTime},  in order for the unperturbed process to reach $\Neighx{\delta}{x}{s'}$, action profile $\alpha'$ needs to be selected for at least $\overline{\uptau}^*_{\delta}$ number of times almost surely (which is the first hitting time when agent $j$ receives the maximum possible utility of $u_j(\alpha')+\snoise$). Thus, due to the Markov property, for any path under the $Q\Pi$ process that reaches $\Neighx{\delta}{x}{s'}$, there exists a subsequence $\{t_k\}$ such that $\alpha(t_k+1)\equiv\alpha'$ while $Z_{t_k}\in \overline{D}_{j,k}(\alpha')^c$ for all $k < \overline{\uptau}_{\delta}^*$.\footnote{Along any path $\{Z_{t_k}\}$ that reaches $\Neighx{\delta}{x}{s'}$, $Z_{t_k}\in\overline{D}_{j,k}(\alpha')^c$ a.s.\,.} 
Thus, 
\begin{eqnarray*}
\breve{P}_{ss'}(\delta,\epsilon,\snoise)  
\leq \Prob_{z'}\left[\exists\{t_k\}:\alpha(t_{k}+1)=\alpha', Z_{t_k}\in \overline{D}_{j,k}(\alpha')^c\,, \forall\, k < \overline{\uptau}_{\delta}^*\right]\,, \quad \mbox{a.s.}\,. 
\end{eqnarray*}
Using the properties of the conditional probability, we also have:
\begin{eqnarray*}
\breve{P}_{ss'}(\delta,\epsilon,\snoise) \leq \Prob_{z'}\left[\alpha(t_k+1)=\alpha' \, | \, \exists\{t_k\}: Z_{t_k}\in \overline{D}_{j,k}(\alpha')^c\,, \forall k < \overline{\uptau}_{\delta}^*\right]. 
\end{eqnarray*}
Using again the Markov property, we have
\begin{eqnarray*}		
\breve{P}_{ss'}(\delta,\epsilon,\snoise) & \leq & 
\prod_{t< \overline{\uptau}_{\delta}^*} \sup_{z\in \overline{D}_{j,t}(\alpha')^c}\Prob_{z'}\left[\alpha(t+1)=\alpha'|Z_{t}=z\right] \cr 
& \leq & \prod_{t < \overline{\uptau}_{\delta}^*}\left(1-\overline{H}_{j}(\snoise)^{t+1}\right). 
\end{eqnarray*}
By applying the natural logarithm on both sides, we get
\begin{eqnarray}
\lefteqn{\log\breve{P}_{ss'}(\delta,\epsilon,\snoise)} \cr 
& \leq & \sum_{t < \uptau_{\delta}^*}\log\left(1-\overline{H}_j(\snoise)^{t+1}\right) - \sum_{t=\overline{\uptau}_{\delta}^*}^{\uptau_{\delta}^*-1}\log\left(1-\overline{H}_j(\snoise)^{t+1}\right)  \cr 
& \leq & \sum_{t < \uptau_{\delta}^*}\log\left(1-\overline{H}_j(\snoise)^{t+1}\right) - \sum_{t=\overline{\uptau}_{\delta}^*}^{\uptau_{\delta}^*-1}\log\left( 1-H_{j}(\alpha')^{t+1} \right)  \label{eq:SM:OneStepTransitionProbabilitySatisfactory_A3} \\ 
& \leq & \sum_{t<\uptau_{\delta}^*}\log\left(1-\overline{H}_j(\snoise)^{t+1}\right) - \sum_{t=\overline{\uptau}_{\delta}^*+1}^{\uptau_{\delta}^*}\log\left( 1-H_{j}(\alpha') \right) \label{eq:SM:OneStepTransitionProbabilitySatisfactory_A4} \\ 
& = & \sum_{t<\uptau_{\delta}^*}\log\left(1-\overline{H}_j(\snoise)^{t+1}\right) - (\uptau_{\delta}^* - \overline{\uptau}_{\delta}^*) \log\left( 1-H_{j}(\alpha') \right) \nonumber
\end{eqnarray}
where in  (\ref{eq:SM:OneStepTransitionProbabilitySatisfactory_A3}), we have used the property that, for sufficiently small $\epsilon$ and $\snoise$, we have $\log\left(1-\overline{H}_j(\snoise)^t\right) \geq \log\left(1-H_j(\alpha')^t\right)$. Furthermore, in (\ref{eq:SM:OneStepTransitionProbabilitySatisfactory_A4}), we have used the property that $\log\left(1-{H}_j(\alpha')^t\right) \geq \log\left(1-H_j(\alpha')\right)$ (given that the natural logarithm is a monotonically increasing function).
According to Proposition~\ref{Pr:SM:ApproximationFirstHittingTime}, as $\snoise\downarrow{0}$, we also have  that $$\lim_{\snoise\downarrow{0}}\overline{\uptau}_{\delta}^*(\epsilon,\snoise) = \lim_{\snoise\downarrow{0}}\uptau_{\delta}^*(\epsilon,\snoise) = \uptau_{\delta}^0(\epsilon)\,, \quad \mbox{a.s.}\,.$$ Thus,
\begin{eqnarray} \label{eq:SM:OneStepTransitionProbabilitySatisfactory_A5}
\lefteqn{\lim_{\snoise\downarrow{0}}\log\breve{P}_{ss'}(\delta,\epsilon,\snoise) \leq \lim_{\snoise\downarrow{0}}\sum_{t < \uptau_{\delta}^*(\epsilon,\snoise)}\log\left(1-\overline{H}_{j}(\snoise)^{t+1}\right)} \cr & = & \sum_{t<\lim_{\snoise\downarrow{0}}\uptau_{\delta}^*(\epsilon,\snoise)}\lim_{\snoise\downarrow{0}}\log\left(1-\overline{H}_{j}(\snoise)^{t+1}\right) = \sum_{t<\uptau_{\delta}^0(\epsilon)}\log(1-H_j(\alpha')^{t+1}),
\end{eqnarray}
where we have used the property that the limit of each of the terms of the series converges as $\snoise\downarrow{0}$ and the number of terms is finite and stabilizes as $\snoise\downarrow{0}$. Finally, note that $\lim_{\snoise\downarrow{0}}\overline{H}_j(\snoise)^{t} = H_{j}(\alpha')^{t}$. Hence, we have that as $\delta\downarrow{0}$ and $\snoise(\delta)<\delta$,
\begin{equation} \label{eq:SM:OneStepTransitionProbabilitySatisfactory_UpperBound}
\breve{P}_{ss'}(\delta,\epsilon,\snoise) \leq \prod_{t<{\uptau}_{\delta}^0(\epsilon)}(1-{H}_j(\alpha')^{t+1})\,, \quad \mbox{a.s.}\,.
\end{equation}

Thus, from Equations~(\ref{eq:SM:OneStepTransitionProbabilitySatisfactory_LowerBound}), (\ref{eq:SM:OneStepTransitionProbabilitySatisfactory_UpperBound}), for $\snoise=\snoise(\delta)<\delta$ and as $\delta\downarrow{0}$,
\begin{eqnarray}		\label{eq:SM:OneStepTransitionProbabilitySatisfactory_A6}
\breve{P}_{ss'}(\delta,\epsilon,\snoise) \approx 
\prod_{t<{\uptau}_{\delta}^0(\epsilon)}\left(1-{H}_{j}(\alpha')^{t+1}\right)\,, \quad \mbox{a.s.}\,,
\end{eqnarray}
which concludes the proof.

(b) We will further approximate the one-step transition probability by investigating the product 
\begin{equation*}	
\OSA{s}{s'}{\delta,\epsilon,\snoise} \approx 
\prod_{t<{\uptau}_{\delta}^0(\epsilon)}\left(1-{H}_{j}(\alpha')^{t+1}\right)\,.
\end{equation*}
Recall that $\uptau^0_\delta$ denotes the first hitting time to $\Neighx{\delta}{x}{s'}$ when playing repeatedly action $\alpha'$ and receiving the nominal utility $u_j(\alpha')$ (i.e., without noise). Note that 
\begin{equation}    \label{eq:SM:OneStepTransitionProbabilitySatisfactory_B1}
\uptau^0_\delta(\epsilon) = \left\lceil{ \frac{\log(\delta)}{\log(H_{j}(\alpha'))} }\right\rceil = \frac{\log(\delta)}{\log(H_{j}(\alpha'))} + \vartheta(\delta,\epsilon),
\end{equation}
for some constant $\vartheta=\vartheta(\delta,\epsilon)\in[0,1)$. 
To simplify the notation, let us temporarily denote $H\df H_j(\alpha')$. We have that 
\begin{equation}		
\lim_{\epsilon\downarrow{0}} \log\left(H^{\uptau_{\delta}^0(\epsilon)}\right) = \lim_{\epsilon\downarrow{0}}\left[\left( \frac{\log(\delta)}{\log(H)}+ \vartheta(\delta,\epsilon) \right) \log(H) \right] = \log(\delta),
\end{equation}
given that $\lim_{\epsilon\downarrow{0}}\log(H)=0$. Furthermore, due to the continuity of the natural logarithm, $$\lim_{\epsilon\downarrow{0}} {H}^{\uptau_{\delta}^0(\epsilon)} = \delta.$$ As a result, for any $\ell\in\mathbb{N}$, $$\lim_{\epsilon\downarrow{0}}{H}^{\ell \uptau_{\delta}^0(\epsilon)}=\delta^\ell.$$ The natural logarithm can be computed by the Mercator series\footnote{Taylor series expansion of the natural logarithm.} (for $|H|<1$ and $H\neq{1}$) \cite[Section~4.1.24]{abramowitz_handbook_2013}, that is
\begin{equation*}
\log\left(1-{H}^{t}\right) = -\sum_{\ell=1}^{\infty}\frac{{H}^{\ell t}}{\ell}.
\end{equation*}
Thus, due to the fact that the natural logarithm is an increasing function, and using the properties of the Geometric series, we have
\begin{eqnarray*}
\lefteqn{\lim_{\snoise\downarrow{0}}(1-{H}) \log\left( \breve{P}_{ss'}(\delta,\epsilon,\snoise) \right)} \cr & \approx & (1-{H})\sum_{t=1}^{{\uptau}^0_\delta(\epsilon)}\log\left(1-{H}^{t}\right) \cr & = & -\sum_{\ell=1}^{\infty}\frac{1}{\ell}\Big[(1-{H})\sum_{t=1}^{{\uptau}^0_\delta(\epsilon)}{H}^{\ell t}\Big] \cr 
& = & -\sum_{\ell=1}^{\infty}\frac{1}{\ell}\Big[(1-{H})\frac{1-{H}^{\ell({\uptau}^0_\delta(\epsilon)+1)}}{1-{H}^\ell}-(1-{H})\Big] \cr
& = & -\sum_{\ell=1}^{\infty}\frac{1}{\ell}\Big[\frac{1-{H}^{\ell({\uptau}^0_\delta(\epsilon)+1)}}{1+{H}+\cdots + {H}^{\ell-1}} -(1-{H})\Big]. 
\end{eqnarray*}
Note that, for any $\ell\in\mathbb{N}$, $H^{\ell}\to{1}$ as $\epsilon\downarrow{0}$, thus $\lim_{\epsilon\downarrow{0}}(1+{H}+\cdots+{H}^{\ell-1}) = \ell$. 

Hence, from Equation~(\ref{eq:SM:OneStepTransitionProbabilitySatisfactory_A6}), we conclude that 
\begin{eqnarray*}
\lefteqn{\lim_{\epsilon\downarrow{0}}\lim_{\snoise\downarrow{0}}(1-{H}) \log\left(\breve{P}_{ss'}(\delta,\epsilon,\snoise) \right) } \cr & \approx & \lim_{\epsilon\downarrow{0}}(1-{H})\sum_{t=1}^{{\uptau}_{\delta}^0(\epsilon)}\log\left(1-{H}^{t}\right) \approx -\sum_{\ell=1}^{\infty}\frac{1}{\ell^2}(1-\delta^{\ell})
\end{eqnarray*}
almost surely. Note that the last series is convergent and let $\eta(\delta)$ denote its limit, i.e., $\eta(\delta) \df \sum_{\ell=1}^{\infty}\frac{1}{\ell^2}(1-\delta^{\ell})$. Hence, using the fact that $1-{H}=\epsilon (u_j(\alpha'))$, we conclude that, as $\epsilon\downarrow{0}$ and $\delta\downarrow{0}$ with $\snoise=\snoise(\delta)<\delta$,
\begin{equation*} 
(1-{H}) \log\left( \breve{P}_{ss'}(\delta,\epsilon, \snoise) \right) \approx -\eta(\delta)\,, \quad \mbox{a.s.}\,.
\end{equation*}
Thus, the conclusion follows.

\section{Proof of Lemma~\ref{Lm:OneStepTransitionProbabilityUnsatisfactory} (Unsatisfactory Transition) 
}  \label{sec:SM:UnsatisfactoryTransition}

Consider the hypotheses of Lemma~\ref{Lm:OneStepTransitionProbabilityApproximation} 
of MD and consider that the nominal utility of agent $j$ at action profile $\alpha'$ is strictly smaller than the corresponding one at $\alpha$, i.e., $u_j(\alpha')<u_j(\alpha)$. Given that agent $j$ selected action $\alpha_j'$ under the t.p.f. $Q$, at any future time under $\Pi$ starting from $z'$ only two action profiles receive positive probability of being selected, namely $\alpha$ and $\alpha'$. 

\textit{In the first part of the proof, we will derive a lower bound for the approximation of the transition probability.} One event under which the strategy vector $x_j(t)$ starts from $s$ and reaches $\Neighx{\delta}{x}{s'}$, under the $Q\Pi$ dynamics, can be realized almost surely when action profile $\alpha'$ is played continuously. Let us define $\hat{\uptau}_{\delta}=\hat{\uptau}_{\delta}(\epsilon)$ to be the first hitting time of the unperturbed process to $\Neighx{\delta}{x}{s'}$ starting from $z'$ until which the aspiration factor satisfies $\phi_j\equiv {h}$. (This is a hypothetical worst-case scenario such that along the path, the aspiration factor is continuously identical to $h$.) We will evaluate later in the proof under which conditions such transition is feasible. We will also introduce some additional notation, similarly to the one of Lemma~\ref{Lm:OneStepTransitionProbabilitySatisfactory}. 
Specifically, let us define $$\hat{D}_{j,\ell}(\alpha')\df \{(\alpha,x,\rho)\in\cZ:x_{j\alpha_j'}>1-\hat{H}^\ell\},$$ where $\hat{H}\df 1-\epsilon{h}$. Define also the set $$\hat{E}_{j,\ell}(\alpha')\df \hat{D}_{j,\ell+1}(\alpha')^c\cap \hat{D}_{j,\ell}(\alpha'),$$ which corresponds to the area in the strategy space of agent $j$ that could be reachable at step $\ell+1$ (under the condition of aspiration factor $\phi_j\equiv{h}$) but it is not reachable at step $\ell$, given that the same action $\alpha'$ is played continuously starting from $z'$ and aspiration factor $h>0$ is applied. Note that, by definition of $\hat{\uptau}_\delta(\epsilon)$, we have
\begin{equation*}    
\hat{\uptau}_\delta(\epsilon) = \left\lceil{ \frac{\log(\delta)}{\log(\hat{H})} }\right\rceil = \frac{\log(\delta)}{\log(\hat{H})} + \hat{\vartheta}(\delta,\epsilon)\,, 
\end{equation*}
for some constant $\hat{\vartheta}=\hat{\vartheta}(\delta,\epsilon)\in[0,1)$. Note that 
\begin{equation}	\label{eq:OneStepTransitionProbabilityUnsatisfactory:Condition_H1}
    \lim_{\epsilon\downarrow{0}} \log\left(\hat{H}^{\hat{\uptau}_{\delta}}\right) = \lim_{\epsilon\downarrow{0}}\left[\left( \frac{\log(\delta)}{\log(\hat{H})}+\hat{\vartheta}(\delta,\epsilon) \right) \log(\hat{H}) \right] = \log(\delta)\,, 
\end{equation}
given that $\lim_{\epsilon\downarrow{0}}\log(\hat{H})=0$. Furthermore, due to the continuity of the natural logarithm, 
\begin{equation} 	\label{eq:OneStepTransitionProbabilityUnsatisfactory:Condition_H2}
\lim_{\epsilon\downarrow{0}} \hat{H}^{\hat{\uptau}_{\delta}} = \delta\,, \quad \lim_{\epsilon\downarrow{0}}\hat{H}^{\ell \hat{\uptau}_{\delta}}=\delta^{\ell} \,, \quad \forall \ell\in\mathbb{N}.
\end{equation}
Note that one event under which a transition from $s$ to $\Neighx{\delta}{x}{s'}$ can be realized is when action profile $\alpha'$ is played continuously for at least $\hat{\uptau}_{\delta}(\epsilon)$ number of times and under the condition that $\snoise=\snoise(\delta)<\delta$ (cf.,~Proposition~\ref{Pr:SM:ApproximationFirstHittingTime} of SM). Using the Markov property, we have:
\begin{equation}		\label{eq:OneStepTransitionProbabilityUnsatisfactory:Eq0} 
\breve{P}_{ss'}(\delta,\epsilon,\snoise) \geq \Prob_{z'}\left[ \alpha(t+1)=\alpha'\,, \forall t < \hat{\uptau}_{\delta}(\epsilon) \right]\cdot \inf_{z\in \Neighx{\delta}{x}{s'}}\Prob_{z}[B_{\infty}]\,.
\end{equation}
Recall that $$B_t \df \{\omega\in\Omega:\alpha(\tau)=\alpha(0)\,, \forall 0\leq\tau\leq{t}\}.$$
Given that actions are selected only based on the strategy vector, and independently of the aspiration level, note that $$\lim_{\delta\downarrow{0}} \inf_{z\in \Neighx{\delta}{x}{s'}}\Prob_{z}[B_\infty] = 1.$$
Using again the Markov property, we have
\begin{eqnarray*}	
\lefteqn{\Prob_{z'}\left[ \alpha(t+1)=\alpha'\,, \forall t < \hat{\uptau}_{\delta}(\epsilon) \right]}\cr 
& \geq & \prod_{t<\hat{\uptau}_{\delta}(\epsilon)} \inf_{z\in \hat{E}_{j,t}(\alpha')}\Prob_{z'}[\alpha(t+1)=\alpha'|Z_t=z] \\ 
& = & \prod_{t<\hat{\uptau}_{\delta}(\epsilon)} (1-\hat{H}_j^{t+1})\,. \nonumber
\end{eqnarray*}
Using the properties of the natural logarithm (similarly to the proof of Lemma~\ref{Lm:OneStepTransitionProbabilitySatisfactory} of MD), 
and by taking $\snoise=\snoise(\delta)<\delta$ and $\delta\downarrow{0}$,
\begin{eqnarray}		\label{eq:OneStepTransitionProbabilityUnsatisfactory:Eq2}
\lefteqn{(1-\hat{H})\log\left(\breve{P}_{ss'}(\delta,\epsilon,\snoise)\right) } \cr 
& \geq & (1-\hat{H}) \sum_{t=1}^{\hat{\uptau}_{\delta}}\log\left(1-\hat{H}^{t} \right) \cr
& = & -\sum_{\ell=1}^{\infty}\frac{1}{\ell}\Big[(1-\hat{H})\sum_{t=1}^{\hat{\uptau}_\delta}\hat{H}^{\ell t}\Big] \cr 
& = & -\sum_{\ell=1}^{\infty}\frac{1}{\ell}\Big[(1-\hat{H})\frac{1-\hat{H}^{\ell(\hat{\uptau}_\delta+1)}}{1-\hat{H}^\ell}-(1-\hat{H})\Big] \cr
& = & -\sum_{\ell=1}^{\infty}\frac{1}{\ell}\Big[\frac{1-\hat{H}^{\ell(\hat{\uptau}_\delta+1)}}{1+\hat{H}+\cdots + \hat{H}^{\ell-1}} -(1-\hat{H})\Big].
\end{eqnarray}
Note that, for any $\ell\in\mathbb{N}$, $\hat{H}^{\ell}\to{1}$ as $\epsilon\downarrow{0}$, thus $\lim_{\epsilon\downarrow{0}}(1+\hat{H}+\cdots+\hat{H}^{\ell-1}) = \ell$. Hence, we have that as $\delta\downarrow{0}$
\begin{eqnarray*}		
\lefteqn{\lim_{\epsilon\downarrow{0}}(1-\hat{H}) \log\left(\breve{P}_{ss'}(\delta,\epsilon,\snoise) \right) } \cr & \geq & \lim_{\epsilon\downarrow{0}}(1-\hat{H})\sum_{t=1}^{\hat{\uptau}_{\delta}(\epsilon)}\log\left(1-\hat{H}^{t}\right) = -\sum_{\ell=1}^{\infty}\frac{1}{\ell^2}(1-\delta^{\ell}) \equiv - \eta(\delta)\,.
\end{eqnarray*}
Using the fact that $1-\hat{H}=\epsilon h$, we conclude that, as $\epsilon\downarrow{0}$ and $\delta\downarrow{0}$, 
\begin{equation*}  
(1-\hat{H}) \log\left( \breve{P}_{ss'}(\delta,\epsilon,\snoise) \right) \geq -\eta(\delta).
\end{equation*}
Equivalently, for sufficiently small $\epsilon>0$ and $\delta>0$ and for $\snoise\leq\delta$, we have (due to the continuity of the natural logarithm)
\begin{equation} \label{eq:OneStepTransitionProbabilityUnsatisfactory:LowerBound} 
\breve{P}_{ss'}(\delta,\epsilon,\snoise) \geq \exp\left(-\frac{\eta(\delta)}{\epsilon h} \right)\,, 
\end{equation}
where the constant $\eta(\delta)$ has the property that $$\eta^* \df \lim_{\delta\downarrow{0}}\eta(\delta) = \lim_{\delta\downarrow{0}}\sum_{\ell=1}^{\infty}\frac{1}{\ell^2}(1-\delta^{\ell}) = \sum_{\ell=1}^{\infty}\frac{1}{\ell^2} < \infty.$$

\textit{In the second part of the proof, we will derive an upper bound for the approximation of the transition probability.} When action profile $\alpha'$ is selected continuously under the unperturbed dynamics starting from $z'$, then the aspiration level evolves as follows (as shown in Proposition~\ref{Pr:SM:ConstantActionSelection} of SM) for any $t>0$:
\begin{equation}		\label{eq:OneStepTransitionProbabilityUnsatisfactory:Condition1}
\rho_j(t) \geq u_j(\alpha') - X(\epsilon)^{t}\cdot\Delta{u}_{j}(\alpha') - Y(\epsilon)^{t}\snoise + \snoise\,, \quad \mbox{a.s.}\,,
\end{equation} 
where $\Delta{u}_j(\alpha) \df u_j(\alpha') - u_j(\alpha) < 0$, $X(\epsilon) \df 1-\epsilon\nu(\epsilon)$ and $Y(\epsilon) \df 1+\epsilon\nu(\epsilon)$. Recall also the definition of $\xi\df \nicefrac{-(1-h)}{\zeta}$. Note that, under the condition that 
\begin{equation}	\label{eq:OneStepTransitionProbabilityUnsatisfactory:Condition2}
X(\epsilon)^t\Delta{u_j}(\alpha')+Y(\epsilon)^{t}\snoise \leq \xi < 0,
\end{equation}
the utility measurement satisfies $$\rewardpert{j}(\alpha')-\rho_j(t)\leq u_j(\alpha')+\snoise-\rho_j(t) \leq \xi <0\,,\quad \mbox{a.s.}\,,$$ which further implies that the aspiration factor of agent $j$ satisfies $\phi_j \equiv h>0$ a.s.\,.

We will derive the maximum time instance $t$ for which the inequality (\ref{eq:OneStepTransitionProbabilityUnsatisfactory:Condition2}) holds. Let us define the following function: $$F(t,\epsilon,\snoise) \df X(\epsilon)^t\Delta{u_j}(\alpha')+Y(\epsilon)^{t}\snoise - \xi <0\,.$$ We first compute the partial derivative of $F$ with respect to $t$, denoted $\partial_{t}F$, that is:
\begin{equation}
\partial_t{F(t,\epsilon,\snoise)} = X(\epsilon)^t \log(X(\epsilon)) \Delta{u}_j(\alpha') + Y(\epsilon)^t\log(Y(\epsilon))\snoise > 0,
\end{equation}
which implies that $F(t,\cdot,\cdot)$ increases with respect to $t$, therefore there will be a time $\tau=\tau(\epsilon,\snoise)$ at which $F(\tau,\epsilon,\snoise)\equiv{0}$.

By considering small $\epsilon>0$, and by Taylor-series expansion of the natural logarithm\footnote{The Taylor-series expansion of the natural logarithm, we have that for small $x>0$, $\log(1+x) \approx x + O(x^2)$, we have $$\log(1+x) \approx - \log(1-x) + O(x^2).$$ As a direct implication, note that for any $t>0$, $$(1+x)^t \approx (1-x)^{-t} + O(t x^2).$$
} we can use the approximation\footnote{We will revisit this approximation when we find how $\tau(\epsilon,\xi)$ scales with $\epsilon$.}
\begin{equation} \label{eq:OneStepTransitionProbabilityUnsatisfactory:Approximation01}
Y(\epsilon)^{\tau} \approx X(\epsilon)^{-\tau} + O\left(\tau( \epsilon\nu(\epsilon))^2\right).
\end{equation}
So, for small $\epsilon$, the condition $F(t,\epsilon,\snoise)=0$ becomes
\begin{equation*}
\Delta{u}_j(\alpha')X(\epsilon)^{\tau}  + \snoise X(\epsilon)^{-\tau} - \xi = 0
\end{equation*}
or equivalently
\begin{equation*}
\Delta{u}_j(\alpha') X(\epsilon)^{2\tau}  - \xi X(\epsilon)^{\tau} + \snoise = 0.
\end{equation*}
The above quadratic identity accepts two solutions, where the meaningful one ($X(\epsilon)^{\tau}>0$) is the largest one (given also that $\Delta{u}_j(\alpha')<0$ and $\xi<0$), that is
\begin{equation*}
X(\epsilon)^{\tau} = \frac{\xi-\sqrt{\xi^2-4\snoise\Delta{u}_j(\alpha')}}{2\Delta{u}_j(\alpha)} \df X^*(\snoise) > 0,
\end{equation*}
which implies that
\begin{equation*}
\tau(\epsilon,\snoise) = \frac{\log X^*(\snoise)}{\log X(\epsilon)} >0.
\end{equation*}
Using the Taylor series expansion of the natural logarithm,\footnote{The Taylor series expansion of the natural logarithm is $$\log(1+x) \approx x - \frac{x^2}{2} + \frac{x^3}{3} - O(x^4).$$} we have the following approximation
\begin{eqnarray*}
\log X(\epsilon) & = & \log(1-\epsilon\nu(\epsilon)) \cr & \approx & -\epsilon\nu(\epsilon) - \frac{(\epsilon\nu(\epsilon))^2}{2} - \frac{(\epsilon\nu(\epsilon))^3}{3} -  O\left((\epsilon\nu(\epsilon))^4\right) \cr
& = & -\epsilon\nu(\epsilon)\left( 1 + \frac{\epsilon\nu(\epsilon)}{2} + \frac{(\epsilon\nu(\epsilon))^2}{3} + O\left((\epsilon\nu(\epsilon))^3\right) \right).
\end{eqnarray*}
and by using the infinite Geometric series by defining $$r \df - \epsilon\nu(\epsilon)/2 - (\epsilon\nu(\epsilon))^2/3 - O\left((\epsilon\nu(\epsilon))^3\right)$$ we can write
\begin{eqnarray*}
\frac{1}{1-r} & = & 1 + r + r^2 + O(r^3) \cr 
& = & 1 + \left(-\frac{\epsilon\nu(\epsilon)}{2} - \frac{(\epsilon\nu(\epsilon))^2}{3} \right) + \left(-\frac{\epsilon\nu(\epsilon)}{2} - \frac{(\epsilon\nu(\epsilon))^2}{3} )\right)^2 + O(r^3) \cr
& = & 1 - \frac{\epsilon\nu(\epsilon)}{2} - \frac{(\epsilon\nu(\epsilon))^2}{12} + O\left((\epsilon\nu(\epsilon))^3\right)\,.
\end{eqnarray*}
Thus, we have 
\begin{eqnarray*}
\frac{1}{\log X(\epsilon)} & \approx & -\frac{1}{\epsilon\nu(\epsilon)}\left(1-\frac{\epsilon\nu(\epsilon)}{2}-\frac{(\epsilon\nu(\epsilon))^2}{12} + O((\epsilon\nu(\epsilon))^3)\right) \cr 
& = & -\frac{1}{\epsilon\nu(\epsilon)} + \frac{1}{2} - \frac{\epsilon\nu(\epsilon)}{12} + O\left((\epsilon\nu(\epsilon))^2\right)
\end{eqnarray*}
Hence, we conclude that 
\begin{equation}  \label{eq:OneStepTransitionProbabilityUnsatisfactory:Approximation02}
\tau(\epsilon,\snoise) \approx -\frac{\log X^*}{\epsilon\nu(\epsilon)} + \frac{\log X^*}{2} - \frac{\epsilon\nu(\epsilon)}{12}\log X^* + O\left((\epsilon\nu(\epsilon))^2\log X^*\right) \quad \mbox{a.s.}\,.
\end{equation}

Given that we have an approximation of the threshold time $\tau(\epsilon,\snoise)$, we can now derive an approximation of the corresponding strategy at that time. Let us denote $\chi \df x_{j\alpha_j'}(\tau(\epsilon,\snoise))$. Under the condition that action profile $\alpha'$ has been played continuously for $\tau(\epsilon,\snoise)$ number of steps, then we have that
\begin{equation}	 \label{eq:OneStepTransitionProbabilityUnsatisfactory:tau2_2}
\tau(\epsilon,\snoise) = \frac{\log(1-\chi)}{\log\hat{H}(\epsilon)}. 
\end{equation}
We again here treat $\tau(\epsilon,\snoise)$ as a continuous variable, given that we want to evaluate its limiting behavior as $\epsilon\downarrow{0}$.  Equivalently, we have that
\begin{equation}
\log(1-\chi) =  \tau(\epsilon,\snoise)\cdot \log\hat{H}(\epsilon) 
\end{equation}
Let us investigate the limiting values of $\chi$ as $\epsilon\downarrow{0}$. By using the Taylor series expansion of the natural logarithm, we have
\begin{equation} \label{eq:OneStepTransitionProbabilityUnsatisfactory:Approximation03}
\log \hat{H}(\epsilon) = \log (1-\epsilon{h}) \approx -\epsilon{h} - \frac{(\epsilon{h})^2}{2} + O(\epsilon^3).
\end{equation}
From (\ref{eq:OneStepTransitionProbabilityUnsatisfactory:Approximation02})--(\ref{eq:OneStepTransitionProbabilityUnsatisfactory:Approximation03}), we have that 
\begin{equation*}
\log(1-\chi) \approx \frac{\log(X^*)}{\nu(\epsilon)}h  - \frac{\log(X^*)}{2}\epsilon{h} + \frac{\log(X^*)}{\nu(\epsilon)}\epsilon{h^2} + O\left(\epsilon^2\log(X^*)\right)\,, \quad \mbox{a.s.}\,.
\end{equation*}
Given also that $\nu(\epsilon)\to{0}$ as $\epsilon\downarrow{0}$, then $\log(1-\chi(\epsilon,\snoise))\to -\infty$ which further implies that $\chi(\epsilon,\snoise)\to{1}$. In other words, independently of the aspiration level update and the values of $\epsilon,h>0$, the strategy value $\chi$ increases with time almost surely (with respect to the noise distribution). 

Recall that $\chi$ has been defined as the strategy of player $j$ such that, when action profile $\alpha'$ is played continuously for $\tau(\epsilon,\snoise)$ steps, then $\phi_j\equiv{h}$ for all $t\leq \tau(\epsilon,\snoise)$ a.s.\,. Furthermore, we have shown that $\tau(\epsilon,\snoise)\to\infty$ and $\chi\to{1}$ as $\epsilon\downarrow{0}$. In other words, from condition (\ref{eq:OneStepTransitionProbabilityUnsatisfactory:tau2_2}), we have that for any $\delta>0$ we may select sufficiently small $\epsilon=\epsilon(\delta)$ such that $\log(1-\chi) \geq \log(\delta)$. Under this condition, (\ref{eq:OneStepTransitionProbabilityUnsatisfactory:tau2_2}) becomes
\begin{equation}	\label{eq:OneStepTransitionProbabilityUnsatisfactory:tau2_4}
\tau(\epsilon,\snoise) > \left\lceil\frac{\log\delta}{\log\hat{H}(\epsilon)} \right\rceil \equiv \hat{\uptau}_{\delta}(\epsilon)\,, \quad \mbox{a.s.}\,.
\end{equation}
From Equations~(\ref{eq:OneStepTransitionProbabilityUnsatisfactory:tau2_2})--(\ref{eq:OneStepTransitionProbabilityUnsatisfactory:tau2_4}) we have that
$$\log\delta > \log(1-\chi) - \hat{\vartheta}(\delta,\epsilon)\cdot\log\hat{H}(\epsilon),$$ for some constant $\hat{\vartheta}(\delta,\epsilon)\in[0,1)$. For a given $\delta>0$, when we take $\epsilon>0$ sufficiently small, we have $\log\hat{H}(\epsilon)\to{0}$ as $\epsilon\downarrow{0}$, and $\delta > 1-\chi$. Under this condition, the strategy reaches $\Neighx{\delta}{x}{s'}$ at time $\hat{\uptau}_{\delta}(\epsilon)$ with the aspiration factor accepting its smallest value $\phi_j\equiv{h}$ a.s.\,. Note that this is always possible for sufficiently small $\epsilon$, given that as $\epsilon\downarrow{0}$, $\chi(\epsilon,\snoise)\to{1}$. 

One possibility of realizing a path from $z'$ to $\Neighx{\delta}{x}{s'}$ is by continuously playing action $\alpha'$ under the above conditions, thus we have 
\begin{eqnarray}		\label{eq:OneStepTransitionProbabilityUnsatisfactory:Proof:Eq1}
\lefteqn{\breve{P}_{ss'}(\delta,\epsilon,\snoise)} \cr & \leq &  \Prob_{z'}\left[ \alpha(t+1)=\alpha'\,, \forall t < \hat{\uptau}_{\delta}(\epsilon) \right] \cr & \leq &
\prod_{t < \hat{\uptau}_\delta} \sup_{z\in \hat{E}_{j,t}(\alpha')}\Prob_{z'}\left[\alpha(t+1)=\alpha'|Z_{t}=z\right] \cr & \leq &  \prod_{t=1}^{\hat{\uptau}_\delta}\left(1-\hat{H}^t\right),
\end{eqnarray}
where in the last inequality we have used Proposition~\ref{Pr:SM:ConstantActionSelection} of SM. 
Following the exact same steps as in the derivation of the lower bound in (\ref{eq:OneStepTransitionProbabilityUnsatisfactory:Eq2}), we can show that as $\epsilon\downarrow{0}$, 
\begin{eqnarray}	
\lefteqn{(1-\hat{H}) \log\left( \breve{P}_{ss'}(\delta,\epsilon,\snoise) \right) } \cr & \leq & (1-\hat{H})\sum_{t=1}^{\hat{\uptau}_\delta(\epsilon)}\log\left(1-\hat{H}^{t}\right)  \approx -\sum_{\ell=1}^{\infty}\frac{1}{\ell^2}\left(1-\delta^{\ell}\right) \equiv - \eta(\delta) < 0 \,.
\end{eqnarray}
almost surely (with respect to the influence of the noise in the definition of the threshold time $\tau$ until which the lower aspiration factor applies, $\phi_j\equiv{h}>0$). Hence, using the fact that $1-\hat{H}(\epsilon)=\epsilon{h}$, we conclude that for sufficiently small $\epsilon$, and for $\snoise\leq\delta$, we have (due to the continuity of the natural logarithm)
\begin{equation} 	\label{eq:OneStepTransitionProbabilityUnsatisfactory:UpperBound}
\breve{P}_{ss'}(\delta,\epsilon,\snoise) \leq \exp\left(-\frac{\eta(\delta)}{\epsilon h} \right)\,.
\end{equation}
Thus, from Equations~(\ref{eq:OneStepTransitionProbabilityUnsatisfactory:LowerBound}), (\ref{eq:OneStepTransitionProbabilityUnsatisfactory:UpperBound}), the conclusion follows.

\section{Proof of Proposition~\ref{Pr:PLAStochasticStability2x2game} (Stochastic Stability of PLA in Coordination Games)} \label{sec:SM:PLAStochasticStability2x2game}

We consider $2$-player $2$-action coordination games of Table~\ref{Tb:CoordinationGame}. When we select $a-c\geq d-b$, then the coordination game corresponds to the so-called Stag-Hunt game (cf.,~\cite{Skyrms_2014}).

\begin{table}[h]
\centering
\begin{game}{2}{2}
      & A    & B\\
A   &$a,a$   &$b,c$\\
B   &$c,b$   &$d,d$
\end{game}
\caption{Coordination game with $a>c>0$, $d>b>0$, and $a>d$.} \label{Tb:CoordinationGame}
\end{table}

In any one-step transition, the minimum hitting time to a $\delta$-neighborhood of a p.s.s. $s$ only depends on the reward received by the single player whose action changes during the transition. For example, in the coordination game of Table~\ref{Tb:CoordinationGame}, the first hitting time to $\Neigh{\delta}{s_{(B,B)}}$ when starting from action $(A,B)$ only depends on the payoff of the row player received at the destination state $(B,B)$, which is equal to $d$. Thus, for the remainder of the proof, and in order to simplify notation, define: 
\begin{equation*}
p(u_{j(s,s')}) \df \breve{P}_{ss'}(\delta,\epsilon,\snoise),
\end{equation*} 
where $j(s,s')$ denotes the agent whose action changed along the transition from $s$ to $s'$. Note that these quantities can directly determine the resistance of the corresponding transition (from $s$ to $s'$) according to Definition~\ref{def:Resistance}. In the remainder of the proof, we will compute the minimum resistance of the four p.s.s.'s of the coordination game, based on which the stochastic stability is determined according to Theorem~\ref{Th:StochasticallyStableStatesMinimumResistance} . 
\begin{figure}[t!]
\centering
\includegraphics[scale=1]{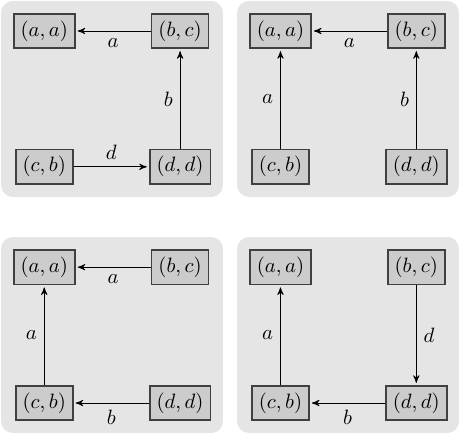}
\caption{One-step $s_{(A,A)}$-graphs and payoff change.}
\label{fig:OneStepPayoffGraph2x2}
\end{figure}

To demonstrate this minimization, let us consider first the computation of the resistance of $s_{(A,A)}$. Figure~\ref{fig:OneStepPayoffGraph2x2} depicts the three (one-step) $\{s_{(A,A)}\}$-graphs and the corresponding payoff change perceived by the player performing the change in action. The minimum resistance of the state $s_{(A,A)}$ is given by the following minimization over all possible $\{s_{(A,A)}\}$-graphs depicted in Figure~\ref{fig:OneStepPayoffGraph2x2},
\begin{equation*}
r_{s_{(A,A)}}^* \df \min\left\{p(a)+p(b)+p(d), 2p(a)+p(b), 2p(a)+p(b)\right\} = 2p(a) + p(b),
\end{equation*}
which is due to the fact that $a>d\Rightarrow p(a)<p(d)$. Accordingly, it is straightforward to check that 
\begin{eqnarray*}
r_{s_{(A,B)}}^* = r_{s_{(B,A)}}^* & \df & p(a) + p(b) + p(c), \cr
r_{s_{(B,B)}}^* & \df & p(a) + p(d) + p(c).
\end{eqnarray*}
By Theorem~\ref{Th:StochasticallyStableStatesMinimumResistance}, we know that for the computation of the stationary distribution $\hat{\pi}$, it suffices to compare the resistances of the p.s.s.'s. 
In particular, we have:
\begin{equation*}
\Delta{r}_1\df r_{s_{(A,B)}}^*-r_{s_{(B,B)}}^* = p(b) - p(d) > 0,
\end{equation*}
\begin{equation*}
\Delta{r}_2\df r_{s_{(A,B)}}^*- r_{s_{(A,A)}}^* = p(c) - p(a) > 0.
\end{equation*}
According to Theorem~\ref{Th:StochasticallyStableStatesMinimumResistance}, we conclude that $s_{(A,B)}$ and $s_{(B,A)}$ are not stochastically stable, since the resistances of $s_{(A,A)}$ and $s_{(B,B)}$ are strictly larger. 

In the remainder of the proof, we compare the minimum resistances of the states $s_{(A,A)}$ and $s_{(B,B)}$. This can be determined by the difference $\Delta{r}_1 - \Delta{r}_2 = r_{s_{(A,A)}}^* - r_{s_{(B,B)}}^*$. If $\Delta{r}_1 - \Delta{r}_2 < 0$, then this implies that $r_{s_{(A,A)}}^* < r_{s_{(B,B)}}^*$, i.e., $s_{(A,A)}$ is the stochastically stable state. If, instead, $\Delta{r}_1 - \Delta{r}_2 > 0$, then $s_{(B,B)}$ will be the stochastically stable state.

(a) Let $a-c < d-b$. We will first investigate the case where $d\leq{c}$. We have that:
\begin{eqnarray*}
\lefteqn{\Delta{r}_1 - \Delta{r}_2}\cr 
& = & p(b)-p(d)-p(c)+p(a) \cr 
& = & p'(y_1)(b-d) - p'(y_2)(c-a)
\end{eqnarray*}
for some $y_1\in(b,d)$ and $y_2\in(c,a)$. The last equality is an implication of the Mean-Value Theorem (cf.,~\cite[Theorem~5.10]{Rudin64}), where
$$p'(y)\df\frac{dp(y)}{dy}=\frac{\log(\delta)}{(1-\epsilon{y})\log(1-\epsilon{y})^2}<0$$ with $p'(y)\to-\infty$ as $\epsilon\to{0}$. Furthermore, note that $p'(y_1)(b-d) > p'(y_1)(c-a)$ given that $a-c<d-b$ and $p'(y_1)<0$. Thus, 
\begin{equation*}
\Delta{r}_1 - \Delta{r}_2 > \left(p'(y_1) - p'(y_2)\right)(c-a)
\end{equation*}
Given that $d\leq{c}$, then $y_1 < y_2$, and by applying again the Mean-Value Theorem, there exists $y_3\in(y_1,y_2)$ such that 
\begin{equation*}
\Delta{r}_1 - \Delta{r}_2 > p''(y_3) (y_1-y_2)(c-a) > 0,
\end{equation*}
since 
\begin{equation*}
p''(y)\df \frac{d^2p(y)}{dy^2}=\frac{\log(\delta)}{(1-\epsilon{y})^2\log(1-\epsilon{y})^2}+ \frac{2\log(\delta)}{(1-\epsilon{y})^2\log(1-\epsilon{y})^3}>0
\end{equation*}
with $p''(y)\to\infty$ as $\epsilon\to{0}$. We conclude that $s_{(B,B)}$ is the unique stochastically stable state as $\delta\downarrow{0}$ and $\epsilon\downarrow{0}$. This was shown under the condition of $d\leq{c}$, however, note that the exact same proof holds if we switch $d$ with $c$.

(b) Let us consider the case where $a-c > d-b$ and $c\leq{b}$. Following similar steps as in the case (a), we have
\begin{eqnarray*}
\lefteqn{\Delta{r}_1 - \Delta{r}_2}\cr 
& = & p(b)-p(d)-p(c)+p(a) \cr 
& = & p'(y_1)(a-d) - p'(y_2)(c-b) \cr
& \leq & p'(y_1)(c-b) - p'(y_2)(c-b) \cr
& = & p''(y_3)(y_1-y_2)(c-b) \cr
& \leq & 0,
\end{eqnarray*}
for some $y_1\in(d,a)$, $y_2\in(c,b)$ and $y_3\in(y_1,y_2)$. According to Theorem~\ref{Th:StochasticallyStableStatesMinimumResistance}, $s_{(A,A)}$ is the unique stochastically stable state.

\section{Simulations in Load-balancing games in CPU-pinning}  \label{sec:SM:SimulationsLoadBalancingGames}

\subsection{Example}

As we discussed in Section~\ref{sec:SM:LoadBalancingGames} of SM, load balancing games are characterized by a multiplicity of pure Nash equilibria. Thus, the question of characterizing the stochastically stable states under either PLA or APLA becomes relevant. Of high interest it is also the evaluation of the complexity in the computation of the stochastically stable states, especially in comparison with the standard PLA dynamics. In this section, we will consider a standard example based on the definitions of Section~\ref{sec:SM:LoadBalancingGames}. 

\begin{example}	\label{ex:PinningExample1}
There are four computing threads or tasks, namely $T_1$, $T_2$, $T_3$ and $T_4$, to be executed by two available CPU cores, namely $b_1$ and $b_2$ with equal speeds. The first two tasks ($T_1$ and $T_2$) are larger tasks of weights (or loads) $w_1=w_2=\nicefrac{3}{2}$, while the other two tasks are smaller of loads $w_3=w_4=1$. 
\end{example}

In Figure~\ref{fig:SM:LoadBalancingExample_WGraphs_Efficient} of SM, we provide a visualization of possible assignments of loads to machines organized in different levels ($A$, $B$, and $C$), reflecting the number of tasks running on the same CPU core. More specifically, at level $A$ all four tasks are allocated on the same core, at level $B$ at maximum three tasks are allocated on the same core, and at level $C$ at maximum two tasks are allocated on the same core. The allocations at level $C$ are of special interest, since (as it will be shown later) they correspond to the pure Nash equilibria of the game. We will denote by $C.1$ the allocation where the large tasks are allocated separately from the small tasks (see, e.g., in Figure~\ref{fig:SM:LoadBalancingExample_WGraphs_Efficient} where $T_1$ and $T_2$ are together allocated in CPU $b_1$). On the other hand, the allocations $C.2$ and $C.3$ are those where in each CPU core a large task is assigned together with a small task.

Note that for each allocation of tasks, there is also a symmetric one, where the CPU cores of the allocation decisions have been switched off. For example, for $C.1$, where $\{T_1,T_2\}$ are allocated to CPU $b_1$ and $T_3,T_4$ are allocated to CPU $b_2$, the symmetric one $C.1'$ is such that $\{T_1,T_2\}$ are allocated to CPU $b_2$ and $\{T_3,T_4\}$ are allocated to CPU $b_1$.

\subsection{Stochastically Stable States}

\begin{figure}[th!]  
\centering
\includegraphics[width=0.7\textheight,keepaspectratio]{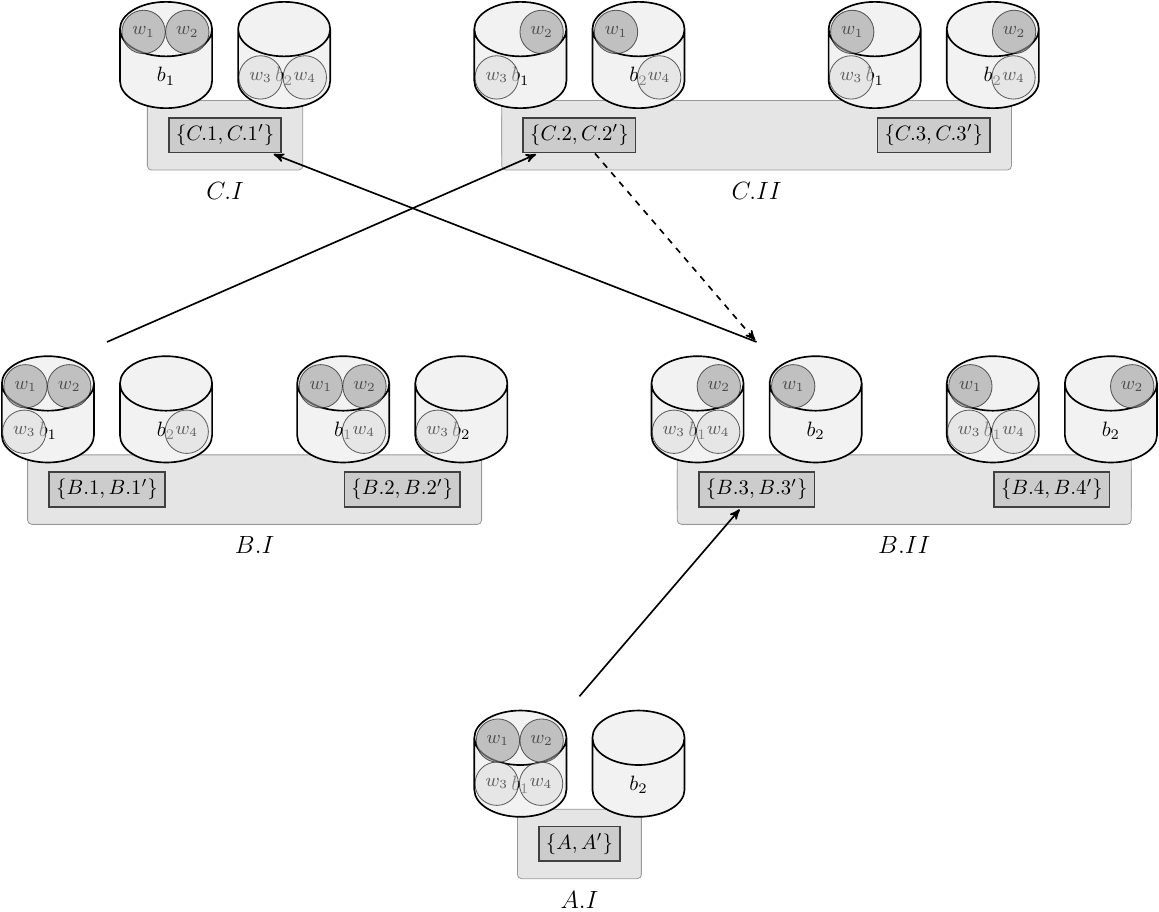}  
\caption{$\{C.I\}$-graph of minimum resistance for the load balancing game.}
\label{fig:SM:LoadBalancingExample_WGraphs_Efficient}
\end{figure}

The visualization of Figure~\ref{fig:SM:LoadBalancingExample_WGraphs_Efficient} also shows a $\{C.1,C.1'\}$-graph (based on the Definition~\ref{def:W-graph_old} of $\mathcal{W}$-graphs of the MD. We will later see that it also corresponds to the $\{C.1,C.1'\}$-graph of minimum-resistance. In this context, the links in this graph indicate transitions between action profiles including a single player making a change in action (moving from one CPU core to another). Note also that in the design of the $\{C.1,C.1'\}$-graph, we have clustered the symmetric allocations (e.g., $C.1$ is grouped with $C.1'$) but also allocations that involve the same type of tasks (e.g., allocations $\{C.2,C.2'\}$ are grouped with allocations $\{C.3,C.3'\}$ since both classes of allocations have the same type of tasks per CPU). Such clustering of allocations is performed in order to simplify the identification the stochastically stable states. It is shown in the following claim that the clustering performed creates an equivalent problem in terms of the calculation of stochastically stable states.

\begin{claim}[Equivalent State Clustering] \label{Cl:SM:EquivalentStateClustering}
Consider the framework of Example~\ref{ex:PinningExample1}. The stochastically stable states of the corresponding load-balancing problem can be equivalently characterized when the p.s.s. states are clustered as follows: 
$C.{I} \df \{C.1,C.1'\}$, $C.{II} \df \{C.2,C.2',C.3,C.3'\}$, $B.{I} \df \{B.1,B.1',B.2,B.2'\}$, $B.{II} \df \{B.3,B.3',B.4,B.4'\}$, $
A.I \df \{A,A'\}$.
\end{claim}
\begin{proof}
According to Lemma~\ref{Lm:StationaryDistribution} of MD, in order to establish equivalence of $\mathcal{W}$-graphs in terms of the characterization of the stochastically stable states, it is sufficient that all transition probabilities between any two given groups of states are the same independently of the source and destination states in those groups. This is due to the fact that stochastically stable states are characterized by state transitions paths of minimum resistance, which implies that if the transition probabilities between any two states of two given groups are the same, then it is sufficient to identify minimum resistance paths with respect to these groups (instead of the independent states). 

For example, when we consider the groups $B.I$ and $C.I$, all possible transitions starting from a state in $B.I$ and ending at a state in $C.I$ will have the same transition probability. (This is due to the fact that any such transition involves moving a small task to a CPU core where only one small task is currently running.) Thus, if such a transition between states of these groups is part of the minimum resistance path, then it implies that only such transitions from the source group $B.I$ to the destination group $C.I$ needs to be considered for structuring the minimum resistance graph. Hence, it is sufficient to define minimum resistance graphs based on these groups of states, instead of the independent states.

To check whether this condition is satisfied for all transitions between groups, note that all possible transitions can be grouped to the following cases:
\begin{itemize}
\item $A.I \to B.I$: Any transition (independently of the destination state) involves a large task moving to an empty CPU core. Thus, the resulting transition probability (which depends only on the utility of the moving task at the destination CPU core) is the same for all possible destination states.
\item $B.I \to C.I$: Any transition (independently of the source and destination state) involves a large task moving to a CPU core where only one large task is currently assigned. Thus, the transition probability is the same for all possible transitions.
\item $B.I \to C.II$: Any transition involves moving a large task to a CPU core where currently only one small task is currently running.
\item $B.II \to C.I$: Any transition involves moving a large task to a CPU core where currently only a large task is currently running.
\item $B.II \to C.II$: Any transition involves moving a small task to another CPU core where currently only a large task is currently running.
\end{itemize}
Given that the transition probabilities from the groups of level $B$ to the groups of level $C$ satisfy this property (of equal transition probabilities between states of the involved groups) the same is also the case for the reversed transitions from the groups of level $C$ to the groups of level $B$. The same is also the same for transitions between the group of level $A$ with the groups of level $B$. As a conclusion, the characterization of the stochastically stable states can be equivalently computed by utilizing the indicated groups of states. 
\end{proof}

\begin{claim}[Nash Equilibria]	\label{Cl:SM:NashEquilibria}
Consider the CPU-pinning Example~\ref{ex:PinningExample1} stated above where four tasks are allocated into two available CPU cores. Let us consider the allocations that correspond to the $C$ level (where maximum two tasks are allocated into each CPU core), i.e., the allocations $C.1$, $C.2$, $C.3$ and their symmetric ones ($C.1'$, $C.2'$ and $C.3'$). All these $C$-level allocations correspond to the pure Nash equilibria of the load balancing game of this example. 
\end{claim}
\begin{proof}
(sketch) Given that the two CPU cores have the same speed, it is straightforward to check that allocations $A$ and $A'$ cannot be Nash equilibria, given that either one of the tasks will have the incentive to move to the empty CPU core. Similar is also the case for the allocations at level $B$. When three tasks are running on the same CPU core, then either one of the tasks will have the incentive to move to the other CPU core (which currently hosts only one task). This is simply because both CPU cores have the same speed and by moving one task (small or big) from the core of 3 tasks to the core of 1 task, the overall machine load that the moving task is experiencing is smaller when it moves to the machine of the single task, i.e., the target CPU core will have smaller overall load compared the overall load of the source CPU core. Finally, at level $C$, where tasks are split into groups of two, when either task moves to the a new core, then it moves to a target CPU node whose overall load grows compared to the source CPU core. As a result, no task will have the incentive to move to the other CPU core. Therefore, all allocations of level $C$ correspond to the pure Nash equilibria of the load balancing game.
\end{proof}

Note that, according to the definition of the resistance of a transition (cf.,~Definition~\ref{def:Resistance} and Lemmas~\ref{Lm:OneStepTransitionProbabilitySatisfactory}--\ref{Lm:OneStepTransitionProbabilityUnsatisfactory} of MD), the agent that performs the change in action determines the resistance of this transition. Furthermore, the resistance of a transition is inversely proportional to the utility of the final assignment (of the involved agent). For example, the resistance of the transition $A \to B.3$ is determined by the utility of task $T_1$ at assignment $B.3$. Having this simple principle in mind, the resistances of the $\mathcal{W}$-graphs can be computed and the graphs with the minimum resistances can be defined. The goal is not to compute the resistances of all possible graphs, rather to characterize the graphs with the minimum resistances, since these will define the set of stochastically stable states as shown by Theorem~\ref{Th:StochasticallyStableStatesMinimumResistance} in MD	. 
As we showed in Claims~\ref{Cl:SM:EquivalentStateClustering}--\ref{Cl:SM:NashEquilibria} above, for the characterization of the stochastically stable states, it is sufficient to consider the minimum resistance graphs of the groups $C.I$ and $C.II$ since these two groups contain the pure Nash equilibria of the load balancing game.

\begin{table}[th!]
\centering
\caption{Utilities experienced under the transitions of the $\{C.I\}$-graph of Figure~\ref{fig:SM:LoadBalancingExample_WGraphs_Efficient} by the involved player when performing the corresponding change in the action profile (from ``source'' $\alpha$ to ``target'' $\alpha'$). }
\begin{tabular}{c *{4}{c}}
\toprule
& \multicolumn{4}{c}{target ($\alpha'$)} \\
\cmidrule(lr){2-5}
source ($\alpha$)
& \shortstack{$B.I$}
& \shortstack{$B.II$}
& $C.I$
& \shortstack{$C.II$} \\
\midrule
$A$ 
    &        & $1$            &        &        \\
\addlinespace
$B.I$ 
    &        &                &       & $\nicefrac{3}{5}$ \\
\addlinespace
$B.II$ 
    &        &                      & $\nicefrac{1}{2}$ & \\
\addlinespace
$C.II$ 
    &    & $\nicefrac{2}{7}$   &        &        \\
\bottomrule
\end{tabular}
\label{tb:SM:ResistancesEfficient}
\end{table}

\begin{claim}[Minimum Resistance Graph of $C.I$] \label{Cl:SM:MinimumResistanceGraph_I}
The minimum resistance graph of the $\{C.I\}$-graphs is given by Figure~\ref{fig:SM:LoadBalancingExample_WGraphs_Efficient} and the corresponding utilities of the involved transitions are depicted in Table~\ref{tb:SM:ResistancesEfficient}.
\end{claim}
\begin{proof}
All transitions from $A$ level to $B$ level have destination utility equal to $1$ independently of the involved task. Thus, the link from level $A$ to $B$ has no influence in the minimum resistance. Regarding the transitions from $B$ level to $C$ level, we need to consider all possible alternative paths leading to $C.I$ such that each state (but $C.I$) is the source node of exactly one link and there is no cycles.. The alternative options (avoiding cycles) will be: a) $B.I\to C.I$, $B.II\to C.I$ and $C.II\to B.I$ with destination utilities $\nicefrac{1}{2}$, $\nicefrac{1}{2}$  and $\nicefrac{3}{8}$, respectively, and total path utility $2\nicefrac{1}{2}+\nicefrac{3}{8}=\nicefrac{11}{8}=1.375$; b) $B.I\to C.I$, $C.II\to B.II$ and $B.II\to C.I$, with utilities $\nicefrac{1}{2}$, $\nicefrac{2}{7}$ and $\nicefrac{1}{2}$, respectively, and with total utility $2\cdot\nicefrac{1}{2}+\nicefrac{2}{7}=\nicefrac{9}{7}=1.286$; c) $B.I\to C.I$, $B.II\to C.II$ and $C.II\to B.I$ with utilities $\nicefrac{1}{2}$, $\nicefrac{2}{5}$ and $\nicefrac{3}{8}$, respectively, and total utility $\nicefrac{1}{2}+\nicefrac{2}{5}+\nicefrac{3}{8}=\nicefrac{51}{40}=1.275$; d) $B.I\to C.II$, $C.II\to B.II$, and $B.II\to C.I$ with utilities $\nicefrac{3}{5}$, $\nicefrac{2}{7}$, and $\nicefrac{1}{2}$, respectively, and total utility $\nicefrac{3}{5}+\nicefrac{2}{7}+\nicefrac{1}{2}=\nicefrac{97}{70}=1.386$. We conclude that option (d) brings the largest total destination utility and defines the minimum resistance path to $C.I$. It is concluded that the graph with the smallest resistance is the one depicted in Table~\ref{tb:SM:ResistancesEfficient} and Figure~\ref{fig:SM:LoadBalancingExample_WGraphs_Efficient}. 
\end{proof}

\begin{figure}[th!]
\centering
\includegraphics[width=0.7\textheight,keepaspectratio]{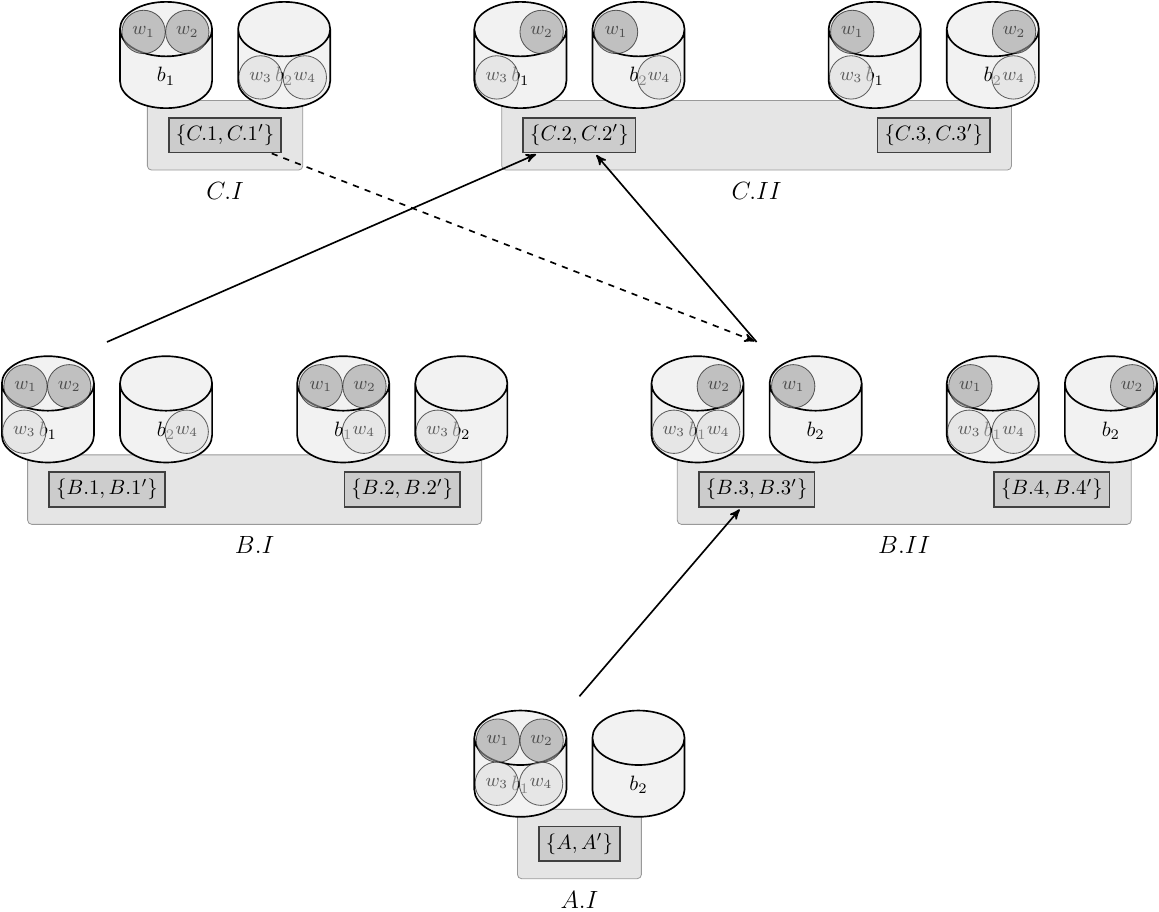}  
\caption{$\{C.II\}$-graph of minimum resistance for the load balancing game.}
\label{fig:SM:LoadBalancingExample_WGraphs_NonEfficient}
\end{figure}

\begin{claim}[Minimum resistance graph for $C.II$] \label{Cl:SM:MinimumResistanceGraph_II}
The minimum resistance graph among the $\{C.II\}$-graphs is given by Figure~\ref{fig:SM:LoadBalancingExample_WGraphs_NonEfficient} and Table~\ref{tb:SM:ResistancesNonEfficient}.
\end{claim}
\begin{proof}
Regarding the transitions from $A$ to $B$ level, we take $A\to B.II$, although any transition from level $A$ to level $B$ brings the exact same destination utility (independently if the involved task is small or large). Regarding the transitions from $B.I$ and $B.II$ to the $C$ level, we need to consider all possible alternative paths leading to $C.II$ such that each state (but $C.II$) is the source node of exactly one link and there is no cycles. The alternative paths include the following cases: a) $C.I\to B.I$, $B.I\to C.II$, and $B.II\to C.II$, with utilities $\nicefrac{2}{8}$, $\nicefrac{3}{5}$, and $\nicefrac{2}{5}$, respectively, and total utility $\nicefrac{2}{8}+\nicefrac{3}{5}+\nicefrac{2}{5}=\nicefrac{50}{40}=1.25$, b) $C.I\to B.I$, $B.I \to C.II$ and $B.II\to C.I$, with utilities $\nicefrac{2}{8}$, $\nicefrac{3}{5}$ and $\nicefrac{1}{2}$, respectively, with total utility $\nicefrac{2}{8}+\nicefrac{1}{2}+\nicefrac{3}{5}=\nicefrac{54}{40}=1.35$, c) $B.I\to C.I$, $C.I\to B.II$ and $B.II\to C.II$, with utilities $\nicefrac{1}{2}$, $\nicefrac{3}{8}$ and $\nicefrac{2}{5}$, respectively, and total utility $\nicefrac{1}{2}+\nicefrac{3}{8}+\nicefrac{2}{5}=\nicefrac{93}{70}=1.329$; d) $C.I\to B.II$, $B.I \to C.II$, and $B.II \to C.II$, with utilities $\nicefrac{3}{7}$, $\nicefrac{3}{5}$, and $\nicefrac{2}{5}$, respectively, with total utilities $\nicefrac{3}{7}+\nicefrac{3}{5}+\nicefrac{2}{5}=\nicefrac{50}{35}=1.428$. Thus, the path with the largest transition utility corresponds to case (d), which corresponds to the graph of Figure~\ref{fig:SM:LoadBalancingExample_WGraphs_NonEfficient} and Table~\ref{tb:SM:ResistancesNonEfficient}.
\end{proof}

\begin{table}[th!]
\centering
\caption{Utilities experienced under the transitions of the $\{C.II\}$-graph of Figure~\ref{fig:SM:LoadBalancingExample_WGraphs_Efficient} by the involved player when performing the corresponding change in the action profile (from ``source'' $\alpha$ to ``target'' $\alpha'$). }
\begin{tabular}{c *{4}{c}}
\toprule
& \multicolumn{4}{c}{target ($\alpha'$)} \\
\cmidrule(lr){2-5}
source ($\alpha$)
& \shortstack{$B.I$}
& \shortstack{$B.II$}
& $C.I$
& \shortstack{$C.II$} \\
\midrule
$A$ 
    &        & $1$            &        &        \\
\addlinespace
$B.I$ 
    &        &                &        & $\nicefrac{3}{5}$ \\
\addlinespace
$B.II$ 
    &        &                &   & $\nicefrac{2}{5}$ \\
\addlinespace
$C.I$ 
    &    &  $\nicefrac{3}{7}$  &        &        \\
\bottomrule
\end{tabular}
\label{tb:SM:ResistancesNonEfficient}
\end{table}

\begin{claim}[Stochastically stable states under PLA] \label{Cl:SM:LoadBalancing:StochasticallyStableStatesPLA}
Under the conditions of Theorem~\ref{Th:StochasticallyStableStatesMinimumResistance} of MD, the Nash equilibria corresponding to the action profiles in the cluster of states $C.II$ are the stochastically stable states under the PLA dynamics (i.e., when the aspiration term satisfies $\phi_i \equiv u_i(\alpha)$ for all players $i\in\cI$ and action profiles $\alpha\in\cA$).
\end{claim}
\begin{proof}
According to Theorem~\ref{Th:StochasticallyStableStatesMinimumResistance}, it is sufficient to compare the aggregate utility of the minimum resistance $\mathcal{W}$-graphs of the Nash equilibria. In the proof of Claims~\ref{Cl:SM:MinimumResistanceGraph_I}--\ref{Cl:SM:MinimumResistanceGraph_II}, we have already calculated the total utilities along the induced transitions of the $\mathcal{W}$-graphs. In the case of the $C.II$ minimum resistance graph, the total utility corresponds to $1.428$ (excluding the transition from level $A$ to level $B$ which is always the same), while in the case of the $C.I$ minimum resistance graph, the total utility of the minimum resistance graph corresponds to $1.386$. In conclusion, the $C.II$ cluster exhibits smaller resistance, and therefore it corresponds to the stochastically stable states under the conditions Theorem~\ref{Th:StochasticallyStableStatesMinimumResistance}.
\end{proof}

The following claim provides the characterization of the stochastically stable states under the APLA dynamics.
\begin{claim}[Stochastically stable states under APLA]	\label{Cl:SM:LoadBalancing:StochasticallyStableStatesAPLA}
Under the conditions of Theorems~\ref{Th:StochasticallyStableStatesMinimumResistance}--~\ref{Th:StationaryDistributionInWeaklyAcyclicGames} of MD, the Nash equilibria corresponding to the action profiles of both the clusters of states $C.I$ and $C.II$ are the stochastically stable states under the APLA dynamics.
\end{claim}
\begin{proof}
Under the APLA dynamics and Lemma~\ref{Lm:OneStepTransitionProbabilityUnsatisfactory} of MD, note that the transition utilities that correspond to unsatisfactory transitions are scaled by $h>0$. This scaling changes the computation of the graph of minimum resistance that we performed in Claims~\ref{Cl:SM:MinimumResistanceGraph_I}--\ref{Cl:SM:MinimumResistanceGraph_II}. In particular, if we take $h>0$ sufficiently small, then unsatisfactory transitions do not influence the computation of the minimum resistance. 

As a result, for the case of $\{C.I\}$-graphs in the proof of Claim~\ref{Cl:SM:MinimumResistanceGraph_I}, the minimum resistance graph corresponds again to case (d) depicted in Figure~\ref{fig:SM:LoadBalancingExample_WGraphs_Efficient}, however the total transition utility is now equal to $\nicefrac{1}{2}+\nicefrac{3}{5} = \nicefrac{11}{10}$ (where we have neglected the transition from $C.II$ to $B.II$ corresponding to the dashed line in Figure~\ref{fig:SM:LoadBalancingExample_WGraphs_Efficient}).

For the case of $\{C.II\}$-graphs in the proof of Claim~\ref{Cl:SM:MinimumResistanceGraph_II}, the minimum resistance graph corresponds to the case (b), which is different than the minimum resistance graph under PLA. Furthermore, the total transition utility in this case, and with sufficiently small $h>0$, is approximated by $\nicefrac{3}{5}+\nicefrac{1}{2}=\nicefrac{11}{10}$ (where we have neglected the transition from $C.I$ to $B.II$ corresponding to the dashed line in Figure~\ref{fig:SM:LoadBalancingExample_WGraphs_Efficient}).

We conclude that in both cases, the total transition utility, under the condition of sufficiently small $h>0$, is the same for both clusters $C.I$ and $C.II$. Therefore, according to Theorem~\ref{Th:StochasticallyStableStatesMinimumResistance}, both clusters $C.I$ and $C.II$ are stochastically stable.    
\end{proof}

\begin{figure}[th!]
\centering
\includegraphics[scale=0.8,keepaspectratio]{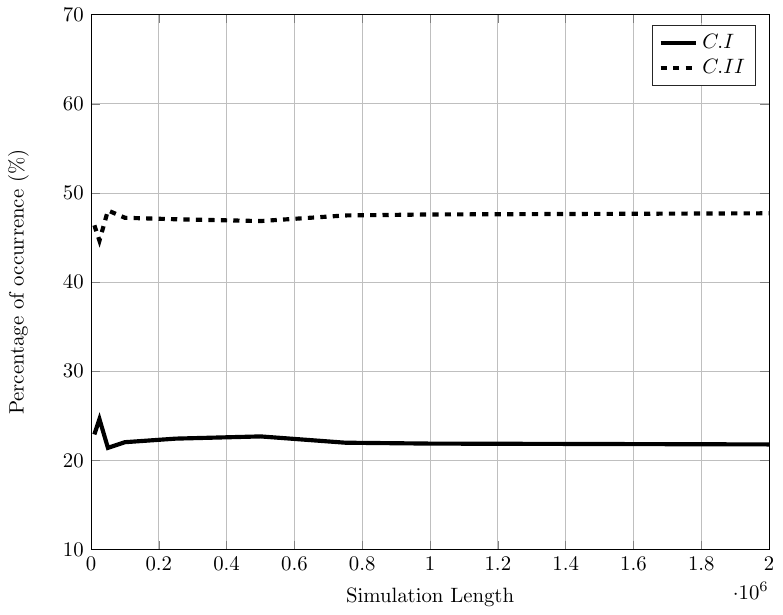}
\caption{Performance of PLA dynamics in load balancing game.}
\label{fig:SM:SimulationLoadBalancing_PLA}
\end{figure}

\begin{figure}[th!]
\centering
\includegraphics[scale=0.8,keepaspectratio]{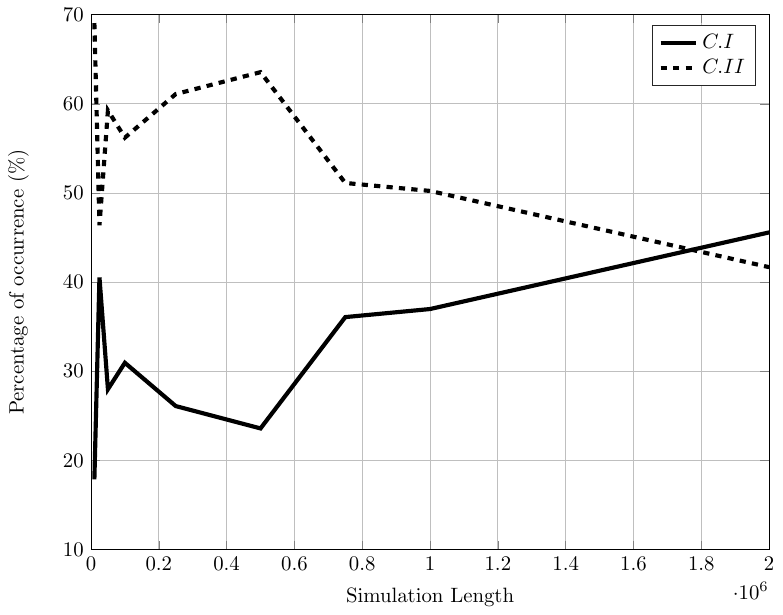}
\caption{Performance of APLA dynamics in load balancing game.}
\label{fig:SM:SimulationLoadBalancing_APLA}
\end{figure}

Claims~\ref{Cl:SM:LoadBalancing:StochasticallyStableStatesPLA}--\ref{Cl:SM:LoadBalancing:StochasticallyStableStatesAPLA} raise several important discussion points. Under the standard PLA dynamics, the set of stochastically stable states correspond to the class $C.{II}$ of Nash equilibria. On the other hand, under the APLA dynamics, both classes $C.I$ and $C.{II}$, which happen to belong to the set of Pareto efficient outcomes, correspond to the set of stochastically stable states. This is because the collection of states in $C.I$ and $C.II$ are absorbing states of the minimum resistance graph within the improvement $\mathcal{W}$-graphs (as shown in Theorem~\ref{Th:AspirationActionFunctionalAndStochasticStability} of MD). Thus, we have both states $C.I$ and $C.II$ equally emerging over time under the APLA dynamics. This is also demonstrated through the simulation results of Figures~\ref{fig:SM:SimulationLoadBalancing_PLA}--\ref{fig:SM:SimulationLoadBalancing_APLA}. In the case of PLA dynamics of Figure~\ref{fig:SM:SimulationLoadBalancing_PLA}, the $C.II$ allocations receive larger weight over time, while in the case of APLA dynamics of  Figure~\ref{fig:SM:SimulationLoadBalancing_APLA}, both $C.I$ and $C.II$ allocations receive equal weight over time.

\subsection{Discussion: Convergence to ``efficient'' profiles}

Note that in this example (Example~\ref{ex:PinningExample1}), all Nash equilibria belong to the absorbing states of the minimum resistance graphs among the improvement $\mathcal{W}$-graphs, $\mathcal{G}_{\rm BR}^*$ (as defined in Theorem~\ref{Th:AspirationActionFunctionalAndStochasticStability} of MD). It also happens that these Nash equilibria also define the set of Pareto profiles. Thus, in this example, there is no immediate benefit with respect to reinforcing convergence to ``desirable'' profiles simply because all Nash equilibria belong to the Pareto profiles and have equal absorbing resistance under APLA. In order for the APLA dynamics to provide an additional benefit in terms of equilibrium selection (versus the standard PLA dynamics), we should have Nash equilibria states with additional utility benefits (as for example we saw in the case of the classical Stag-Hunt game).

It is also worth noting that, as we also discussed in Section~\ref{sec:StochasticStabilityWeaklyAcyclicGames} of MD, the analysis of the computation of stochastically stable outcomes under the APLA dynamics is based on a transience notion, rather than a static notion. For a Nash equilibrium to be stochastically stable under APLA, it would need to be absorbing along the better-reply (improvement) paths. In games with sufficient alignment of interests between agents, this would imply that the Pareto efficient Nash equilibria are stochastically stable (as it was the case in the Stag-Hunt game). However, in general, stochastic stability will need to be evaluated based on the utilities experienced along the improvement paths.

\subsection{Discussion: Computational complexity}

There is however an indirect benefit coming from the implementation of the APLA dynamics, as also discussed in Section~\ref{sec:StochasticStabilityWeaklyAcyclicGames} of MD. The characterization of the stochastically stable states through the minimization of the aspiration action-functional significantly simplifies the standard minimum-resistance optimization. This is due to the fact that standard stochastic stability derivations would require a minimization over all possible $\mathcal{W}$-graphs. However, under the APLA dynamics this computation is restricted to the improvement paths only (which can be characterized through only the better reply transitions). Hence, it provides a framework for characterizing stochastic stability that significantly simplifies the design of games (both with respect to the criteria as well as the computational effort).

\bibliographystyle{abbrv} 

\bibliography{APLA_bibliography}

\end{document}

%% file: macros.tex
\newcommand{\Cc}{C_b}

\newcommand{\cA}{{\mathcal{A}}}
\newcommand{\cX}{{\mathcal{X}}}

\newcommand{\cZ}{{\mathcal{Z}}}
\newcommand{\cS}{{\mathcal{S}}}

\newcommand{\Bor}{{\mathfrak{B}}}
\newcommand{\cI}{\mathcal{I}}
\newcommand{\df}{\doteq}
\newcommand{\reward}[1]{u_{#1}}
\newcommand{\rewardpert}[1]{\tilde{u}_{#1}}

\newcommand{\Prob}{{\mathbb{P}}} 
\newcommand{\Exp}{{\mathbb{E}}} 
\newcommand{\sF}{{\mathfrak{F}}}    

\newcommand{\supnorm}[1]{{\lVert}#1{\rVert}_{\infty}}

\newcommand{\Dirac}[1]{\boldsymbol{\delta}_{#1}}

\newcommand{\ess}{\mathrm{ess}}

\newcommand{\noise}{\upsilon}
\newcommand{\dnoise}{\Upupsilon}
\newcommand{\snoise}{\overline{\upsilon}}

\newcommand{\Neigh}[2]{\mathcal{B}_{#1}(#2)}
\newcommand{\Neighx}[3]{\mathcal{B}_{#1}^{#2}(#3)}

\newcommand{\Res}[3]{r_{#1 #2}(#3)}

\newcommand{\ResGraph}[3]{r_{#1|#2}(#3)}
\newcommand{\ResGraphRed}[2]{r_{#1|#2}}
\newcommand{\MinResGraph}[2]{r_{#1}^*(#2)}
\newcommand{\MinResGraphRed}[1]{r_{#1}^*}

\newcommand{\OSA}[3]{\breve{P}_{#1 #2}(#3)}
\newcommand{\OSAs}[2]{\breve{P}_{#1 #2}}

\newcommand{\tr}{^{\mathrm T}}

\newcommand{\magn}[1]{\left\vert #1 \right\vert}

\newtheorem{property}{Property}

\newtheorem{definition}{Definition}
\newtheorem{proposition}{Proposition}
\newtheorem{theorem}{Theorem}
\newtheorem{lemma}{Lemma}

\newenvironment{proof}{\textit{Proof.}}{\hfill$\square$}
\newtheorem{claim}{Claim}
\newtheorem{example}{Example}